\documentclass[11pt,a4paper]{article}
\usepackage{jheppub}
\usepackage{amsmath}
\usepackage{verbatim}
\usepackage{amssymb}
\usepackage{inputenc, array}
\usepackage{textcomp}
\usepackage{appendix}
\usepackage{amsfonts}
\usepackage{amsthm}
\usepackage{mathrsfs}
\usepackage{tikz}
\usetikzlibrary{decorations.markings}
\usepackage{mathtools}
\usepackage{bm}
\allowdisplaybreaks
\usepackage{relsize}
\usepackage{graphicx}
\usepackage{xifthen}
\usepackage{pict2e}
\allowdisplaybreaks

\makeatletter
\newcommand{\loplus}{\mathbin{\mathpalette\dog@lsemi{+}}}
\newcommand{\dog@lsemi}[2]{\dog@semi{#1}{#2}{270,90}}
\newcommand{\dog@semi}[3]{%
  \begingroup
  \sbox\z@{$\m@th#1#2$}%
  \setlength{\unitlength}{\dimexpr\ht\z@+\dp\z@\relax}%
  \makebox[\wd\z@]{\raisebox{-\dp\z@}{%
    \begin{picture}(1,1)
    \linethickness{\variable@rule{#1}}
    \roundcap
    \put(0.5,0.5){\makebox(0,0){\raisebox{\dp\z@}{$\m@th#1#2$}}}
    \put(0.5,0.5){\arc[#3]{0.5}}
    \end{picture}%
  }}%
  \endgroup
}
\newcommand{\variable@rule}[1]{%
  \fontdimen8  
  \ifx#1\displaystyle\textfont3\else
    \ifx#1\textstyle\textfont3\else
      \ifx#1\scriptstyle\scriptfont3\else
        \scriptscriptfont3\relax
  \fi\fi\fi
}
\makeatother

\newcommand{\oneline}[1]{%
  \newdimen{\namewidth}%
  \setlength{\namewidth}{\widthof{#1}}%
  \ifthenelse{\lengthtest{\namewidth < \textwidth}}%
  {#1}
  {\resizebox{\textwidth}{!}{#1}}
}

\numberwithin{equation}{section}

\def\bea{\begin{eqnarray}}
\def\eea{\end{eqnarray}}
\newcommand{\beq}{\begin{eqnarray}}
\newcommand{\eqq}{\end{eqnarray}}
 \newcommand{\badat}{\begin{alignedat}}
 \newcommand{\eadat}{\end{alignedat}}

\newcommand{\eal}[1]{\be \begin{aligned} #1 \end{aligned}\end{equation}} 
\newcommand{\eqn}[1]{\be #1 \end{equation}} 
\newcommand{\eqa}[1]{\bea  #1\end{eqnarray}}

\newcommand{\bz}{\bar{z}}

\newcommand{\mc}{\mathcal{C}}

\newcommand{\zb}{\bar{z}}

\long\def\new#1\endnew{{\bf #1}}		
\long\def\del#1\enddel{}

\def\del{\partial}

\newcommand{\be}{\begin{eqnarray}}
\newcommand{\en}{\end{eqnarray}}

\def\bz{{\bar z}}

\numberwithin{equation}{section} 

\title{\oneline{Holographic Carrollian Currents for  Massless Scattering}}

\author[a]{Romain Ruzziconi,}
\author[b]{Amartya Saha}
\date{}
\affiliation[a]{Mathematical Institute, University of Oxford, Oxford, OX2 6GG, U.K. } 
\affiliation[b]{Indian Institute of Technology Kanpur, Kanpur 208016, INDIA} 
\emailAdd{romain.ruzziconi@maths.ox.ac.uk} \emailAdd{amartyas@iitk.ac.in}
\abstract{We show that the Ward identities of a Carrollian CFT stress tensor at null infinity reproduce the leading and subleading soft graviton theorems for massless scattering in the bulk. We deduce the expressions of the stress tensor components in terms of the bulk radiative modes, and these components turn out to be local at $\mathscr{I}$ in terms of the twistor potentials. This analysis makes the correspondence between the large-time limit of Carrollian amplitudes and the soft limit of momentum space amplitudes manifest. We then construct Carrollian CFT currents from the ascendants of the hard graviton operator, which satisfy the $Lw_{1+\infty}$ algebra. We show that the large-time limit of their Ward identities implies an infinite tower of projected soft graviton theorems in the bulk, while their finite-time OPEs encode the collinear limit of scattering amplitudes.
}
\begin{document}
\maketitle
\flushbottom
\section{Introduction}

Celestial holography proposes that gravity in four-dimensional (4D) asymptotically flat spacetime is dual to a 2D CFT living on the celestial sphere \cite{deBoer:2003vf,He:2015zea,Pasterski:2016qvg,Cheung:2016iub,Pasterski:2017kqt,Strominger:2017zoo,Pasterski:2017ylz}. One of the main successes of this approach to flat space holography is the reformulation of the symmetries of the bulk $\mathcal{S}$-matrix in terms of Ward identities in the celestial CFT \cite{Adamo:2019ipt,Puhm:2019zbl,Guevara:2019ypd,Pate:2019mfs}. Lorentz invariance in the bulk implies global conformal invariance at the boundary, while the subleading soft graviton theorem implies local conformal invariance \cite{Cachazo:2014fwa,Kapec:2014opa} (it was crucially noticed early on that the 2D celestial stress tensor has a non-local expression in terms of the bulk metric \cite{Kapec:2016jld}). Similarly, supertranslation invariance in the bulk \cite{Strominger:2013jfa,He:2014laa}, encoding the leading soft graviton theorem \cite{Weinberg:1965nx}, also induces Ward identities at the boundary. However, supertranslations act as internal symmetries on the celestial primaries \cite{Stieberger:2018onx} in a somewhat unnatural way from the point of view of a standard 2D CFT. In particular, translation invariance implies distributional low-point correlation functions \cite{Pasterski:2017ylz}. Remarkably, the interplay between the celestial symmetries and the infrared sector of the bulk $\mathcal{S}$-matrix has been pushed even further by reformulating the collinear limit of scattering amplitudes in terms of celestial OPEs \cite{Fan:2019emx,Pate:2019lpp,Fotopoulos:2019vac}. The OPE algebra of conformally soft operators was shown to coincide with $Lw_{1+\infty}$ \cite{Guevara:2021abz,Strominger:2021lvk}. These symmetries have been identified in the radiative phase space of gravity at null infinity \cite{Freidel:2021ytz,Geiller:2024bgf,Cresto:2024fhd,Cresto:2024mne} and find a natural geometric interpretation on twistor space \cite{Adamo:2021lrv,Mason:2022hly,Bu:2022iak,Kmec:2024nmu}.

A seemingly different but nonetheless equivalent approach to flat space holography, called Carrollian holography \cite{Arcioni:2003xx,Dappiaggi:2005ci,Barnich:2006av,Barnich:2010eb,Bagchi:2010zz,Barnich:2012xq,Barnich:2012rz,Bagchi:2012xr,Bagchi:2015wna,Bagchi:2016bcd,Ciambelli:2018wre,Donnay:2022aba,Bagchi:2022emh,Donnay:2022wvx,Saha:2023hsl}, suggests that gravity in 4D asymptotically flat spacetime can be reformulated in terms of a 3D Carrollian CFT living at null infinity. In contrast with the celestial approach, all the BMS/conformal Carrollian symmetries (i.e., superrotations and supertranslations) act as spacetime symmetries in the putative dual theory. Global conformal Carrollian invariance of the boundary correlators, referred to as Carrollian amplitudes \cite{Donnay:2022wvx,Mason:2023mti}, is induced by Poincaré invariance in the bulk. However, the precise role of pure supertranslations and superrotations in encoding the information on bulk soft graviton theorems in the dual Carrollian CFT is still under investigation. A proposal has been made in \cite{Donnay:2022aba,Donnay:2022wvx} to understand the soft theorems via sourced Carrollian CFT Ward identities involving a stress tensor associated with the non-radiative subsector of gravity. Moreover, Carrollian stress tensor Ward identities have been written in \cite{Saha:2023hsl} for the radiative sector and shown to encode patterns of the soft theorems, with a natural extension to $Lw_{1+\infty}$ symmetries \cite{Saha:2023abr}. Soft photon theorems for Carrollian amplitudes have also been recently derived in \cite{Kim:2023qbl,Kraus:2024gso} from a Carrollian partition function at null infinity, using a flat space analogue of the GKP/W dictionary \cite{Gubser:1998bc,Witten:1998qj}.

In this paper, we clarify the symmetry matching between the 3D Carrollian CFT at null infinity and massless scattering in the 4D bulk. First, we show that the Carrollian CFT stress tensor Ward identities, encoding invariance under the full BMS symmetries (including supertranslations and superrotations) in the dual theory, reproduce the leading and subleading soft graviton theorems for massless scattering in flat spacetime. In particular, the soft limit in the bulk is implemented by a large-time limit at the level of the Carrollian amplitudes. This discussion allows us to deduce an explicit expression for the Carrollian CFT stress tensor at null infinity in terms of the bulk radiative modes. This expression can be made local at $\mathscr{I}$ when written in terms of twistor potentials. We comment on the central charge of the Carrollian CFT that arises from this identification. We then push the analysis forward and construct Carrollian CFT currents, whose OPE algebra is shown to coincide with the $Lw_{1+\infty}$ algebra. These currents form an infinite tower of Carrollian ascendants of the hard (finite-energy) graviton operator. They consistently reduce to the celestial soft currents in the large-time limit, and the corresponding large-time behavior of their Ward identities encodes the information on an infinite hierarchy of projected soft graviton theorems in the bulk. Finally, we show that the finite-time OPEs involving the $Lw_{1+\infty}$ currents reproduce the collinear limit of bulk scattering amplitudes.

The rest of the paper is organized as follows. In Section \ref{sec:Basics of Carrollian CFT}, we review the basics of Carrollian CFTs, including the definition of conformal Carrollian primaries and the Carrollian stress tensor Ward identities. In Section \ref{sec:Carrollian stress tensor for massless scattering}, we show that the large-time limit of the Carrollian CFT stress tensor Ward identities reproduces the bulk soft theorems and deduce explicit expressions for the stress tensor components in terms of bulk radiative data. In Section \ref{sec:Lw(1+infinity) Carrollian CFT currents}, we extend the discussion to the infinite tower of $Lw_{1+\infty}$ Carrollian CFT currents and discuss their consequences for massless scattering in the bulk. In Section \ref{sec:Discussion}, we relate our results to previous discussions and offer some perspectives. 

The main text is also complemented by several appendices: Appendix \ref{A} reviews the classical derivation of the Carrollian CFT stress tensor, Appendix \ref{sec:OPE formulae} discusses OPE formulae for Carrollian CFT currents $S_0^+$ and $S_1^+$ with generic operators, Appendix \ref{sec:Full Lorentz symmetries} discusses the operator $T$ generating holomorphic superrotations, Appendix \ref{sec:Central charge} presents the two-point function of the stress tensor in relation with the central charge, Appendix \ref{sec:Infinite tower of currents} reviews how the whole tower of $Lw_{1+\infty}$ currents can be generated recursively in the Carrollian CFT. Finally, Appendix \ref{sec:Applications to bulk amplitudes} illustrates the results obtained in this paper by discussing explicit examples of computations on Carrollian amplitudes.

\section{Basics of Carrollian CFT}
\label{sec:Basics of Carrollian CFT}

In this section, we review some basic features on Carrollian CFT and introduce notations and conventions. We discuss the notion of conformal Carrollian primaries, and present the Ward identities satisfied by a Carrollian CFT stress tensor.

\subsection{Conformal Carrollian primaries}

Consider a three-dimensional (3D) manifold $\mathscr{I} \simeq \mathbb{R} \times \mathcal{S}$ (where $\mathcal{S}$ is a celestial Riemann surface, which is usually taken to be the celestial sphere $S^2$) with coordinates $\mathbf{x} = (u,z, \bar{z})$ and $\vec{x} = (x^i)= (z, \bar{z})$. A Carrollian geometry \cite{Henneaux:1979vn} consists of a degenerate metric $q_{ab} dx^a dx^b = 0 du^2 + 2 dz d\bar{z}$ and a vector field $n^a \partial_a = \partial_u$ in the kernel of the metric, $q_{ab} n^b = 0$. This is the natural geometric structure arising at null infinity \cite{1977asst.conf....1G,Ashtekar:2014zsa,Duval:2014lpa} through the conformal compactification \cite{Penrose:1962ij,Penrose:1985bww}. The conformal symmetries of a Carrollian geometry, called the conformal Carrollian symmetries \cite{Duval:2014uva}, satisfy
\begin{equation}
    \mathcal{L}_{\zeta} q_{ab} = 2 \alpha q_{ab}, \qquad \mathcal{L}_{\zeta} n^a = - \alpha n^a
\end{equation} and coincide with the BMS symmetries in four dimensions \cite{Bondi:1962px,Sachs:1962wk,Sachs:1962zza}. The solutions are generated by vector fields 
\begin{equation}
    \zeta = (\mathcal{T} (z, \bar{z}) + u \alpha ) \partial_u + \mathcal{Y} \partial + \bar{\mathcal{Y}} \bar{\partial}, \qquad \alpha = \frac{1}{2} (\partial\mathcal{Y} + \bar{\partial} \bar{\mathcal{Y}})
\end{equation} where $\mathcal{T}(z, \bar{z})$ are the supertanslation parameters, $(\mathcal{Y}(z), \bar{\mathcal{Y}}(\bar{z}))$ are the superrotation parameters, and $\partial \equiv \partial_z$ and $\bar{\partial} \equiv \partial_{\bar{z}}$. These vector fields satisfy the algebra
\begin{equation}
    [ \zeta (\mathcal{T}_1, \mathcal{Y}_1, \bar{\mathcal{Y}}_1 ) , \zeta (\mathcal{T}_2, \mathcal{Y}_2, \bar{\mathcal{Y}}_2 )] = \zeta (\mathcal{T}_{12}, \mathcal{Y}_{12}, \bar{\mathcal{Y}}_{12} )
\end{equation} where 
\begin{equation}
    \begin{cases}
       \mathcal{T}_{12} = \mathcal{Y}_1 \partial \mathcal{T}_2 - \frac{1}{2} \partial \mathcal{Y}_2 \mathcal{T}_2 - (1 \leftrightarrow 2) + c.c. , \\
      \mathcal{Y}_{12} = \mathcal{Y}_1 \partial \mathcal{Y}_2 - (1 \leftrightarrow 2)   , \\
      \bar{\mathcal{Y}}_{12} = \bar{\mathcal{Y}}_1 \bar{\partial} \bar{\mathcal{Y}}_2 - (1 \leftrightarrow 2)   .
    \end{cases}     
\end{equation} This algebra is called the (extended \cite{Barnich:2009se,Barnich:2010eb,Barnich:2010ojg}) BMS$_4$ algebra, $\mathfrak{bms}_4 \simeq (\text{Witt} \oplus \text{Witt}) \loplus \mathfrak{s}$ (here $\mathfrak{s}$ denotes the abelian supertranslation algebra), and corresponds to the asymptotic symmetry algebra for gravity in four-dimensional asymptotically flat spacetimes \cite{Bondi:1962px,Sachs:1962wk,Sachs:1962zza}. Equivalently, from an intrinsic perspective at $\mathscr{I}$, this corresponds to the conformal Carrollian algebra in three dimensions, $\mathfrak{CCarr}(3)$, and we have $\mathfrak{bms}_4 \simeq \mathfrak{CCarr}(3)$ \cite{Duval:2014uva}. The global conformal Carrollian subalgebra is obtained by setting $\partial^2 \mathcal{T} = 0 = \bar{\partial}^2 \mathcal{T}$ and $\partial^3 \mathcal{Y} = 0 = \bar{\partial}^3 \bar{\mathcal{Y}}$ and is generated by 
\begin{equation}
    \mathcal{T} = 1, z, \bar{z}, z\bar{z}, \qquad \mathcal{Y} = 1, z, z^2, \qquad \bar{\mathcal{Y}} = 1, \bar{z}, \bar{z}^2 .
\end{equation} These coincide with the four bulk translation generators and six Lorentz generators of the Poincaré algebra in four dimensions, and we have an isomorphism between the global subalgebra of $\mathfrak{CCarr}(3)$ and $\mathfrak{iso}(3,1)$. These symmetries are naturally obtained from an In\"on\"u-Wigner contraction of the conformal algebra $\mathfrak{so}(3,2)$ implementing the Carrollian limit, which formally consists of taking the speed of light to zero \cite{Levy1965}. 

A conformal Carrollian (quasi-)primary multiplet ${\Phi} (u,z,\bar{z})$ transforms as follows under the action of the (global) conformal Carrollian symmetries: 
\begin{equation}
\begin{split}
&\delta_{\mathcal{T}}\Phi=\left[\mathcal{T}\partial_u+\partial\mathcal{T}\bm{\xi}+\bar{\partial}\mathcal{T}\bar{\bm{\xi}}\right]\cdot\Phi , \\
&\delta_{\mathcal{Y}}\Phi=\left[\mathcal{Y}\partial +\partial \mathcal{Y} \left(\mathbf{h}+\frac{u}{2}\partial_u\right)+\partial^2 \mathcal{Y}\frac{u}{2}\bm{\xi}\right]\cdot\Phi ,\\
&\delta_{\bar{\mathcal{Y}}}\Phi=\left[\bar{\mathcal{Y}}\bar{\partial}+\bar{\partial}\bar{\mathcal{Y}}\left(\bar{\mathbf{h}}+\frac{u}{2}\partial_u\right)+\bar{\partial}^2\bar{\mathcal{Y}}\frac{u}{2}\bar{\bm{\xi}}\right]\cdot\Phi .\label{70}
\end{split}
\end{equation}
Here $\mathbf{h},\bar{\mathbf{h}}=\frac{\Delta\mathbf{I}\pm\mathbf{J}}{2}$, with $\Delta$ the conformal Carrollian dimension, $\mathbf{I}$ the identity matrix, $\mathbf{J}$ the spin representation matrix, and $\bm{\xi}$ and $\bar{\bm{\xi}}$ furnishing a finite-dimensional indecomposable but reducible representation of the two Carrollian boosts in three dimensions \cite{Saha:2023hsl}. If the multiplet $\Phi$ transforms under the 3D Carrollian spin-boost irreducible representation, it must have $\bm{\xi}\cdot\Phi=\bar{\bm{\xi}}\cdot\Phi=0$. In that case, we can consider singlets to recover the definition of conformal Carrollian (quasi-)primary field $\Phi_{(h,\bar{h})}(u,z,\bar{z})$ given in \cite{Donnay:2022aba,Donnay:2022wvx,Nguyen:2023vfz}:
\begin{equation}    \delta_{(\mathcal{T},\mathcal{Y}, \bar{\mathcal{Y}})} \Phi_{(h,\bar{h})} = \Big[\Big(\mathcal{T} + \frac{u}{2}(\partial \mathcal{Y} + \bar \partial \bar{\mathcal{Y}}) \Big)  \partial_u + \mathcal{Y} \partial + \bar{\mathcal{Y}} \bar \partial + h \partial \mathcal{Y} + \bar{h}  \bar \partial\bar{\mathcal{Y}} \Big] \Phi_{(h,\bar{h})} 
\label{carrollian primary}
\end{equation} where $h = \frac{\Delta + s}{2}$ and $\bar{h} = \frac{\Delta - s}{2}$ are the conformal Carrollian weights. An important property of a conformal Carrollian primary singlet is that its $\partial_u$-descendants are still primaries, but with shifted weights. Indeed, one can check that $\partial_u^m \Phi_{(h,\bar{h})}$ is a primary with weights $(h+\frac{m}{2} , \bar{h}+\frac{m}{2})$. Similarly, one could define $\partial_u$-ascendants of a primary singlet $\Phi_{(h,\bar{h})}$ through
\begin{equation}
    \partial_u^{-k} \Phi_{(h,\bar{h})} (u, z, \bar{z}) = \frac{1}{k!}\int\limits_{-\infty}^u du' (u-u')^k \partial_{u^\prime} \Phi_{(h,\bar{h})} (u', z, \bar{z})
    \label{ascendant}
\end{equation} and check that this field transforms as a conformal Carrollian primary singlet \eqref{carrollian primary} with weights $(h-\frac{k}{2}, \bar{h}-\frac{k}{2})$, provided $\lim\limits_{u \to - \infty}\Phi_{(h,\bar{h})} \sim o(u^{-k})$,\footnote{\label{ascendant operator}As we shall see, the fall-off condition $\lim\limits_{u \to - \infty}\Phi_{(h,\bar{h})} \sim o(u^{-k})$ will not be satisfied in general due to soft contributions, and the definition of ascendant \eqref{ascendant} shall be adapted accordingly.} see Section \ref{sec:Carrollian CFT currents and radiative data at} for details.

\subsection{Stress tensor Ward identities}

The classical constraints implied on the 3D Carrollian CFT stress tensor have been discussed e.g. in \cite{deBoer:2017ing,Ciambelli:2018xat,deBoer:2021jej,Petkou:2022bmz}. This derivation is reviewed in Appendix \ref{A}. The stress tensor constitutes a particular example of conformal Carrollian primary multiplet \eqref{70} with conformal Carrollian dimension $\Delta = 3$. Considering a multiplet with the relevant components\footnote{In this work, we consider ${T^u}_u$ to be complex and denote by ${{\bar{T}}^u}_{\hspace{1.5mm}u}$ its complex conjugate.} $\Phi = ({T^u}_u \, {{\bar{T}}^u}_{\hspace{1.5mm}u} \, {T^u}_z \, {T^u}_{\bar z})^t$, the spin and boost matrices are respectively given by 
\begin{equation}
    \bm{J} = \begin{pmatrix}
        0 & 0 & 0 & 0 \\
        0 & 0 & 0 & 0 \\
        0 & 0 & 1 & 0 \\
        0 & 0 & 0 & -1
    \end{pmatrix}  \, , \qquad  \bm{\xi} = \begin{pmatrix}
        0 & 0 & 0 & 0\\
        0 & 0 & 0 & 0\\
        0 & \frac{3}{2} & 0 & 0\\
        0 & 0 & 0 & 0
    \end{pmatrix} \, , \qquad  \bm{\bar{\xi}} = \begin{pmatrix}
        0 & 0 & 0 & 0\\
        0 & 0 & 0 & 0\\
        0 & 0 & 0 & 0\\
        \frac{3}{2} & 0 & 0 & 0
    \end{pmatrix} 
    \label{stress tensor rep}
\end{equation} (see also Section \ref{sec:Carrollian CFT currents and radiative data at}). In particular, when ${T^u}_u$ is real, the transformation of the stress tensor coincides with the coadjoint representation of $\mathfrak{bms}_4$ constructed in \cite{Barnich:2021dta,Barnich:2022bni}.


At a quantum level,\footnote{All correlators and OPEs considered in this work are implicitly covariant time ordered (as defined in Section $6.1.4.$ of \cite{Itzykson:1980rh}). Covariant time-ordering commutes with spacetime differentiation and integration.} the Carrollian stress tensor Ward identities have been derived explicitly in \cite{Donnay:2022wvx,Saha:2023hsl,Chen:2023pqf}, and hold up to derivative of temporal contact terms:\footnote{\label{contact term}Derivative of temporal contact terms may appear when going from the integrated to the local version of the stress tensor Ward identities (see Appendix \ref{A}). The extension to $Lw_{1+\infty}$ discussed in Section \ref{sec:Lw(1+infinity) Carrollian CFT currents} will completely fix these contact terms, which play a crucial role in the collinear limit.}
\begin{align}
\partial_\mu\langle T^\mu_{\hspace{1.5mm}\nu}(\mathbf{x})X\rangle &=-i\sum\limits_{p=1}^n\text{ }{{\partial_{\nu_p}}\langle X\rangle}\text{ }\delta(u-u_p)\delta^2(\vec{x}-\vec{x}_p)\, ,\label{eq:2}\\
\langle T^i_{\hspace{1.5mm}u}(\mathbf{x})X\rangle &=-i\sum\limits_{p=1}^n\text{ }{(\bm{\xi_i})}_p\cdot{\langle X\rangle}\text{ }\delta(u-u_p)\delta^2(\vec{x}-\vec{x}_p)\, ,\label{eq:3}\\
\langle T^\mu_{\hspace{1.5mm}\mu}(\mathbf{x})X\rangle &=-i\sum\limits_{p=1}^n\text{ }{\Delta}_p{\langle X\rangle}\text{ }\delta(u-u_p)\delta^2(\vec{x}-\vec{x}_p)\, ,\label{eq:4}\\
\langle T^z_{\hspace{1.5mm}z}(\mathbf{x})X\rangle-\langle T^{\bar{z}}_{\hspace{1.5mm}\bar{z}}(\mathbf{x})X\rangle &=-i\sum\limits_{p=1}^n\text{ }s_p{\langle X\rangle}\text{ }\delta(u-u_p)\delta^2(\vec{x}-\vec{x}_p)\, .\label{eq:5}
\end{align}
Here $X = \Phi_1 \ldots \Phi_n$ is a string of conformal Carrollian primary multiplets as defined in \eqref{70}, ${(\bm{\xi_i})}_p$ is the Carrollian boost representation-matrix in the $i^{\text{th}}$ spatial direction, $s_p$ is the spin (i.e. the eigenvalue of the spatial rotation) and $\Delta_p$ is the conformal Carrollian dimension of the $p^{\text{th}}$ primary multiplet.

Let us emphasize that the above writing of the Carrollian CFT Ward identities for the stess tensor requires the symmetries $\mathfrak{so}(3,1) \loplus \mathfrak{s}$ (i.e. the original BMS symmetries uncovered in \cite{Bondi:1962px,Sachs:1962wk,Sachs:1962zza}), and not just the global conformal Carrollian subalgebra (see Appendix \ref{A} for details). It is then easy to show that the existence of a Carrollian CFT stress tensor implies the invariance of the theory under the full Virasoro superrotations \cite{Barnich:2009se,Barnich:2010eb,Barnich:2010ojg} due to the conservation of the currents $j^\mu_{(\mathcal{Y},\bar{\mathcal{Y}})} = {T^\mu}_\nu \zeta^\nu_{(\mathcal{Y},\bar{\mathcal{Y}})}$, where $\zeta_{(\mathcal{Y},\bar{\mathcal{Y}})} = \frac{u}{2}(\partial \mathcal{Y} +\bar{\partial} \bar{\mathcal{Y}}) \partial_u + \mathcal{Y} \partial + \bar{\mathcal{Y}} \bar{\partial}$ are Witt$\oplus$Witt symmetry generators. Hence, analogously to 2D CFT, the original BMS algebra \cite{Bondi:1962px,Sachs:1962wk,Sachs:1962zza} is enhanced into the extended BMS algebra \cite{Barnich:2009se,Barnich:2010eb,Barnich:2010ojg}.

\section{Carrollian stress tensor for massless scattering}
\label{sec:Carrollian stress tensor for massless scattering}

In this section, we show that the conformal Carrollian Ward identities \eqref{eq:2}-\eqref{eq:5} for the stress tensor at $\mathscr{I}$ imply the bulk leading and subleading soft graviton theorems for a massless scattering. We deduce the explicit expression for the Carrollian stress tensor in terms of bulk radiative fields and twistor potentials. 

\subsection{Soft graviton theorems for Carrollian amplitudes}
\label{sec:Soft graviton theorems for Carrollian amplitudes}

First we briefly review how the bulk $\mathcal{S}$-matrix for a massless scattering can be encoded in terms of Carrollian correlators at $\mathscr{I}$ \cite{Donnay:2022wvx,Mason:2023mti}, and how the soft theorems can be expressed in this framework. Consider a massless scattering amplitude in Minkowski space, 
\begin{equation}
    \mathcal{A}_n (\{p_1 \}^{\epsilon_1}_{J_1}, \ldots , \{p_n \}^{\epsilon_n}_{J_n}), \qquad p^\mu_i \equiv \omega_i \Big(1+z_i\bar z_i,z_i+\bar z_i,-i(z_i-\bar z_i),1-z_i\bar z_i\Big)
    \label{momentum space amplitude}
\end{equation} where $n$ is the number of particles involved, $p^\mu_i$ is the momentum of the $i^{\text{th}}$ particle, $\omega_i$ its energy, $J_i$ its physical helicity, $\epsilon_i = \pm 1$ tells us whether the particle is outgoing/incoming so that $s_i=\epsilon_iJ_i$ is its helicity in the `outgoing convention' and $(z_i,\bar{z}_i)$ are coordinates on the celestial sphere. The amplitudes satisfy the leading \cite{Weinberg:1965nx} and subleading \cite{Cachazo:2014fwa} soft graviton theorems which, in the above parametrization, read as 
\begin{equation}
  \lim_{\omega \to 0} \omega \mathcal{A}_{n+1} (\{\omega, z, \bar{z}\}, \{\omega_i, z_i, \bar{z}_i\}^{\epsilon_i}_{J_i})  = -\left( \sum_{i=1}^n \frac{\epsilon_i \omega_i 
(\bar z -  \bar z_i)}{z - z_i}  \right) \mathcal{A}_{n}  ( \{\omega_i, z_i, \bar{z}_i\}^{\epsilon_i}_{J_i} ) 
\end{equation}
and
\begin{multline}
\lim_{\omega \to 0} (1 + \omega \partial_\omega) \mathcal{A}_{n+1} (\{\omega, z, \bar{z}\}, \{\omega_i, z_i, \bar{z}_i\}^{\epsilon_i}_{J_i})  \\
= -\left( \sum_{i=1}^n \frac{(\bar z- \bz_i)^2 \bar \partial_i+ ( \epsilon_i J_i + \omega_i \partial_{\omega_i}) (\bz -\bz_i) }{z - z_i}  \right)  \mathcal{A}_{n}  ( \{\omega_i, z_i, \bar{z}_i\}^{\epsilon_i}_{J_i} ) \, .
\end{multline}  

One can always rewrite the amplitude \eqref{momentum space amplitude} in position space at $\mathscr{I}$ via a Fourier transform over the energies of the particles. 
\begin{multline}
\label{eq:AtoC}
    \mathcal{C}_n\left(\left\lbrace u_1, z_1, \bar{z}_1\right\rbrace_{J_1}^{\epsilon_1}, \dots , \left\lbrace u_n, z_n, \bar{z}_n\right\rbrace_{J_n}^{\epsilon_n}\right) \\
    =  \prod_{i=1}^n \left( \int_0^{+\infty}  \frac{d\omega_i}{2\pi} \, e^{-i\epsilon_i \omega_i u_i} \right)  \mathcal{A}_n \left(\left\lbrace \omega_1, z_1, \bar{z}_1\right\rbrace_{J_1}^{\epsilon_1}, \dots , \left\lbrace \omega_n, z_n, \bar{z}_n\right\rbrace_{J_n}^{\epsilon_n}\right) \, .
\end{multline} This object, called Carrollian amplitude \cite{Donnay:2022wvx,Mason:2023mti}, can be re-interpreted as a correlator of Carrollian primaries \eqref{carrollian primary} at $\mathscr{I}$,
\begin{equation}
\label{carrollian identification}
    \mc_n\left(\left\lbrace u_1, z_1, \zb_1\right\rbrace_{J_1}^{\epsilon_1}, \dots , \left\lbrace u_n, z_n, \zb_n\right\rbrace_{J_n}^{\epsilon_n}\right) \equiv \langle \Phi^{\epsilon_1}_{(h_1,\bar{h}_1)} (u_1, z_1, \bz_1) \ldots \Phi^{\epsilon_n}_{(h_n,\bar{h}_n)} (u_n, z_n, \bz_n)  \rangle \ , 
\end{equation} where the Carrollian weights are fixed in terms of the helicity
\begin{equation}
    h_i = \frac{1+\epsilon_i J_i}{2} \ , \qquad \bar h_i = \frac{1- \epsilon_i J_i}{2} \ .
    \label{fixed Carrollian weights}
\end{equation}  Indeed, one can show that this correlator satisfies the global conformal Carrollian Ward identities. We refer to \cite{Donnay:2022aba,Bagchi:2022emh,Donnay:2022wvx,Salzer:2023jqv,Saha:2023abr,Nguyen:2023vfz,Nguyen:2023miw,Mason:2023mti,Bagchi:2023cen,Liu:2024nfc,Stieberger:2024shv,Adamo:2024mqn,Alday:2024yyj,Ruzziconi:2024zkr,Jorstad:2024yzm} for recent developments on Carrollian amplitudes. 

The soft graviton theorems can also be expressed in position space at $\mathscr{I}$. Defining the leading and the subleading soft graviton operators \cite{Strominger:2017zoo}, 
\begin{equation}
    {S}_0^+(z,\bar{z}) = \lim_{\omega \to 0^+} \frac{\omega}{2} [ a^{\text{out}}_+ (\omega) - a^{\text{in}}_- (\omega)^\dagger  ] \, , \quad S_1^+(z,\bar{z}) = \lim_{\omega \to 0^+} \frac{i}{2} (1 + \omega \partial_{\omega}) [a^{\text{out}}_+ (\omega) + a^{\text{in}}_- (\omega)^\dagger] \, ,
\label{leading and subleading soft operator}
\end{equation}
the leading soft theorem reads as 
\begin{equation}
    \langle i S^+_0 (z,\bar{z}) \prod_{i=1}^n \Phi^{\epsilon_i}_{(h_i,\bar{h}_i)} (u_i, z_i, \bz_i)   \rangle = \left(\sum_{i=1}^n    \frac{\bar z- \bar z_i}{z- z_i} \partial_{u_i} \right)  \langle  \prod_{i=1}^n \Phi^{\epsilon_i}_{(h_i,\bar{h}_i)} (u_i, z_i, \bz_i)  \rangle  
    \label{leading soft them position}
\end{equation} and the subleading soft theorem as
\begin{multline} \label{subleading soft graviton theorem}
    \langle i S_1^+ (z,\bar{z})  \prod_{i=1}^n \Phi^{\epsilon_i}_{(h_i,\bar{h}_i)} (u_i, z_i, \bz_i)   \rangle  \\
    = \left( \sum_{i=1}^n \frac{(\bar z-\bar z_i)^2 \bar \partial_i+ (2 \bar h_i + u_i \partial_{u_i}) (\bar z_i- \bar z) }{z - z_i}  \right) \langle \prod_{i=1}^n \Phi^{\epsilon_i}_{(h_i,\bar{h}_i)} (u_i, z_i, \bz_i)   \rangle 
\end{multline} with $h_i$ fixed as in \eqref{fixed Carrollian weights}. We refer to Appendix \ref{G.1} for the derivation of these Carrollian position-space soft theorems starting from their (bulk) momentum-space counterparts. In the rest of this Section, we show that the said relations also follow from the Carrollian stress tensor Ward identities \eqref{eq:2}-\eqref{eq:5} in the soft limit, which in position space at $\mathscr{I}$ corresponds to the $u \to \infty$ limit.

\subsection{Leading soft graviton theorem}

Let us first reproduce \eqref{leading soft them position}. Subtracting the spatial divergence of \eqref{eq:3} from \eqref{eq:2}$_{\nu=u}$, we obtain: 
\begin{align}
\partial_u\langle T^u_{\hspace{1.5mm}u}(u,\vec{x})X\rangle=-i\sum\limits_{p=1}^n\text{ }\delta(u-u_p)\left[\delta^2(\vec{x}-\vec{x}_p)\partial_{u_p}-\left(\vec{\bm{\xi}}_p\cdot\bm{\partial}_{\vec{x}}\right)\delta^2(\vec{x}-\vec{x}_p)\right]\langle X\rangle 
\end{align} where $\bm{\partial}_{\vec{x}} = (\partial_x, \partial_y)$ and $\vec{\bm{\xi}} = (\bm{\xi}_x, \bm{\xi}_y)$.
Choosing the `retarded' initial condition\footnote{\label{foot:adv}Alternatively, one could choose the `advanced' initial condition $\langle T^u_{\hspace{1.5mm}u}(u\rightarrow+\infty,\vec{x})X\rangle=0$, which would affect the signs.}
\begin{align}
\langle T^u_{\hspace{1.5mm}u}(u\rightarrow-\infty,\vec{x})X\rangle=0 \, ,\label{eq:6}
\end{align}
the solution to the above temporal partial differential equation is found to be
\begin{align}
\langle T^u_{\hspace{1.5mm}u}(u,\vec{x})X\rangle=-i\sum\limits_{p=1}^n\text{ }\theta(u-u_p)\left[\delta^2(\vec{x}-\vec{x}_p)\partial_{u_p}-\left(\vec{\bm{\xi}}_p\cdot\bm{\partial}_{\vec{x}}\right)\delta^2(\vec{x}-\vec{x}_p)\right]\langle X\rangle \, \label{eq:7}
\end{align} where $\theta(u)$ is the temporal  Heaviside function. We wish to convert all the $S^2$ contact-term singularities in \eqref{eq:7} into pole singularities while avoiding branch-cuts. For this purpose, we note that
\begin{align}
\langle {T}^u_{\hspace{1.5mm}u}(u,\vec{x})X\rangle&=-\frac{i}{\pi}\sum\limits_{p=1}^n\text{ }\theta(u-u_p)\text{ }{\bar{\partial}}^2\left[\frac{\bar{z}-\bar{z}_p}{z-z_p}\partial_{u_p}+\frac{\bar{z}-\bar{z}_p}{{(z-z_p)}^2}\bm{\xi}_p-\frac{\bar{{\bm{\xi}}}_p}{z-z_p}\right]\langle X\rangle\label{eq:8}\\
&=-\frac{i}{\pi}\sum\limits_{p=1}^n\text{ }\theta(u-u_p)\text{ }{{\partial}}^2\left[\frac{z-z_p}{\bar{z}-\bar{z}_p}\partial_{u_p}+\frac{z-z_p}{({\bar{z}-\bar{z}_p})^2}\bar{\bm{\xi}}_p-\frac{{{\bm{\xi}}}_p}{\bar{z}-\bar{z}_p}\right]\langle X\rangle\label{eq:9}
\end{align}
where in these equations, $\bm{\xi},\bar{\bm{{\xi}}}\equiv\bm{\xi}_x\pm i\bm{\xi}_y$ and we used $\partial_z (\frac{1}{\bar{z}}) = \pi \delta^2 (z)$ and its complex conjugate. It is then suggestive to introduce the Carrollian operators $S^\pm_0 (u,\vec{x})$ \cite{Saha:2023hsl} (where $\pm$ will coincide later with the bulk helicity) by respectively inverting the ${\bar{\partial}}^2$ operator in \eqref{eq:8} and the $\partial^2$ operator in \eqref{eq:9} as
\begin{align}
&S_0^+(u,z,\bar{z}):=\int\limits_{S^2} d^2\vec{x}^\prime\text{ }\frac{\bar{z}-\bar{z}^\prime}{z-z^\prime}\text{ }{T}^u_{\hspace{1.5mm}u}(u,\vec{x}^\prime)\hspace{2.5mm}\Longrightarrow\hspace{2.5mm}\bar{\partial}^2S_0^+=\pi {T}^u_{\hspace{1.5mm}u} \, ,\label{Tuu component}\\
&S_0^-(u,z,\bar{z}):=\int\limits_{S^2} d^2\vec{x}^\prime\text{ }\frac{z-z^\prime}{\bar{z}-\bar{z}^\prime}\text{ }{T}^u_{\hspace{1.5mm}u}(u,\vec{x}^\prime)\hspace{2.5mm}\Longrightarrow\hspace{2.5mm}{\partial}^2S_0^-=\pi {T}^u_{\hspace{1.5mm}u} \, .
\end{align}
The scaling dimension $\Delta$ and the spin $s$ of the fields $S^\pm_0$ are $(\Delta,s)=(1,\pm 2)$. So, by construction, the fields $S^\pm_0$ are the 2D shadow-transformations (on $S^2$) of each other. Since, a field and its shadow
can not both be treated as local fields in a theory \cite{Banerjee:2022wht}, only one among $S^\pm_0$ is to be chosen as a local field. In this work, we opt to treat $S^+_0$ as the local field while relegating $S^-_0$ as its non-local shadow. We are thus investigating the holomorphic sector of the 3D Carrollian CFT. 


Let us discuss the Ward identities in terms of $S_0^+$ (those in terms of $S^-_0$ would have been obtained in a completely analogous way). As a consequence of \eqref{eq:8}, we have 
\begin{align}
&\langle S^+_0(u,z,\bar{z})X\rangle=-i\sum\limits_{p=1}^n\text{ }\theta(u-u_p)\text{ }\left\{\frac{\bar{z}-\bar{z}_p}{z-z_p}\partial_{u_p}+\frac{\bar{z}-\bar{z}_p}{{(z-z_p)}^2}\bm{\xi}_p-\frac{\bar{{\bm{\xi}}}_p}{z-z_p}\right\}\langle X\rangle\label{3}\\
\Longrightarrow\hspace{2.5mm}&\langle\bar{\partial}^2 S^+_0(u,z,\bar{z})X\rangle=-i\sum\limits_{p=1}^n\text{ }\theta(u-u_p)\text{ }\left[\text{contact terms on $S^2$}\right]\nonumber\\
\text{and}\hspace{2.5mm}&{\partial_u}\langle S^+_0(u,z,\bar{z})X\rangle=\left[\text{temporal contact terms}\right] \, . \label{2}
\end{align}

Taking $u\to+\infty$ in \eqref{3}, and considering primaries in $X$ with $\bm{\xi}=\bar{\bm{{\xi}}}=0$ \cite{Saha:2023hsl}, i.e. primary fields \eqref{carrollian primary}, we reproduce exactly the leading soft graviton theorem \eqref{leading soft them position} written in position space at $\mathscr{I}$, provided 
\begin{equation}
    \lim_{u \to +\infty} S^+_0 (u,z,\bar{z}) = {S}_0^+ (z,\bar{z})
    \label{identific1}
\end{equation} where the leading soft graviton operator $S_0^+ (z,\bar{z})$ was defined in \eqref{leading and subleading soft operator}. A similar relation would hold for $S^-_0$. This identification will allow us to deduce the expression of the ${T^u}_u$ component of the Carrollian stress tensor in terms of the bulk radiative modes, see Section \ref{sec:Carrollian CFT stress tensor}. This shows explicitly that the soft limit $\omega \to 0$ coincides with the $u\to +\infty$ limit via the Fourier transform \eqref{eq:AtoC} .

As mentioned before (see Footnote \ref{contact term}), the $S^+_0$ Ward identity \eqref{3} is valid up to temporal contact terms, all of which vanish when $u\rightarrow + \infty$ since the insertions of Carrollian primaries take place at finite $u_p$. Clearly, the temporal Fourier transformation of each of these contact terms also vanishes in the $\omega\rightarrow0$ soft limit. Thus, for the purpose of providing a 3D Carrollian CFT description of the soft sector of 4D gravity in flat space, we can safely ignore these temporal contact terms. Obviously, this is no longer the case if we are to describe the bulk physics at a non-zero $\omega$, since these contact terms produce non-vanishing contributions upon temporal Fourier transformation.

We end this section by a technical digression. To discuss the conformal Carrollian OPEs (see e.g. Section \ref{sec:OPE algebra of Carrollian CFT currents}), it will be convenient to convert the temporal step-function appearing in the correlators like \eqref{3} into a $j\epsilon$-prescription, with $j$ being a second complex unit, following \cite{Saha:2023hsl,Saha:2022gjw}. The starting point is the following hyper-complexification of the $(u,z,\bar{z})$ coordinates: 
\begin{equation}
\hat{z}:=z+ju\hspace{2.5mm};\hspace{2.5mm}\hat{\bar{z}}:=\bar{z}+ju\hspace{2.5mm};\hspace{2.5mm}\hat{u}:=u \, . \label{jepsilon}
\end{equation}
While $z$ is a complex number on the $x-y$ plane, it suffices for the computational purpose to restrict $\hat{z}$ to be a complex number on a $y=ax+b$ plane of the 3D $(u,x,y)$ space. $u>0$ is the upper-half of any such plane. Since all of $\frac{\partial\hat{z}}{\partial\hat{u}},\frac{\partial\hat{\bar{z}}}{\partial\hat{u}},\frac{\partial\hat{\bar{z}}}{\partial\hat{z}}$ vanish, we can treat $\hat{u},\hat{z},\hat{\bar{z}}$ as independent variables. Thus, in most cases, the point of insertion of a Carrollian field is chosen to be at $(\hat{u},\hat{z},\hat{\bar{z}})=(u,z,\bar{z})$. E.g. the $j\epsilon$-form of the $S^+_0$ Ward identity \eqref{3} then is 
\begin{equation}
\langle S^+_0(u,{z},{\bar{z}})X\rangle=\lim\limits_{\epsilon\rightarrow0^+}-i\sum\limits_{p=1}^n\left\{\frac{\bar{z}-\bar{z}_p}{(\Delta\tilde{z}_p)}\partial_{u_p}+\frac{\bar{z}-\bar{z}_p}{{(\Delta\tilde{z}_p)}^2}\bm{\xi}_p-\frac{\bar{{\bm{\xi}}}_p}{(\Delta\tilde{z}_p)}\right\}\langle X\rangle
\end{equation} with $\Delta\tilde{z}_p:=z-z_p-j\epsilon(u-u_p)$. The $j\epsilon$-prescription was introduced to directly relate the Carrollian CFT OPEs and the corresponding commutation relations while facilitating a straightforward utilisation of the OPE commutativity property.\footnote{See Section 4.3 of \cite{Saha:2023hsl} and Section 5.1 of \cite{Saha:2022gjw} for more details.} An important point to remember is that $\frac{1}{\Delta\tilde{z}_p}$ reduces to $\frac{1}{z-z_p}$ only when $u-u_p>0$ and to $0$ when $u-u_p<0$ in the distributional sense \cite{Saha:2022gjw} inheriting the property of the temporal step-function. Thus, $\frac{1}{\Delta\tilde{z}_p}\equiv0$ when $u\rightarrow-\infty$, justifying the initial condition \eqref{eq:6}, and $\frac{1}{\Delta\tilde{z}_p}\equiv \frac{1}{z-{z}_p}$ when $u\rightarrow\infty$.  

Using this $j\epsilon$-prescription, we provide in Appendix \ref{sec:OPE formulae} OPE formulae between $S_0^+$ and a general Carrollian operator. These results will be useful to extend the analysis to the whole tower of $Lw_{1+\infty}$ currents in Section \ref{sec:Lw(1+infinity) Carrollian CFT currents}.

\subsection{Subleading soft graviton theorem}

We now show that the large time behaviour of the Carrollian CFT stress tensor Ward identities also reproduces \eqref{subleading soft graviton theorem}. The Ward identities \eqref{eq:4}, \eqref{eq:5} and \eqref{eq:7} can be linearly combined into the following form: 
\begin{equation}
\langle T^z_{\hspace{1.5mm}z}(\mathbf{x})X\rangle+\frac{1}{2}\langle T^u_{\hspace{1.5mm}u}(\mathbf{x})X\rangle=-i\sum\limits_{p=1}^n\text{ }{h}_p{\langle X\rangle}\text{ }\delta(u-u_p)\delta^2(\vec{x}-\vec{x}_p) \, ,
\end{equation} which implies
\begin{align}
& i\langle T^z_{\hspace{1.5mm}z}(\mathbf{x})X\rangle \\
&=\sum\limits_{p=1}^n\left[\delta(u-u_p)\delta^2(\vec{x}-\vec{x}_p)h_p-\frac{\theta(u-u_p)}{2}\left\{\delta^2(\vec{x}-\vec{x}_p)\partial_{u_p}-\left(\vec{\bm{\xi}}_p\cdot\bm{\partial}_{\vec{x}}\right)\delta^2(\vec{x}-\vec{x}_p)\right\}\right]{\langle X\rangle} \, .  \nonumber
\end{align}
Thus, subtraction of $\partial_z\langle T^z_{\hspace{1.5mm}z}(\mathbf{x})X\rangle$ from \eqref{eq:2}$_{\nu=z}$ leads to
\begin{align*}
\partial_{\bar{z}}\langle T^{\bar{z}}_{\hspace{1.5mm}z}(\mathbf{x})X\rangle+\partial_u\langle T^u_{\hspace{1.5mm}z}(\mathbf{x})X\rangle=&-i\sum\limits_{p=1}^n\Big[\delta(u-u_p)\left\{\delta^2(\vec{x}-\vec{x}_p)\partial_{z_p}-h_p\partial_{z}\delta^2(\vec{x}-\vec{x}_p)\right\}\nonumber\\
&+\frac{\theta(u-u_p)}{2}\partial_z\left\{\delta^2(\vec{x}-\vec{x}_p)\partial_{u_p}-\left(\vec{\bm{\xi}}_p\cdot\bm{\partial}_{\vec{x}}\right)\delta^2(\vec{x}-\vec{x}_p)\right\}\Big]{\langle X\rangle} \, .
\end{align*}
Choosing a `retarded' initial condition similar to \eqref{eq:6},
\begin{align}
\langle T^u_{\hspace{1.5mm}z}(u\rightarrow-\infty,\vec{x})X\rangle=0 \, ,  \label{retarded initial}
\end{align}
we obtain the following solution to the above temporal partial differential equation:
\begin{equation}
\begin{split}
&\langle T^u_{\hspace{1.5mm}z}(\mathbf{x})X\rangle+\int\limits^u_{-\infty}du^\prime\text{ }\partial_{\bar{z}}\langle T^{\bar{z}}_{\hspace{1.5mm}z}(u^\prime,\vec{x})X\rangle \\
&\qquad\qquad\qquad\qquad=-i\sum\limits_{p=1}^n\theta(u-u_p)\Big[\delta^2(\vec{x}-\vec{x}_p)\partial_{z_p}-h_p\partial_{z}\delta^2(\vec{x}-\vec{x}_p)\\
&\qquad\qquad\qquad\qquad\qquad+\frac{u-u_p}{2}\partial_z\left\{\delta^2(\vec{x}-\vec{x}_p)\partial_{u_p}-\left(\vec{\bm{\xi}}_p\cdot\bm{\partial}_{\vec{x}}\right)\delta^2(\vec{x}-\vec{x}_p)\right\}\Big]{\langle X\rangle} \, . \label{a13}
\end{split}
\end{equation}
The `complex-conjugated' version can be found in an exactly similar way.

\medskip

Next, we extract a $\partial^3$- (or, $\bar{\partial}^3$-) derivative from the RHS of \eqref{a13} (or, its complex-conjugate) to convert all the $S^2$ contact terms into pole singularities. The conformal Carrollian fields $S^\pm_1$ were defined by inverting these derivatives in \cite{Saha:2023hsl} as below: 
\begin{align}
&S_1^-(u,z,\bar{z})=\int\limits_{S^2} d^2{\vec{x}^\prime}\text{ }\frac{{(z-z^\prime)}^2}{\bar{z}-\bar{z}^\prime}\left[T^u_{\hspace{1.5mm}z}(u,\vec{x}^\prime)+\int\limits_{-\infty}^u du^\prime\partial_{\bar{z}^\prime}T^{\bar{z}}_{\hspace{1.5mm}z}(u^\prime,\vec{x}^\prime)\right]\label{47} \, ,\\
&S_1^+(u,z,\bar{z})=\int\limits_{S^2} d^2{\vec{x}^\prime}\text{ }\frac{{(\bar{z}-\bar{z}^\prime)}^2}{{z}-{z}^\prime}\left[T^u_{\hspace{1.5mm}\bar{z}}(u,\vec{x}^\prime)+\int\limits_{-\infty}^u du^\prime\partial_{{z}^\prime} T^{{z}}_{\hspace{1.5mm}\bar{z}}(u^\prime,\vec{x}^\prime)\right] \, .\label{46}
\end{align}
The dimensions of the fields $S^\pm_1$ are $(\Delta,s)=(0,\pm2)$. It was shown in \cite{Saha:2023hsl}, following \cite{Banerjee:2022wht}, that the fields $S^+_0$ and $S^-_1$ are not mutually local, while $S^+_0$ and $S^+_1$ can be treated as local simultaneously (see also Appendix \ref{sec:OPE formulae}).

The $S^+_1$ Ward identity with a string $X$ of mutually local conformal Carrollian primaries then, up to temporal contact terms, is:
\begin{align}
&\langle S_1^+(u,z,\bar{z}) X\rangle=-i\sum\limits_{p=1}^n\theta(u-u_p)\Big[\frac{(\bar{z}-\bar{z}_p)^2}{z-{z}_p}\partial_{\bar{z}_p}-2\bar{h}_p\frac{\bar{z}-\bar{z}_p}{z-{z}_p} \label{10}\\
&\hspace{46mm}+(u-u_p)\Big(\frac{\bar{z}-\bar{z}_p}{z-{z}_p}\partial_{u_p}+\frac{\bar{z}-\bar{z}_p}{{(z-{z}_p)}^2}\bm{\xi}_p-\frac{\bar{{\bm{\xi}}}_p}{z-{z}_p}\Big)\Big]\langle X\rangle \nonumber\\
\Longrightarrow\hspace{5mm}&\langle\bar{\partial}^3 S^+_1(u,z,\bar{z})X\rangle=-i\sum\limits_{p=1}^n\text{ }\theta(u-u_p)\text{ }\left[\text{contact terms on $S^2$}\right]\nonumber\\
\text{and}\hspace{5mm}&{\partial_u}\langle S^+_1(u,z,\bar{z})X\rangle-\langle S^+_0(u,z,\bar{z})X\rangle=\left[\text{temporal contact terms}\right] \, . \label{11}
\end{align}
As discussed in \cite{Saha:2023hsl}, temporal-Fourier transforming the $S^+_1$ Ward identity \eqref{10} and then imposing the soft $\omega\rightarrow0$ limit only for $S^+_1$, one obtains a Laurent expansion around $\omega=0$. To isolate the subleading soft graviton theorem, and motivated by \eqref{2} and \eqref{11}, we re-express the $S^+_1$ field inside the correlator as 
\begin{align}
&S^+_1(u,z,\bar{z})=S^+_{1e}(u,z,\bar{z})+u S^+_0(u,z,\bar{z})\label{12}\\
\Longrightarrow\hspace{5mm}&{\partial_u}\langle S^+_{1e}(u,z,\bar{z})X\rangle=\left[\text{temporal contact terms}\right]\nonumber\\
\text{and}\hspace{5mm}&\langle\bar{\partial}^3 S^+_{1e}(u,z,\bar{z})X\rangle=-i\sum\limits_{p=1}^n\text{ }\theta(u-u_p)\text{ }\left[\text{contact terms on $S^2$}\right] \, . \nonumber
\end{align} Using this definition for $S_{1e}^+$, we deduce from \eqref{3} and \eqref{10} that 
\begin{multline}
    \langle S_{1e}^+(u,z,\bar{z}) X\rangle=-i\sum\limits_{p=1}^n\theta(u-u_p)\Big[\frac{(\bar{z}-\bar{z}_p)^2}{z-{z}_p}\partial_{\bar{z}_p}-2\bar{h}_p\frac{\bar{z}-\bar{z}_p}{z-{z}_p} \\
-u_p\Big(\frac{\bar{z}-\bar{z}_p}{z-{z}_p}\partial_{u_p}+\frac{\bar{z}-\bar{z}_p}{{(z-{z}_p)}^2}\bm{\xi}_p-\frac{\bar{{\bm{\xi}}}_p}{z-{z}_p}\Big)\Big]\langle X\rangle \, . 
\end{multline} Taking the soft limit in position space at $\mathscr{I}$, corresponding to $u \to + \infty$, assuming $\bm{\xi}=\bar{\bm{{\xi}}}=0$, and identifying the Carrollian weights as in \eqref{fixed Carrollian weights}, one reproduces exactly the subleading soft graviton theorem \eqref{subleading soft graviton theorem}, provided one makes the following identification: 
\begin{equation}
    \lim_{u \to +\infty} S_{1e}^+ (u, z , \bar{z}) = S_{1}^+ (z , \bar{z})
\label{identif2}
\end{equation} where the subleading soft graviton operator $S^+_{1} (z , \bar{z})$ was defined in \eqref{leading and subleading soft operator}.

Again, using the $j\epsilon$-prescription, we provide in Appendix \ref{sec:OPE formulae} OPE formulae between $S_1^+$ and a general Carrollian operator.

\subsection{Comments on the symmetries}
\label{sec:Comments on the symmetries}

In the above discussion, the superrotations were defined as the standard local conformal transformations forming the Witt$\oplus$Witt algebra. However, similarly to what was discussed in \cite{Banerjee:2020zlg}, the $S^+_{1e}$ operator identified with the subleading soft graviton in the $u \to \infty$ limit generates $\overline{\mathfrak{sl}_2}$ transformations corresponding to the following infinitesimal Carrollian diffeomorphisms at $\mathscr{I}$ \cite{Saha:2023hsl} (see Appendix \ref{A} for details):
\begin{align}
&z\rightarrow z\hspace{2.5mm};\hspace{2.5mm}\bar{z}\rightarrow \bar{z}+\epsilon z^n\bar{z}^{q+1}\hspace{2.5mm};\hspace{2.5mm}u\rightarrow u+\epsilon\frac{u}{2}(q+1)z^n\bar{z}^q\hspace{5mm} \label{9} \\
&\text{(with $q=0,\pm1$ and $n\in\mathbb{Z}$)}. \nonumber
\end{align} 
These transformations are generated by the superrotation parameters whose mode expansion reads as 
\begin{equation}
    \mathcal{Y}(z, \bar{z}) = 0, \qquad \bar{\mathcal{Y}}(z, \bar{z}) = \sum_{n=0}^2 y_n(z) \bar{z}^{n}
\end{equation} where $y_n(z)$ is a meromorphic function. Unlike $S^+_1$, $S^+_{1e}$ is not a conformal Carrollian primary field (see Section \ref{sec:Carrollian CFT currents and radiative data at}, and in particular Equation \eqref{transfoS0} for the precise transformation law of $S^+_{1e}$); it just contains some modes that do not appear in the $S^+_0$ field. On the other hand, the modes of the $S^+_0$ field generate the following supertranslations:
\begin{align}
z\rightarrow z^\prime=z\hspace{2.5mm};\hspace{2.5mm}\bar{z}\rightarrow \bar{z}^\prime=\bar{z}\hspace{2.5mm};\hspace{2.5mm}u\rightarrow u^\prime=u+ \sum_{n=0}^1 t_n(z) \bar{z}^{n}\label{13}
\end{align}
where $t_n(z)$ is a meromorphic function, and the corresponding supertranslation parameters are 
\begin{equation}
    \mathcal{T}(z,\bar{z}) = \sum_{n=0}^1 t_n(z) \bar{z}^{n} \, . 
\label{wedge spin 0}
\end{equation}

As noted in \cite{Banerjee:2020zlg}, the Lorentz transformations $\mathcal{Y}=1,z,z^2$ do not seem to be captured by the above discussion. In Appendix \ref{sec:Full Lorentz symmetries}, we discuss how the missing symmetries can be generated by a Carrollian operator $T(u,z,\bar{z})$ mutually local with $S_0^+$ and $S_1^+$, and satisfying 
\begin{equation}
    2\bar{\partial}T=\partial^3S^-_1 \, . 
    \label{relation S1T}
\end{equation}
Indeed, this operator captures all the modes associated with holomorphic superrotations. Hence, at the end, the symmetry algebra generated by the mutually local operators $S^+_0$, $S_1^+$ and $T$ is given by \cite{Saha:2023hsl} 
\begin{equation}
    \text{Witt} \loplus ( \overline{\mathfrak{sl}_2} \loplus {\mathfrak{s}}_\wedge)
    \label{algebra S0S1T}
\end{equation} where ${\mathfrak{s}}_\wedge$ denotes the supetranslations \eqref{wedge spin 0} obeying the anti-holomorphic wedge condition $\bar{\partial}^2 \mathcal{T} = 0$. This algebra constitutes a quantum version of the previously considered $\mathfrak{bms}_4$ algebra \cite{Banerjee:2020zlg,Pano:2023slc}. As we shall discuss in Section \ref{sec:Lw(1+infinity) Carrollian CFT currents}, the above algebra naturally extends to $\text{Witt} \loplus Lw_{1+\infty}$. Furthermore, as discussed in Appendix \ref{A}, this algebra imposes exactly the same constraints on the Carrollian CFT stress tensor as the $\mathfrak{bms}_4$ symmetries, plus the extra condition ${T^z}_{\bar{z}} = 0$. It is also shown that ${T^{\bar{z}}}_z = 0$ if we further extend the superrotations to all diffeomorphisms on the sphere \cite{Campiglia:2014yka,Campiglia:2015yka,Compere:2018ylh}. This suggests that, for a massless scattering, these components of the holographic Carrollian stress tensor can be set to zero, consistently with soft theorems.

\subsection{Carrollian CFT stress tensor at \texorpdfstring{$\mathscr{I}$}{Scri}}
\label{sec:Carrollian CFT stress tensor}

We now deduce from the above analysis the explicit expression of the non-trivial Carrollian stress tensor components in terms of the bulk radiative modes. We begin by recalling the mode expansion of the asymptotic shear in terms of the graviton creation and annihilation operators \cite{Ashtekar:1981hw, Ashtekar:1981sf,Kapec:2014opa}:
\begin{align}
    C_{zz}(u,z,\bar{z})&=-\frac{i}{2\pi}\int\limits_{0}^\infty \frac{d\omega}{2\pi}\left[a_+^{\text{out}}(\omega,z,\bar{z})e^{-i\omega u}-a_-^{\text{out}}(\omega,z,\bar{z})^\dagger e^{i\omega u}\right] \qquad(\mathscr{I}^+)\, , \label{66prime}\\
    D_{zz}(u,z,\bar{z})&=\frac{i}{2\pi}\int\limits_{0}^\infty \frac{d\omega}{2\pi}\left[a_+^{\text{in}}(\omega,z,\bar{z})e^{-i\omega u}-a_-^{\text{in}}(\omega,z,\bar{z})^\dagger e^{i\omega u}\right] \qquad(\mathscr{I}^-) \, .\label{66}
\end{align} Here we consider the planar Bondi coordinates (see Appendix A of \cite{Donnay:2022wvx}) which allow us to describe both $\mathscr{I}^+$ and $\mathscr{I}^-$ with the same $(u,z,\bar{z})$ coordinates\footnote{We define the boundary value at $\mathscr{I}^-$, $D_{zz}$, with a $-$ sign with respect to the conventions of \cite{Donnay:2022wvx}. The conventions of the antipodal matching conditions and the extrapolate dictionary will be modified accordingly.} (the index $\epsilon=\pm 1$ introduced in Section \ref{sec:Soft graviton theorems for Carrollian amplitudes} keeps track of incoming/outgoing). Using the extrapolate dictionary \cite{Donnay:2022aba,Donnay:2022wvx,Mason:2023mti}, the operators \eqref{66prime}-\eqref{66} correspond to conformal Carrollian quasi-primaries \eqref{carrollian primary} with weights \eqref{fixed Carrollian weights} and $\epsilon J= 2$ (see Section \ref{sec:Carrollian CFT currents and radiative data at}), inserted inside of the Carrollian amplitudes. More explicitly, we have 
\begin{equation}
    \Phi^{\epsilon = +1}_{J=+2} = C_{zz}\, ,\qquad \Phi^{\epsilon = -1}_{J=-2} = D_{zz}
\label{extrapolate}
\end{equation} and the hermitian conjugate expressions for the opposite helicity $\epsilon J = -2$. Hence, the expressions of the stress tensor given below can also be interpreted as expressions in terms of Carrollian CFT operators associated with hard (finite-energy) gravitons.

First, the deduction of the leading soft graviton theorem in the large-time limit of the Carrollian stress tensor Ward identities required the identification \eqref{identific1}, where the leading soft operator defined in \eqref{leading and subleading soft operator} can be re-expressed in terms of the shear as \cite{Strominger:2013jfa,Kapec:2014opa}
\begin{align}
  -\frac{1}{2\pi}  S_0^+ (\infty, z,\bar{z})  
  &= C_{zz} (+\infty,z,\bar{z} ) - C_{zz} (-\infty,z,\bar{z} ) - D_{zz}(-\infty,z,\bar{z}) +  D_{zz}(+\infty,z,\bar{z}) \nonumber \\
  &= \int\limits_{-\infty}^{+\infty} du \, \partial_u [C_{zz} (u,z,\bar{z}) + D_{zz}(u,z,\bar{z})] \,.
\label{S0andC}
\end{align} Notice that the the second and the last term in the RHS of the first line cancel because of the antipodal matching condition $C_{zz}(-\infty,z,\bar{z})=D_{zz}(\infty,z,\bar{z})$ \cite{Kapec:2014opa}. To define the Carrollian operator $S_0^+(u,z,\bar{z})$, we just propose to extend this expression to all finite values of $u$ as below:
\begin{align}
    -\frac{1}{2\pi}S^+_0(u,z,\bar{z})&=C_{zz}(u,z,\bar{z})-C_{zz}(-\infty,z,\bar{z})-D_{zz}(-\infty,z,\bar{z})+D_{zz}(u,z,\bar{z})\nonumber\\
    &=\int\limits_{-\infty}^udu^\prime\text{ }\partial_{u^\prime}\left[C_{zz}(u^\prime,z,\bar{z})+D_{zz}(u^\prime,z,\bar{z})\right] \, . \label{61}
    \end{align}  This expression already tells us that $S^+_0(u,z,\bar{z})$ is directly related to the hard graviton operator \eqref{extrapolate} (see Section \ref{sec:More on the collinear limit} for more details). From \eqref{Tuu component}, we deduce the ${T^u}_u$ component of the Carrollian stress tensor in terms of the asymptotic shear
    \begin{equation}
    \begin{split}
     {T^u}_u (u,z,\bar{z})
 &= - 2\bar{\partial}^2\left[C_{zz}(u,z,\bar{z})+D_{zz}(u,z,\bar{z})-C_{zz}(-\infty,z,\bar{z})-D_{zz}(-\infty,z,\bar{z})\right] \, . 
 \end{split} \label{Tuucomp}
    \end{equation} In particular, this expression satisfies the retarded initial condition \eqref{eq:6}. Notice that the antipodal matching condition together with the usual electricity condition on the shear, $\partial^2 C_{\bar{z}\bar{z}}(\infty, z, \bar{z}) = \bar \partial^2 C_{{z}{z}}(\infty, z, \bar{z})$ and $\partial^2 D_{\bar{z}\bar{z}}(-\infty, z, \bar{z}) = \bar \partial^2 D_{{z}{z}}(-\infty, z, \bar{z})$, implies ${T^u}_u (\infty, z, \bar{z})  = {{\bar{T}}^u}_{\hspace{2.0mm}u}(\infty, z, \bar{z})$. 
    
Second, we also found the identification \eqref{identif2} in the derivation of the subleading soft graviton theorem from the Carrollian CFT Ward identities. The subleading soft graviton operator defined in \eqref{leading and subleading soft operator} can be rewritten in terms of the shear as 
\begin{equation}
   \frac{1}{2\pi} S^+_{1}(z,\bar{z})=\int\limits_{-\infty}^\infty du^\prime\text{ }{u^\prime}\text{ }\partial_{u^\prime}\left[C_{zz}(u^\prime,z,\bar{z})+D_{zz}(u^\prime,z,\bar{z})\right] \, .
\end{equation} Following the same procedure as above, we propose to define $S_{1e}^+(u,z,\bar{z})$ as 
\begin{equation}
    \frac{1}{2\pi} S_{1e}^+(u,z,\bar{z}) = \int\limits_{-\infty}^u du^\prime\text{ }{u^\prime}\text{ }\partial_{u^\prime}\left[C_{zz}(u^\prime,z,\bar{z})+D_{zz}(u^\prime,z,\bar{z})\right] \, .
\end{equation} From \eqref{12}, we infer
\begin{equation}
\begin{split}
     -\frac{1}{2\pi}S^+_1(u,z,\bar{z})
   &=\int\limits_{-\infty}^udu^\prime\text{ }(u-u^\prime)\text{ }\partial_{u^\prime}\left[C_{zz}(u^\prime,z,\bar{z})+D_{zz}(u^\prime,z,\bar{z})\right] \, .
   \end{split} \label{equ}
\end{equation}  Using \eqref{47}, \eqref{46} and \eqref{relation S1T}, we find the Carrollian stress tensor components in terms of the shear,
\begin{align}
    {T^u}_{\bar{z}} (u,z, \bar{z}) &=  \frac{1}{2\pi} \bar{\partial}^3 S_1^+ = - \int\limits_{-\infty}^udu^\prime\text{ } (u-u^\prime)\text{ }\partial_{u^\prime}
 \bar{\partial}^3\left[C_{zz}(u^\prime,z,\bar{z})+D_{zz}(u^\prime,z,\bar{z})\right]
 \label{Tubzcomp}
\end{align} and 
\begin{align}
    {T^u}_z  (u,z, \bar{z})  &=  \frac{1}{\pi} \bar\partial  T  =  - \int\limits_{-\infty}^udu^\prime\text{ }(u-u^\prime)\text{ }\partial_{u^\prime}\partial^3\left[C_{\bar{z}\bar{z}}(u^\prime,z,\bar{z})+D_{\bar{z}\bar{z}}(u^\prime,z,\bar{z})\right] \, . \label{Tuzcomp}
\end{align} Again, these expressions are compatible with the retarded initial condition \eqref{retarded initial}. However, they are non local with respect to the $u$ coordinate. A way to circumvent this problem is to use twistor potentials \cite{Newman:1976gc,Hansen:1978jz,Eastwood:1982,Adamo:2022mev} instead of spacetime metric to describe the gravitational field:
\begin{equation}
\begin{split}
    h (u, z, \bar{z}) &= \int\limits^u_{-\infty} du' \, [C_{zz} (u',z,\bar{z}) - C_{zz} (-\infty,z,\bar{z})  + D_{zz}(u',z,\bar{z}) -D_{zz}(-\infty,z,\bar{z})] \\
    &=  \int\limits^u_{-\infty} du' \, (u-u') \, \partial_{u'}[C_{zz} (u',z,\bar{z})   + D_{zz}(u',z,\bar{z}) ]
    \, . 
\end{split}
    \label{twistor potential}
\end{equation} This field corresponds to a deformation of the complex structure on twistor space and can be seen as a Carrollian ascendent of the shear, as defined in \eqref{ascendant}. Furthermore, it has recently been argued in \cite{Kmec:2024nmu} that $h$ is the natural symplectic variable of the radiative phase space at null infinity. Using this potential, the stress tensor components are local at $\mathscr{I}$ and read as
 \begin{equation}
 \begin{split}
      {T^u}_u &= - 2\bar{\partial}^2\partial_u h \, ,  \\
      {T^u}_{\bar z} &= - \bar{\partial}^3 h \, , \\
      {T^u}_z &= - \partial^3 \bar{h} \, .
    \label{twistor comp}
\end{split}
\end{equation}
It would be interesting to understand the role played by the field $h$ in the formulation of the holographic Carrollian CFT. Notice that the above stress tensor formally satisfies the classical Carrollian evolution/constraint equations $\partial_u  {T^u}_{\bar z}  = \frac{1}{2}\bar{\partial}  {T^u}_u$ and $\partial_u {T^u}_{z}  = \frac{1}{2}\partial  {{\bar{T}}^u}_{\hspace{2.0mm}u}$. 

Finally, the possible central extensions of $\mathfrak{bms}_4$ have been discussed in Appendix 2 of \cite{Barnich:2010ojg}. In Appendix \ref{sec:Central charge}, we compute the two-point function of the Carrollian CFT stress tensor and show that it vanishes, hence suggesting there is no central charge. The absence of central charge for the massless scattering in 4D flat space is in agreement with related analyses in \cite{Fotopoulos:2019vac,Donnay:2021wrk,Banerjee:2022wht,Fiorucci:2023lpb}.

\section{\texorpdfstring{$Lw_{1+\infty}$}{Lw(1+infinity)} Carrollian CFT currents}
\label{sec:Lw(1+infinity) Carrollian CFT currents}

In this section, we extend the above discussion to an infinite tower of Carrollian CFT currents at null infinity satisfying the $Lw_{1+\infty}$ algebra. We discuss how the existence of these currents in the holographic Carrollian CFT implies an infinite tower of projected soft graviton theorems in the bulk. Then we show that, at finite $u$, the OPEs of the Carrollian CFT currents encode the collinear limit of scattering amplitudes. We also show that these currents correspond to Carrollian ascendants of the hard graviton operator.

\subsection{OPE algebra of Carrollian CFT currents}\label{s3.1}
\label{sec:OPE algebra of Carrollian CFT currents}

The conformal Carrollian symmetries hitherto discussed are part of the defining properties of Carrollian CFTs. In particular, as described above, the leading \cite{Weinberg:1965nx} and subleading \cite{Cachazo:2014fwa} soft graviton theorems can be obtained from the generic stress tensor Ward identities of any such 3D Carrollian CFT. In light of the Carrollian holography proposal, this is perfectly consistent with the fact that the leading and the subleading soft graviton theorems are universal in a generic theory of quantum gravity in 4D asymptoptically flat spacetime \cite{Laddha:2017ygw}.

On the other hand, the subsubleading soft graviton theorem that occurs, e.g. at the tree-level Einstein gravity \cite{Cachazo:2014fwa}, is not universal \cite{Laddha:2017ygw,Elvang:2016qvq}. Equivalently in the dual Carrollian CFT description, as shown in \cite{Saha:2023abr}, we require an additional conformal Carrollian field $S^+_2$, besides the Carrollian generator fields $S^+_0$, $S^+_1$ and $T$ (and their $S^2$ shadows) inherited from the stress tensor, in order to probe the non-universal subsubleading (positive helicity) soft graviton theorem. This $S^+_2$ field, taken to be mutually local with $S^+_0$, $S^+_1$, and $T$,  was postulated, in the OPE limit, to exactly satisfy:
\begin{align}
\left(\partial_u S^+_2-S^+_1\right)(u,z,\bar{z})\Phi(\mathbf{x}_p)\sim0  \, . \label{20}
\end{align}
Together with \eqref{12}, this suggests the following decomposition of $S^+_2$ inside a correlator:
\begin{align}
&S^+_2(u,z,\bar{z})=S^+_{2e}(u,z,\bar{z})+uS^+_{1e}(u,z,\bar{z})+\frac{u^2}{2}S^+_0(u,z,\bar{z})\label{21}\\
\Longrightarrow\hspace{5mm}&{\partial_u}\langle S^+_{2e}(u,z,\bar{z})X\rangle=\left[\text{temporal contact terms}\right]\nonumber
\end{align}
where the $S^+_{2e}$ part consists of new modes not appearing in any of the $S^+_0$, $S^+_1$, $T$ fields. Clearly, the $S^+_2$ field\footnote{Since an analogous field $S^-_2$ with spin $s=-2$ to probe the negative helicity soft theorem can not be mutually local with $S^+_0$, $S^+_1$, $T$, we refrain from introducing any such field here.} has dimensions $(\Delta,s)=(-1,2)$.

In Appendix \ref{sec:Infinite tower of currents}, we sketch the outlines of the derivation, detailed in \cite{Saha:2023abr}, of the singular parts of the mutual OPEs between the fields $S^+_2$, $S^+_0$ and $S^+_1$. We then review there how the existence of the operator $S^+_2$ implies the \textit{automatic} existence of an infinite tower of mutually local Carrollian CFT currents $S^+_{k}$ ($k\geq3$) satisfying the following conditions:
\begin{align}
&\left(\partial_u S^+_{k}-S^+_{k-1}\right)(u,z,\bar{z})\Phi(\mathbf{x}_p)\sim0\hspace{15mm}\text{(OPE exact)}\label{27}\\
\text{and}\hspace{5mm}&\text{ }\bar{\partial}^{k+2} S^+_k(u,z,\bar{z})\Phi(\mathbf{x}_p)\sim0 \, .\label{28}
\end{align} 
Looking at the relation \eqref{27}, the field $S^+_k$ is decomposed inside a correlator as below:
\begin{align}
&S^+_k(u,z,\bar{z})=\sum_{r=1}^k \frac{u^{k-r}}{(k-r)!}\text{ } S^+_{r(e)}(u,z,\bar{z})+\frac{u^k}{k!}S^+_0(u,z,\bar{z})\label{29}\\
\Longrightarrow\hspace{5mm}&{\partial_u}\langle S^+_{r(e)}(u,z,\bar{z})X\rangle=\left[\text{temporal contact terms}\right]\nonumber
\end{align}
while the restriction \eqref{28} implies the following further decomposition of the object $S^+_{k(e)}$ inside a correlator:
\begin{align}
&S^+_{k(e)}(u,z,\bar{z})=\frac{1}{(k+1)!}\sum^{k+1}_{s=0}{k+1 \choose s}(-)^{k+1-s}\bar{z}^{s}\text{ } H^k_{\frac{k+1}{2}-s}(u,z,\bar{z})\label{30}\\
&S^+_0(u,z,\bar{z})=\bar{z}H^0_{-\frac{1}{2}}(u,z,\bar{z})-H^0_{\frac{1}{2}}(u,z,\bar{z})\nonumber\\
\text{such that}\hspace{2.5mm}&\langle\bar{\partial} H^k_{{\frac{k+1}{2}-s}}(u,z,\bar{z})X\rangle=-i\sum\limits_{p=1}^n\text{ }\theta(u-u_p)\text{ }\left[\text{contact terms on $S^2$}\right] \nonumber
\end{align} (this naturally extends the decomposition provided in Appendix \ref{sec:OPE formulae} for the cases $k=0,1$). We reiterate that the objects $S^+_{k(e)}$ are not local Carrollian fields but consist of modes. Since the weights of the $S^+_k$ field are $(\Delta,s)=(1-k,2)$, the holomorphic weight of both $S^+_k$ as well as $H^k_{{\frac{k+1}{2}-s}}$ is $h=\frac{3-k}{2}$.

Following the intrinsic, conformal Carrollian symmetry based derivation of the $S^+_2S^+_k$ OPE \eqref{35}, one obtains the most general $S^+_kS^+_l$ OPE\footnote{The two Carrollian operators whose product is to be expanded are inserted at different spatial-positions as well as at different times \cite{Saha:2022gjw}. Hence, no contact term survives in the OPE limit.} in the $j\epsilon$-form, using a very similar iterative algorithm, as \cite{Saha:2023abr}:
\begin{multline}
S_k^+(\mathbf{x})S^+_{l}(\mathbf{x}_p) \\
\sim\lim\limits_{\epsilon\rightarrow0^+}-i\sum_{r=0}^k
\frac{\left(u-u_p\right)^{k-r}}{(k-r)!}\sum_{m=0}^r\frac{\left(\bar{z}-\bar{z}_p\right)^{m+1}}{{(\Delta\tilde{z}_p)}}\text{ }\frac{(l+1)_{r-m}}{(r-m)!\cdot m!}\text{ }\bar{\partial}^m_p S^+_{r+l-1}(\mathbf{x}_p)  \, . \label{31}
\end{multline}
In Appendix \ref{finite u check}, we have verified this OPE starting from the graviton collinear splitting in the 4D Einstein-Yang-Mills theory.

Recalling the decomposition \eqref{29}, we may compare the $\mathcal{O}(u^0u_p^0)$ terms from the both sides of the OPE \eqref{31} to find that:
\begin{multline}
S_{k(e)}^+(u,z,\bar{z})S^+_{l(e)}(u_p,z_p,\bar{z}_p) \\
\sim\lim\limits_{\epsilon\rightarrow0^+}-i\sum_{m=0}^k\frac{\left(\bar{z}-\bar{z}_p\right)^{m+1}}{{(\Delta\tilde{z}_p)}}\text{ }\frac{(k+1-m)_l}{l!\cdot m!}\text{ }\bar{\partial}^m_p S^+_{k+l-1(e)}(u_p,z_p,\bar{z}_p)  \, .\label{32}
\end{multline}
This reproduces the OPE of celestial conformally soft positive-helicity gravitons at the tree level of Einstein gravity in the bulk \cite{Guevara:2021abz}, if we explicitly take the $u\rightarrow\infty$ limit so that $\frac{1}{\Delta\tilde{z}_p}$ reduces to $\frac{1}{z-z_p}$ for any\footnote{First we take the $u\rightarrow\infty$ limit, and then we remember that $S^+_{l(e)}(u_p,z_p,\bar{z}_p)$ is independent of $u_p$ in the OPE limit.} $u_p$. Indeed, our analysis establishes that the 2D celestial conformally soft graviton primary $H^k(z,\bar{z})$ of \cite{Guevara:2021abz} is the object $-iS_{1-k(e)}^+(\infty,z,\bar{z})$ that appears in the 3D conformal Carrollian primary field $-iS_{1-k}^+(u\rightarrow\infty,z,\bar{z})$.

To make the symmetry algebra manifest in the conformal Carrollian OPEs \eqref{31}, we first use the decomposition \eqref{30} on both sides of \eqref{32} and compare order-by-order in $\bar{z}$ and $\bar{z}_p$ to obtain \cite{Saha:2023abr}:
\begin{align}
H^k_{a}(u,z,\bar{z})H^l_{b}(u_p,z_p,\bar{z}_p)\sim\lim\limits_{\epsilon\rightarrow0^+}-i[a(l+1)-b(k+1)]\text{ }\frac{H^{k+l-1}_{a+b}(u_p,z_p,\bar{z}_p)}{\Delta\tilde{z}_p}\label{33}
\end{align}
with $-\frac{k+1}{2}\leq a\leq\frac{k+1}{2}$ and $-\frac{l+1}{2}\leq b\leq\frac{l+1}{2}$ ($2a,2b\in\mathbb{Z}$). Next, we write down the following mode-expansion of the object $H^k_a(u,z,\bar{z})$, keeping in mind that $H^k_a$ is independent of both $u$ and $\bar{z}$ in the OPE limit and that it has $h=\frac{3-k}{2}$:
\begin{align}
H^k_a(u,z,\bar{z})=\sum_{n\in\mathbb{Z}+\{\frac{k+1}{2}\}}H^k_{a;n} z^{-n-\frac{3-k}{2}}\, , \hspace{20mm}\{p\}\equiv\text{frac}(p)\,.\label{37}
\end{align}
Finally, applying the correspondence between 3D Carrollian CFT OPE and commutators (discussed e.g. in \cite{Saha:2023hsl}) on \eqref{33}, we extract the following mode-algebra:
\begin{align}
\left[H^k_{a;n}\text{ },\text{ }H^l_{b;m}\right]=-i\left[a(l+1)-b(k+1)\right]H^{k+l-1}_{a+b;n+m} \,.\label{34}
\end{align}
After the relabelling of modes as $w^p_{a;n}\equiv\frac{i}{2} H^{2p-3}_{a;n}$ with $1-p\leq a\leq p-1$, $p\in\frac{\mathbb{N}}{2}+1$, and $n+p\in\mathbb{Z}$, this algebra is recognized to be the (level $0$) Ka\v c-Moody algebra of the wedge subalgebra of $w_{1+\infty}$:
\begin{align*}
\left[w^p_{a;n}\text{ },\text{ }w^q_{b;m}\right]=\left[a(q-1)-b(p-1)\right]w^{p+q-2}_{a+b;n+m}
\end{align*}
which is the symmetry algebra at the level of the 2D celestial conformally soft graviton OPEs \cite{Strominger:2021lvk}. Only the modes of the two `universal' (i.e. present in any 3D Carrollian CFT) fields $S^+_0$ and $S^+_1$  among the infinite tower $\left\{S^+_k\right\}$ can form a subalgebra of the symmetry-algebra \eqref{34}, that is a semi-direct product of the $\overline{\mathfrak{sl}_2}$ Ka\v c-Moody algebra with the abelian supertranslations \cite{Saha:2023hsl}. Finally, the modes of the remaining field $T$ generate a Witt algebra so that the complete quantum symmetry of a 3D holographic Carrollian CFT possessing the local field $S^+_2$ is $\text{Witt}\loplus 
 Lw_{1+\infty}$ \cite{Saha:2023abr} (here $Lw_{1+\infty}$ denotes the loop of the wedge algebra of $w_{1+\infty}$). This naturally enhances the algebra previously considered in \eqref{algebra S0S1T}.

\subsection{Infinite tower of projected soft graviton theorems}
\label{sec:Infinite tower of soft graviton theorems}

We shall now revisit the argument presented in \cite{Saha:2023abr} to establish the direct connection between the infinite tower of the conformal Carrollian primaries $\left\{S^+_k\right\}$ and the infinite hierarchy of soft graviton theorems obeyed by amplitudes at tree-level after certain projections \cite{Hamada:2018vrw,Li:2018gnc}. In particular, we provide a symmetry-based justification of the crucial OPE formula \eqref{38} and extend the relation between the $u\to \infty$ limit in the Carrollian CFT and the soft limit in the bulk discussed in Section \ref{sec:Carrollian stress tensor for massless scattering}.

The OPE formulae for $S^+_2$ with a conformal Carrollian primary operator $\Phi$, the corresponding Ward identities and the relation with the subsubleading soft graviton theorem are reviewed in Appendix \ref{sec:S2 formulae}. Using a $\text{Witt}\loplus Lw_{1+\infty}$ symmetry argument, there we have also provided a partial derivation of the $S^+_2\Phi$ OPE \eqref{36}, originally postulated in \cite{Saha:2023abr}, hence providing a stronger case for the said OPE formula. As outlined towards the end of Appendix \ref{sec:S2 formulae}, beginning with the $S^+_2\Phi$ OPE \eqref{36} one can then recursively deduce the $S^+_{k\geq3}\Phi$ OPEs for a conformal Carrollian primary $\Phi$ with dimensions $(h,\bar{h})$ and $\left(\bar{{\bm{\xi}}}\cdot \Phi\right)=0=\left({{\bm{\xi}}}\cdot \Phi\right)$ as below: 
\begin{multline}
S_k^+(\mathbf{x})\Phi(\mathbf{x}_p) \\
\sim\lim\limits_{\epsilon\rightarrow0^+}-i\sum_{r=0}^k
\frac{\left(u-u_p\right)^{k-r}}{(k-r)!}\sum_{m=0}^r\frac{\left(\bar{z}-\bar{z}_p\right)^{m+1}}{{(\Delta\tilde{z}_p)}}\text{ }\frac{(-2\bar{h})_{r-m}}{(r-m)!\cdot m!}\text{ }\bar{\partial}^m_p \Phi_{r-1}(\mathbf{x}_p)\label{38}
\end{multline}
where the unique local ascendant fields $\left\{\Phi_r\right\}$ satisfy (in the OPE limit):
\begin{align}
\partial_u^r\Phi_r\sim\Phi \text{ ($r\geq1$)}\hspace{5mm},\hspace{5mm}\Phi_0=\Phi\hspace{5mm}\text{and}\hspace{5mm}\Phi_{-1}=\dot{\Phi} \, .\label{44}
\end{align} 
It is reassuring to note that, besides the symmetry-derived $S^+_kS^+_l$ OPEs \eqref{31}, there exist other examples, found in \cite{Mason:2023mti} by explicitly analysing the bulk Einstein-Yang-Mills theory, consistent with the above $S^+_k\Phi$ OPE \eqref{38}. To see this, we first apply the decomposition \eqref{29} to extract the $S^+_{1-k(e)}\Phi$ part from the $S^+_{1-k}\Phi$ OPE and then impose the $u\rightarrow\infty$ limit so that $\frac{1}{\Delta\tilde{z}_p}$ becomes $\frac{1}{z-z_p}$. Substituting appropriate values for $\bar{h}$ next, one finally recovers the OPEs in \cite{Mason:2023mti} of the (celestial) soft graviton operators with the Carrollian graviton and gluon operators. This observation is again consistent with the expectation that the 2D celestial conformally soft graviton primary $H^k(z,\bar{z})$ is nothing but the object $-iS^+_{1-k(e)}(\infty,z,\bar{z})$. We will come back to this identification in Section \ref{sec:Carrollian CFT currents and radiative data at}. More importantly, we have also obtained the OPE \eqref{38} at finite $u$ directly from the Einstein-Yang-Mills collinear splitting \eqref{86} in Appendix \ref{finite u check}.


We now discuss how the infinite number of `projected' soft graviton theorems \cite{Hamada:2018vrw,Li:2018gnc} can be encoded into the Ward identities of the infinite tower of conformal Carrollian primaries $S^+_k$. We note that the $S_{k\geq3}^+$ Ward identity below can not be uniquely fixed (even up to temporal contact terms) by demanding finiteness (at $z\neq z_p$) of the $\langle S_{k\geq3}^+(u,z,\bar{z})X\rangle$ correlators; they are rather determined\footnote{For a primary singlet $\Phi$ with holomorphic weight $h$, its correlators satisfy: $\lim\limits_{z\rightarrow\infty}z^{2h}\left\langle \Phi(u,z,\bar{z})X\right\rangle=\text{finite}$, just as in a 2D CFT. Since the field $S^+_{k}$ has $h=\frac{3-k}{2}$, the correlators $\left\langle S^+_{k\geq3}(u,z,\bar{z})X\right\rangle$ grow polynomially as $z^{k-3}$ near $z\rightarrow\infty$. These polynomials can not be fixed from symmetry alone. In contrast, the $\left\langle S^+_{k\leq2}(u,z,\bar{z})X\right\rangle$ correlators vanish at $z\rightarrow\infty$.} up to a $(k-3)$-th degree polynomial in $z$: 
\begin{multline}
i\langle S_k^+(u,z,\bar{z}) X\rangle \\
=\sum_{p=1}^n\theta(u-u_p)\sum_{r=0}^k
\frac{\left(u-u_p\right)^{k-r}}{(k-r)!}\sum_{m=0}^r\frac{\left(\bar{z}-\bar{z}_p\right)^{m+1}}{{z-z_p}}\text{ }\frac{(-2\bar{h}_p)_{r-m}}{(r-m)!\cdot m!}\text{ }\bar{\partial}^m_p \partial_{u_p}^{1-r}\langle X\rangle \, .\label{39}
\end{multline}
A very similar situation arises for the 4D graviton amplitudes. The (energetically) soft graviton factorization extends beyond the subsubleading order only for a certain projected part of the amplitude; such a projection leaves out a homogeneous piece that can not be fixed by the Ward identities of the large diffeomorphisms \cite{Hamada:2018vrw} or by gauge invariance \cite{Li:2018gnc}. It is then plausible that this homogeneous piece of the $\text{(sub)}^k$-leading soft graviton amplitude corresponds to the undetermined polynomial-in-$z$ part of the $\langle S_{k(e)}^+(\infty,z,\bar{z})X\rangle$ `correlator' for $k\geq3$ with both vanishing for $k\leq2$.

In any case, similarly to what was found in Section \ref{sec:Carrollian stress tensor for massless scattering} for the cases $k=0,1$, the $\langle S_{k(e)}^+(u,z,\bar{z}) X\rangle$ part of the above $S^+_k$ Ward identity is equivalent to the projected part of the $\text{(sub)}^k$-leading energetically soft graviton theorem in the $u\rightarrow\infty$ limit when all the primaries in $X$ possess $\Delta=1$ and $\left(\bar{{\bm{\xi}}}\cdot \Phi\right)=0=\left({{\bm{\xi}}}\cdot \Phi\right)$. This can be directly inferred from the observation that, by temporal-Fourier transforming the `partial' $S^+_k$ Ward identity \eqref{39} and then explicitly making the energy of the field $S^+_k$ soft and substituting all $\Delta_p=1$, one obtains the `shifted' soft graviton expansion of the projected amplitude \cite{Saha:2023abr}:
\begin{align}
\lim\limits_{\omega\rightarrow0^+}\left\langle \tilde{S}_k^+(\epsilon\omega,z,\bar{z}) \tilde{X}_{\text{out}}\tilde{X}_{\text{in}}\right\rangle=
\lim\limits_{\omega\rightarrow0^+}\lim\limits_{\delta\rightarrow0^+}i^{k-1}\left[\sum_{r=0}^k\frac{F^{(r)}}{(\epsilon\omega+i\delta)^{k+1-r}}+\mathcal{O}(\omega^0)\right] \left\langle\tilde{X}_{\text{out}}\tilde{X}_{\text{in}}\right\rangle\label{40}
\end{align} where $\tilde{S}^+_k(\epsilon\omega,z,\bar{z})$ is the temporal-Fourier transformation of the field $S^+_k(u,z,\bar{z})$,
\begin{align}
    \label{temporal fourier}\tilde{S}^+_k(\epsilon\omega,z,\bar{z})=\int\limits_{-\infty}^\infty du\text{ }e^{i\epsilon\omega u}S^+_k(u,z,\bar{z}) \hspace{10mm}\text{(with $\omega>0$)} \, , 
\end{align} similar for the operators in $\tilde{X}$ and we have used the fact that:
\begin{align}
  \int\limits_{-\infty}^\infty du\text{ }e^{i\epsilon\omega u}\text{ }\theta\left(u-u_p\right)\frac{\left(u-u_p\right)^n}{n!}=\lim\limits_{\delta\rightarrow0^+}\frac{e^{i\epsilon\omega u_p}}{\left(-i\epsilon\omega+\delta\right)^{n+1}} \, .\label{127}  
\end{align}
The $i\delta$-prescription above bears the imprint of the retarded initial conditions, e.g. \eqref{eq:6}. In \eqref{40}, $F^{(0)}$ is the Weinberg leading soft factor \cite{Weinberg:1965nx,He:2014laa}, $F^{(1)}$ is the Cachazo-Strominger subleading soft factor \cite{Cachazo:2014fwa,Kapec:2014opa} and $F^{(2)}$ is the subsubleading soft factor \cite{Cachazo:2014fwa,Conde:2016rom}. The quantity $F^{(k)}$ has a $(k+1)$-th degree polynomial in $(\bar{z}-\bar{z}_p)$ in the numerator, and $(z-z_p)$ in the denominator and depends on $\{\omega_p\}$ and $\{s_p\}$. The terms at $\mathcal{O}(\omega^0)$ and beyond can not be completely
determined from the Ward identity \eqref{39} alone for finite $k$, because of the finite contributions coming from the yet-unknown temporal contact terms. On the other hand, the 
$\omega\rightarrow0$ Fourier counterpart of the relation \eqref{27} which is exact at $u\rightarrow\infty$:
\begin{align}
\lim\limits_{\delta\rightarrow0^+}\lim\limits_{\omega\rightarrow0^+}(-i\epsilon\omega+\delta)\left\langle \tilde{S}_{k+1}^+(\epsilon\omega,z,\bar{z}) \tilde{X}_{\text{out}}\tilde{X}_{\text{in}}\right\rangle=\lim\limits_{\omega\rightarrow0^+}\left\langle \tilde{S}_{k}^+(\epsilon\omega,z,\bar{z}) \tilde{X}_{\text{out}}\tilde{X}_{\text{in}}\right\rangle\label{42}
\end{align}
implies that the $\mathcal{O}(\omega^0)$ term of the soft factor in \eqref{40} must be $F^{(k+1)}$ and, by induction, the $\mathcal{O}(\omega^p)$ term, $F^{(k+1+p)}$ with $p\geq0$. Thus, multiplying both sides of \eqref{40} with $(\epsilon\omega)^k$, recognizing that: 
\begin{align*}
\lim\limits_{\delta\rightarrow0^+}\lim\limits_{\omega\rightarrow0^+}(-i\epsilon\omega+\delta)^k\left\langle \tilde{S}_{k}^+(\epsilon\omega,z,\bar{z}) \tilde{X}_{\text{out}}\tilde{X}_{\text{in}}\right\rangle=\lim\limits_{\omega\rightarrow0^+}\left\langle \tilde{S}_{0}^+(\epsilon\omega,z,\bar{z}) \tilde{X}_{\text{out}}\tilde{X}_{\text{in}}\right\rangle
\end{align*}
and finally setting $k\rightarrow\infty$, we recover the complete standard soft graviton expansion of the projected amplitudes \cite{Li:2018gnc} in 4D flat space as an insertion of the soft operator $\tilde{S}_{0}^+(\epsilon\omega\rightarrow0,z,\bar{z})$:
\begin{align}
\lim\limits_{\omega\rightarrow0^+}i\left\langle \tilde{S}_0^+(\epsilon\omega,z,\bar{z}) \tilde{X}_{\text{out}}\tilde{X}_{\text{in}}\right\rangle=
\lim\limits_{\delta\rightarrow0^+}\lim\limits_{\omega\rightarrow0^+}\left[\frac{F^{(0)}}{\epsilon\omega+i\delta}+\sum\limits_{k=0}^\infty{F^{(k+1)}}{(\epsilon\omega+i\delta)^{k}}\right] \left\langle\tilde{X}_{\text{out}}\tilde{X}_{\text{in}}\right\rangle \,.\label{41}
\end{align}
That we needed to explicitly take the energetically soft limit for the $i\tilde{S}_0^+(\epsilon\omega,z,\bar{z})$ field to obtain the above soft graviton expansion, suggests that the insertion of the 3D conformal Carrollian field $iS^+_0(u,z,\bar{z})$ in a correlator may give us interesting information on graviton scattering amplitudes in flat space at finite energy. The central goal of the present work is to establish that this is indeed the case.

\subsection{Collinear limit of scattering amplitudes}

As we have reviewed, the $\langle S_{k(e)}^+(u,z,\bar{z}) X\rangle$ part of the $\langle S_{k}^+(u,z,\bar{z}) X\rangle$ Ward identity \eqref{39} provides, exclusively in the $u\rightarrow\infty$ limit, a 3D Carrollian holographic description of the tree-level projected $(\text{sub})^k$-leading soft graviton theorem in 4D. We now wish to uncover the 4D bulk physics at non-zero finite $\omega$ that is encoded by these Ward identities or the corresponding OPEs \eqref{38} when $u$ is kept finite.

Clearly, the yet unknown temporal contact terms in the Ward identities \eqref{39} will produce, upon temporal Fourier transformation, non-zero contributions at generic $\omega$. So, we have to first determine these temporal contact terms. For this purpose, taking cue from the above-described derivation of the tree-level projected soft graviton expansion \eqref{41}, we demand that the following relation holds exactly (for all $k\geq0$): 
\begin{align}
    \partial_u S^+_{k+1}(u,z,\bar{z})=S^+_k(u,z,\bar{z})\label{43}
\end{align}
by extending the validity of \eqref{27} which was only an OPE-level statement (i.e. valid up to temporal contact terms). Hence, the primaries $S_k^+(u,z,\bar{z})$ correspond to an infinite tower of $\partial_u$-ascendants of $S^+_0(u,z,\bar{z})$, as defined in \eqref{ascendant}. Meanwhile, $\partial_u S^+_{0}(u,z,\bar{z})$ must still vanish up to temporal contact terms as before. The main motivation behind demanding \eqref{43} is to extend the soft-relation \eqref{42} to any non-zero $\omega$. Similarly, the OPE-level relations \eqref{44} are also extended into the following exact conditions:
\begin{align}
\partial_u^r\Phi_r=\Phi \text{ ($r\geq1$)}\hspace{5mm},\hspace{5mm}\Phi_0=\Phi\hspace{5mm}\text{and}\hspace{5mm}\Phi_{-1}=\dot{\Phi} \,.\label{129}
\end{align}
By virtue of the conditions \eqref{43}, the temporal contact terms in an $\langle S_{k}^+(u,z,\bar{z}) X\rangle$ correlator are all related to those in any other $\langle S_{l}^+(u,z,\bar{z}) X\rangle$.

With the exact constraints \eqref{43} in possession, we are now ready to take temporal Fourier transformation of the $\langle S_{\{k\}}^+(u,z,\bar{z}) X\rangle$ Ward identities. But, since the $\langle S_{k}^+(u,z,\bar{z}) X\rangle$ Ward identity (up to temporal contact terms) for $k\geq3$ can be fixed by symmetry arguments alone only up to a $(k-3)$-th degree polynomial in $z$, it will be judicial to temporal Fourier transform the corresponding OPE \eqref{38} instead, by keeping track of the temporal contact terms, to reliably recover the complete physics of the 4D massless scattering amplitudes at the leading-order in the collinear singularity as that undetermined polynomial is clearly subleading in the OPE/collinear limit. Restoring the temporal step-function in the OPE \eqref{38}, keeping in mind that there are yet unknown temporal contact terms, performing the temporal Fourier transformation \eqref{temporal fourier} and finally using \eqref{127}, one obtains (with $\omega,\omega_p\geq0$):
\begin{align}
\tilde{S}_k^+&(\epsilon\omega,z,\bar{z})\tilde{\Phi}(\epsilon_p\omega_p,z_p,\bar{z}_p) \label{45} \\
\sim\lim\limits_{\delta\rightarrow0^+}&-i(-i\epsilon\omega+\delta)^{-k}\text{ }\sum_{r=0}^k
\left(\frac{\epsilon\omega+i\delta}{\epsilon_p\omega_p+\epsilon\omega}\right)^{r-1}\sum_{m=0}^r\Big[\frac{\left(\bar{z}-\bar{z}_p\right)^{m+1}}{{z-z_p}}\text{ }\frac{(-2\bar{h}_p)_{r-m}}{(r-m)!\cdot m!}\nonumber\\
&\times\text{ } \bar{\partial}^m_p\tilde{\Phi}(\epsilon_p\omega_p+\epsilon\omega,z_p,\bar{z}_p)\Big]+f_k(\epsilon\omega,z,\bar{z};\epsilon_p\omega_p+\epsilon\omega,z_p,\bar{z}_p) \nonumber
\end{align}
where the function $f_k$ collects the contributions from the temporal contact terms and $\sim$ denotes `modulo terms regular in the holomorphic collinear limit'. It is noteworthy that the shift in energy $\omega_p$ of the field $\tilde\Phi$ by the amount $\omega$ has its origin in the temporal step-function hidden in the $j\epsilon$-prescription of the OPE \eqref{38}.

Iterating the relation \eqref{43} $l$-times, it is readily observed that the coefficient of $\frac{(u-u_p)^n}{n!}$ with $0\leq n\leq l$ in the $S^+_{k+l+1}\Phi$ OPE is the same as that of the $\partial_u^{l-n}\delta(u-u_p)$ temporal contact term in the $S^+_k\Phi$ OPE.  Moreover, the $S^+_k\Phi$ OPE \eqref{38} is a $k$-th degree polynomial in $(u-u_p)$. Thus, we expect the $S^+_{k=\infty}\Phi$ OPE to generate the temporal contact terms in the $S^+_l\Phi$ OPEs for finite $l$ via the iteration of \eqref{43}, while containing none itself. This amounts to setting $f_{k=\infty}=0$ identically in the momentum space that, from \eqref{45}, leads to:
\begin{align}
&\lim\limits_{\delta\rightarrow0^+}i\left[(-i\epsilon\omega+\delta)^{k}\tilde{S}_k^+(\epsilon\omega,z,\bar{z})\right]_{k=\infty}\tilde{\Phi}(\epsilon_p\omega_p,z_p,\bar{z}_p)\label{92}\\
\sim&\lim\limits_{\delta\rightarrow0^+} \sum_{r=0}^\infty
\left(\frac{\epsilon\omega+i\delta}{\epsilon_p\omega_p+\epsilon\omega}\right)^{r-1}\sum_{m=0}^r\left[\frac{\left(\bar{z}-\bar{z}_p\right)^{m+1}}{{z-z_p}}\text{ }\frac{(-2\bar{h}_p)_{r-m}}{(r-m)!\cdot m!}\times\text{ }\bar{\partial}^m_p \tilde{\Phi}(\epsilon_p\omega_p+\epsilon\omega,z_p,\bar{z}_p)\right] \, .\nonumber
\end{align}
Finally, remembering that:
\begin{align}
    \lim\limits_{\delta\rightarrow0^+}&(-i\epsilon\omega+\delta)^{k}\tilde{S}_k^+(\epsilon\omega)=\tilde{S}_0^+(\epsilon\omega)\label{102}\\
    &(-i\epsilon_p\omega_p)^{r}\tilde{\Phi}_r(\epsilon_p\omega_p)=\tilde{\Phi}(\epsilon_p\omega_p)\nonumber
\end{align}
where the first line is now valid for any $\omega\geq0$, $k\geq0$ as a consequence of the exact relation \eqref{43} with the retarded initial condition, while the second one for any $\omega_p\geq0$, $r\geq-1$ due to the relations \eqref{129} with the temporal boundary condition unspecified, and performing the summation,\footnote{Though this summation converges only when $\left\vert\frac{\epsilon\omega}{\epsilon_p\omega_p+\epsilon\omega}\right\vert<1$, one can analytically continue to any $\epsilon_p\omega_p\neq-\epsilon\omega$.} we find:
\begin{align} 
&i\tilde{S}_0^+(\epsilon\omega,z,\bar{z})\tilde{\Phi}(\epsilon_p\omega_p,z_p,\bar{z}_p)\label{48} \\
\sim&\lim\limits_{\delta\rightarrow0^+} \frac{\bar{z}-\bar{z}_p}{{z-z_p}} \text{ }\frac{\epsilon_p\omega_p-i\delta}{\epsilon\omega+i\delta} 
\left(\frac{\epsilon_p\omega_p+\epsilon\omega}{\epsilon_p\omega_p-i\delta}\right)^{{s}_p}\text{ }\tilde{\Phi}\left(\epsilon_p\omega_p+\epsilon\omega,z_p,\frac{\left(\epsilon_p\omega_p-i\delta\right)\bar{z}_p+(\epsilon\omega+i\delta)\bar{z}}{\epsilon_p\omega_p+\epsilon\omega}\right)\nonumber
\end{align}
after substituting $\Delta_p=1$ since we want the conformal Carrollian primary $\Phi$ to correspond to a 4D massless scattering field with `outgoing' helicity $s_p$.

Intriguingly, the RHS above is recognised to be displaying the leading order universal holomorphic collinear splitting property involving at least one positive helicity graviton $h_{+2}$, of the 4D Einstein-Yang-Mills amplitudes \cite{Bern:1998sv,Pate:2019lpp}. More precisely, the `OPE' \eqref{48} formally resembles the following collinear splitting (in the limit ${z\approx z_p}$ and $\omega,\omega_p>0$):
\begin{align}
{\Phi}^*_{s_p}\left(\epsilon_p\omega_p+\epsilon\omega,z_p,\frac{\epsilon_p\omega_p\bar{z}_p+\epsilon\omega\bar{z}}{\epsilon_p\omega_p+\epsilon\omega}\right)\longrightarrow h_{+2}(\epsilon\omega,z,\bar{z}){\Phi}_{s_p}(\epsilon_p\omega_p,z_p,\bar{z}_p)\label{86}
\end{align}
where the energies and helicities are expressed in the `all outgoing' convention and $\Phi^*$ denotes a virtual particle. This observation provides yet another strong evidence in support of the direct correspondence between the conformal Carrollian primary $S^+_0(u,z,\bar{z})$ and the 4D positive helicity (hard) graviton $h_{+2}$. The explicit correspondence was introduced in Section \ref{sec:Carrollian CFT stress tensor} and will be further discussed in Section \ref{sec:Carrollian CFT currents and radiative data at} for all $S^+_k(u,z,\bar{z})$. The $S^+_k\Phi$ OPEs \eqref{38} at finite $u$ thus collectively encode the aforementioned collinear splitting property of the 4D bulk Einstein-Yang-Mills amplitudes.

It is worth emphasizing that we have reached the bulk null momentum space holomorphic collinear splitting property \eqref{48} by temporal Fourier transforming the $S^+_{k=\infty}\Phi$ OPE that was itself obtained from purely Carrollian physics, without any hint from the bulk theory. We only needed to demand that a field $S^+_2$ obeying the relation \eqref{43}$_{k=1}$ exists in the Carrollian CFT and that the two assumptions stated in Appendix \ref{sec:Infinite tower of currents} hold. Using only the Carrollian symmetry arguments under those two assumptions, the general $S^+_kS^+_l$ OPE \eqref{31} and the corresponding symmetry algebra $Lw_{1+\infty}$ were derived first and then, using that algebra, all the $S^+_{k}\Phi$ OPEs \eqref{38} could be inferred (as explained in Appendix \ref{sec:S2 formulae}).

In celestial holography, the above sequence was derived in quite the opposite way. The bulk collinear splitting \eqref{86} is itself the starting point (usually) for deriving, via Mellin transformations, the general OPE involving at least one celestial conformal primary graviton \cite{Pate:2019lpp} from which the 2D celestial conformally soft graviton OPE \cite{Guevara:2021abz}, that contains the $ Lw_{1+\infty}$ symmetry \cite{Strominger:2021lvk}, is obtained. Though one can utilize the constraints arising from the 2D celestial subsubleading conformally soft graviton theorem to derive the former OPE \cite{Pate:2019lpp}, include contributions from all the anti-holomorphic descendants \cite{Guevara:2021abz} and finally perform the inverse Mellin transformations to obtain the bulk collinear splitting \eqref{48}, the first step itself explicitly requires some inputs from the bulk Einstein-Yang-Mills theory,\footnote{However in \cite{Banerjee:2024hvb}, working in a light-transformed basis, the leading gluon OPE singularity along with the coefficients was uniquely fixed using only symmetry arguments, without any input from the bulk.} unlike the purely holographic Carrollian derivation described above.

\subsection{Carrollian CFT currents at \texorpdfstring{$\mathscr{I}$}{Scri}}
\label{sec:Carrollian CFT currents and radiative data at}

We now extend the discussion presented in Section \ref{sec:Carrollian CFT stress tensor} to express the whole tower of Carrollian CFT currents $S^+_k(u,z,\bar{z})$ in terms of bulk radiative modes \eqref{66prime}-\eqref{66}. First, the relation \eqref{43} can be inverted as
\begin{align}
    S^+_{k+1}(u,z,\bar{z})=\int\limits_{-\infty}^udu^\prime\text{ }S^+_k(u^\prime,z,\bar{z}) \, .
\end{align}
 Starting from the proposal \eqref{61} for the field $S^+_0$, and using the above formula iteratively, we can find the following relations between the fields $S^+_{k\geq0}$ and the asymptotic shear:   
\begin{align}   
   -\frac{1}{2\pi}S^+_k(u,z,\bar{z})
   &=\frac{1}{k!}\int\limits_{-\infty}^udu^\prime\text{ }(u-u^\prime)^k\text{ }\partial_{u^\prime}\left[C_{zz}(u^\prime,z,\bar{z})+D_{zz}(u^\prime,z,\bar{z})\right]\label{62}\\
   &=\sum_{r=0}^k \frac{u^{k-r}}{(k-r)!}\cdot\frac{(-)^r}{r!}\int\limits_{-\infty}^udu^\prime\text{ }{u^\prime}^r\text{ }\partial_{u^\prime}\left[C_{zz}(u^\prime,z,\bar{z})+D_{zz}(u^\prime,z,\bar{z})\right] \, . \nonumber
\end{align}
Comparing the second line above with the decomposition \eqref{29}, we infer that (with $S^+_{0(e)}\equiv S^+_0$):
\begin{align}
-\frac{1}{2\pi}S^+_{r(e)}(u,z,\bar{z})=\frac{(-)^r}{r!}\int\limits_{-\infty}^udu^\prime\text{ }{u^\prime}^r\text{ }\partial_{u^\prime}\left[C_{zz}(u^\prime,z,\bar{z})+D_{zz}(u^\prime,z,\bar{z})\right] \, .\label{63}  
\end{align}
As suggested above, we can identify the celestial conformally soft graviton primary $H^{1-k}(z,\bar{z})$ with the object $-iS^+_{k(e)}(\infty,z,\bar{z})$. This leads to the following well-known expression of the former in terms of the shear \cite{Freidel:2021ytz}: 
\begin{align}
H^{1-k}(z,\bar{z})&\equiv-iS^+_{k(e)}(u\rightarrow\infty,z,\bar{z}) \nonumber\\
&=\frac{2\pi i(-)^{k}}{k!}\int\limits_{-\infty}^\infty du^\prime\text{ }{u^\prime}^k\text{ }\partial_{u^\prime}\left[C_{zz}(u^\prime,z,\bar{z})+D_{zz}(u^\prime,z,\bar{z})\right] \, . 
\end{align}

It was shown in \cite{Saha:2023abr} that all the fields $S^+_k(u,z,\bar{z})$ are 3D conformal Carrollian primaries of the form \eqref{carrollian primary} i.e. singlets. In view of the relation \eqref{62}, this should also follow from the transformation properties of the asymptotic shear under the bulk supertranslations and superrotations if the proposed correspondence \eqref{61} is to be valid. We check this explicitly. Under the supertranslation with parameter $\mathcal{T}(z,\bar{z})$, the variation of the shear is given by (see e.g. \cite{Barnich:2010eb}): 
\begin{equation}
    \begin{split}
 &\delta_\mathcal{T} C_{zz}(u,z,\bar{z})=\mathcal{T}(z,\bar{z})\partial_uC_{zz}(u,z,\bar{z})-2\partial_z^2\mathcal{T}(z,\bar{z}) \, , \\
 &\delta_\mathcal{T}D_{zz}(u,z,\bar{z})=\mathcal{T}(z,\bar{z})\partial_uD_{zz}(u,z,\bar{z})+2\partial_z^2\mathcal{T}(z,\bar{z}) \, ,
\end{split} \label{shear transfo}
\end{equation}
with the antipodal matching condition \cite{Kapec:2014opa, Strominger:2017zoo} already taken care of. Accordingly, the variation of the field $S^+_k$ \eqref{62} is found to be:
\begin{align}
  -\frac{1}{2\pi}\delta_\mathcal{T} S^+_k(u,z,\bar{z})
   &=\frac{1}{k!}\int\limits_{-\infty}^udu^\prime\text{ }(u-u^\prime)^k\text{ }\partial_{u^\prime}\left[\delta_\mathcal{T}C_{zz}(u^\prime,z,\bar{z})+\delta_{\mathcal{T}}D_{zz}(u^\prime,z,\bar{z})\right] \nonumber\\
   &=\mathcal{T}(z,\bar{z})\cdot\frac{1}{k!}\int\limits_{-\infty}^udu^\prime\text{ }(u-u^\prime)^k\text{ }\partial_{u^\prime}^2\left[C_{zz}(u^\prime,z,\bar{z})+D_{zz}(u^\prime,z,\bar{z})\right]\nonumber\\
   &=-\frac{1}{2\pi}\mathcal{T}(z,\bar{z})S^+_{k-1}(u,z,\bar{z}) \\
   &=-\frac{1}{2\pi}\mathcal{T}(z,\bar{z})\partial_u S^+_{k}(u,z,\bar{z})\nonumber
\end{align}
where the last line follows from \eqref{43}. This is the correct transformation law of a conformal Carrollian primary singlet under supertranslation. To go from the second line to the third line, we have assumed that\footnote{Schwartzian falloffs in $u$ are usually assumed in the phase space approach of $Lw_{1+\infty}$, see e.g. \cite{Freidel:2021ytz,Freidel:2022skz,Geiller:2024bgf,Kmec:2024nmu}.} 
\begin{align}
   \lim_{u \to-\infty}u^k\partial_{u}\left[C_{zz}(u,z,\bar{z})+D_{zz}(u,z,\bar{z})\right]=0\hspace{5mm}\text{for any $k\geq0$} \, .\label{64} 
\end{align} Thus, even though the shear individually transforms inhomogeneously as in \eqref{shear transfo} under supertranslation, the fields $S^+_k$ all have homogeneous transformation properties since the inhomogeneous pieces cancel each other due to the combination \eqref{62}.

It is worth mentioning that none of the objects $S^+_{k(e)}(u,z,\bar{z})$ \eqref{63} for $k\geq1$ transforms as a conformal Carrollian primary under the supertranslations since
\begin{align}
\label{transfoS0}
 \delta_\mathcal{T} S^+_{k(e)}(u,z,\bar{z})= \mathcal{T}(z,\bar{z})\left[\partial_u S^+_{k(e)}+S^+_{k-1(e)}\right](u,z,\bar{z}) \, . 
\end{align}
The $u\to \infty$ limit of these transformations is however consistent with results previously found in the literature for the subleading soft graviton, see e.g. \cite{Donnay:2022hkf,Agrawal:2023zea}.

Let us now consider a holomorphic superrotation with parameter $\mathcal{Y}(z)$, under which the variation of the shear is \cite{Barnich:2010eb}
\begin{align}
 &\delta_\mathcal{Y}C_{zz}(u,z,\bar{z})=\left[\partial \mathcal{Y}(z)\left(\frac{3}{2}+\frac{u}{2}\partial_u\right)+\mathcal{Y}(z)\partial\right]C_{zz}(u,z,\bar{z})-u\partial^3\mathcal{Y}(z) \, , \\
 &\delta_\mathcal{Y}D_{zz}(u,z,\bar{z})=\left[\partial \mathcal{Y}(z)\left(\frac{3}{2}+\frac{u}{2}\partial_u\right)+\mathcal{Y}(z)\partial\right]D_{zz}(u,z,\bar{z})+u\partial^3\mathcal{Y}(z) 
  \, , \nonumber
\end{align}
with the anti-podal matching condition \cite{Kapec:2014opa,Strominger:2017zoo} already implemented. The variation of the field $S^+_k$ is then obtained as
\begin{align}
  \frac{-1}{2\pi}\delta_\mathcal{Y} S^+_k(u,z,\bar{z})
   =&\frac{1}{k!}\int\limits_{-\infty}^udu^\prime\text{ }(u-u^\prime)^k\text{ }\partial_{u^\prime}\left[\delta_\mathcal{Y}C_{zz}(u^\prime,z,\bar{z})+\delta_{\mathcal{Y}}D_{zz}(u^\prime,z,\bar{z})\right] \nonumber\\
   =&\frac{1}{k!}\int\limits_{-\infty}^udu^\prime\text{ }(u-u^\prime)^k\text{ }\partial_{u^\prime}\left[\left\{\partial \mathcal{Y}(z)\left(\frac{3}{2}+\frac{u^\prime}{2}\partial_{u^\prime}\right)+\mathcal{Y}(z)\partial\right\}C_{zz}(u^\prime,z,\bar{z})\right.\nonumber\\
   &\left. -u^\prime\partial^3\mathcal{Y}(z)  +\left\{\partial \mathcal{Y}(z)\left(\frac{3}{2}+\frac{u^\prime}{2}\partial_{u^\prime}\right)+\mathcal{Y}(z)\partial\right\}D_{zz}(u^\prime,z,\bar{z})+u^\prime\partial^3\mathcal{Y}(z)\right]\nonumber\\
   =&\frac{-1}{2\pi}\mathcal{Y}(z)\partial S^+_k(u,z,\bar{z})+2\cdot\frac{-1}{2\pi}\partial \mathcal{Y}(z)S^+_k(u,z,\bar{z})\nonumber\\
   &+\partial \mathcal{Y}(z)\cdot\frac{1}{2\cdot k!}\int\limits_{-\infty}^udu^\prime\text{ }u^\prime(u-u^\prime)^k\text{ }\partial_{u^\prime}^2\left[C_{zz}(u^\prime,z,\bar{z})+D_{zz}(u^\prime,z,\bar{z})\right] \nonumber\\
   =&\frac{-1}{2\pi}\mathcal{Y}(z)\partial S^+_k(u,z,\bar{z})+\partial \mathcal{Y}(z)\left(\frac{3-k}{2}+\frac{u}{2}\partial_u\right)\frac{-1}{2\pi}S^+_k(u,z,\bar{z}) \, ,
\end{align}
exactly as required for a conformal Carrollian primary singlet $S^+_k$ with holomorphic weight $h=\frac{3-k}{2}$, see \eqref{carrollian primary}. Here, to reach the last line from the third one, we have used the assumption \eqref{64} and the relation \eqref{43}. Again, the inhomogeneous terms in the variations of the individual shears get mutually cancelled inside the combination \eqref{62}, leaving the variations of the fields $S^+_k$ under the superrotations homogeneous. In a similar way, from the variation of the shear under the anti-holomorphic superrotations, we can also verify that the field $S^+_k$ \eqref{62} transforms as a conformal Carrollian primary with $\bar{h}=-\frac{1+k}{2}$.

We may also consider the following variations of the shears under the Carrollian diffeomorphisms \eqref{9} expressed as $\zeta=\bar{V}(z,\bar{z})\bar{\partial}+\frac{u}{2}\bar{\partial}\bar{V}(z,\bar{z})\partial_u$ \cite{Campiglia:2015yka,Compere:2018ylh} (with $C_{z\bar{z}}=0$ initially):
\begin{align}
 &\delta_{\bar{V}}C_{zz}(u,z,\bar{z})=\left[\bar{\partial}\bar{V}\left(\frac{u}{2}\partial_u-\frac{1}{2}\right)+\bar{V}(z,\bar{z})\bar{\partial}\right]C_{zz}(u,z,\bar{z})-u\partial^2\bar{\partial}\bar{V} \, , \\
 &\delta_{\bar{V}}D_{zz}(u,z,\bar{z})=\left[\bar{\partial}\bar{V}\left(\frac{u}{2}\partial_u-\frac{1}{2}\right)+\bar{V}(z,\bar{z})\bar{\partial}\right]D_{zz}(u,z,\bar{z})+u\partial^2\bar{\partial}\bar{V}\nonumber
 \end{align}
to find, upon comparison with \eqref{108}, that the fields $S^+_k$ \eqref{62} transform as primary singlets with $\bar{h}=-\frac{1+k}{2}$ under the diffeomorphisms \eqref{9}: 
\begin{align}
\delta_{\bar{V}}S^+_k(u,z,\bar{z})=\bar{V}(z,\bar{z})\bar{\partial} S^+_k(u,z,\bar{z})+\bar{\partial}\bar{V}(z,\bar{z})\left(\frac{u}{2}\partial_u-\frac{1+k}{2}\right)S^+_k(u,z,\bar{z})
\end{align}
which confirms that $\left\{S^+_k\right\}$ are primary fields also under the $\text{Witt} \loplus ( \overline{\mathfrak{sl}_2} \loplus {\mathfrak{s}}_\wedge)$ symmetry. (See Appendix \ref{sec:Full Lorentz symmetries} for the defining conditions of a primary multiplet under the said symmetry.)

Specifying the above discussion to $k=0,1$ and remembering that $\partial_uS^+_0\sim0$ inside an OPE \eqref{7}, one can verify explicitly that the components of the Carrollian CFT stress tensor identified in \eqref{Tuucomp} and \eqref{Tubzcomp} transform as a conformal Carrollian primary multiplet (see discussion around Equation \eqref{stress tensor rep}) under the algebra \eqref{algebra S0S1T}, i.e.
\begin{equation}
    \begin{split}
\delta_{(\mathcal{T}_{\wedge},\mathcal{Y}, \bar{{V}})} {T^u}_u =& \Big[\mathcal{Y} \partial + \bar{V} \bar \partial + \frac{3}{2} \partial \mathcal{Y} + \frac{3}{2}  \bar \partial\bar{V} \Big] {T^u}_u \hspace{10mm}\left(\text{since }\partial_u{T^u}_u\sim0\right)\, , \\
\delta_{(\mathcal{T}_{\wedge},\mathcal{Y}, \bar{V})} {T^u}_{\bar z} =& \Big[\Big(\mathcal{T}_{\wedge} + \frac{u}{2}(\partial \mathcal{Y} + \bar \partial \bar{V}) \Big)  \partial_u + \mathcal{Y} \partial + \bar{V} \bar \partial +  \partial \mathcal{Y} + 2 \bar \partial\bar{V} \Big] {T^u}_{\bar z}\\
&+ \frac{3}{2} \bar\partial \Big(\mathcal{T}_{\wedge} + \frac{u}{2} \bar \partial \bar{V} \Big) {T^u}_u \, , 
    \end{split}
\end{equation}
where $\mathcal{T}_{\wedge}$ and $\bar{V}$ satisfy the following `wedge conditions': $\bar{\partial}^2\mathcal{T}_{\wedge}=0=\bar{\partial}^3\bar{V}$. However, under a supertranslation $\mathcal{T}(z,\bar{z})$ not obeying the above wedge condition and an anti-holomorphic superrotation $\bar{\mathcal{Y}}(\bar{z})$, the components ${T^u}_u$ and ${T^u}_{\bar z}$ do not transform as primaries.

Similarly, the component ${T^u}_{z}$ given by \eqref{Tuzcomp} does not transform as a primary under the general supertranslations $\mathcal{T}(z,\bar{z})$ and the holomorphic superrotations $\mathcal{Y}(z)$. But, under the wedge supertranslations $\bar{\mathcal{T}}_{\wedge}(z,\bar{z})$ obeying $\partial^2\bar{\mathcal{T}}_{\wedge}=0$, diffeomorphisms $V(z,\bar{z})$ conjugate to \eqref{9} and satisfying $\partial^3V=0$, and arbitrary anti-holomorphic superrotations $\bar{\mathcal{Y}}(\bar{z})$, it transforms as a primary multiplet: 
\begin{align}
    \delta_{(\bar{\mathcal{T}}_{\wedge},V,\bar{\mathcal{Y}})} {T^u}_z =& \Big[\Big(\bar{\mathcal{T}}_{\wedge} + \frac{u}{2}(\partial V + \bar \partial \bar{\mathcal{Y}}) \Big)  \partial_u + V\partial + \bar{\mathcal{Y}} \bar \partial + 2 \partial V +\bar \partial\bar{\mathcal{Y}} \Big] {T^u}_z  \nonumber\\
&+ \frac{3}{2} \partial \Big(\bar{\mathcal{T}}_{\wedge} + \frac{u}{2}\partial V \Big) {{\bar{T}}^u}_{\hspace{2.0mm}u} \, .
\end{align}

Importantly, the intersection of all the different kinds of transformations discussed above consists of precisely the ten global conformal Carrollian transformations. Thus, the stress-tensor components \eqref{Tuucomp} (and its conjugate), \eqref{Tubzcomp}, and \eqref{Tuzcomp} form a conformal Carrollian quasi-primary multiplet with scaling dimension $\Delta=3$ and spin-boost matrices given by \eqref{stress tensor rep}.

\subsection{Carrollian CFT currents in momentum space}

To make the relation with 4D scattering amplitudes more explicit, it is useful to present the expressions of $S_k^+$ in momentum space, in terms of creation and annihilation operators. Let us start with $S^+_0$. Adapting the mode decomposition \eqref{66prime}-\eqref{66} to the retarded initial condition \eqref{eq:6}, inserting those into \eqref{61}, and subtracting the non-Fourier soft contributions at $u=-\infty$,\footnote{The subtractions of $C_{zz}(-\infty, z, \bar{z})$ and $D_{zz}(-\infty,z, \bar{z})$ in \eqref{61} cancel the non-Fourier soft contributions, thus allowing for a well-defined Fourier-mode expansion; see e.g. \cite{Kraus:2024gso, Jorstad:2024yzm} for related discussions.} we find
\begin{align}
S^+_0(u,z,\bar{z})=\lim\limits_{\delta\rightarrow0^+} i\int\limits_{0}^\infty \frac{d\omega}{2\pi}&\left[e^{-i\omega u}\left\{a_+^{\text{out}}-{a_+^{\text{in}}}\right\}(\omega+i\delta,z,\bar{z})\right.\nonumber\\
&\left.\qquad\qquad-\text{ } e^{i\omega u}\left\{{a_-^{\text{out}}}^\dagger-{a_-^{\text{in}}}^\dagger\right\}(\omega-i\delta,z,\bar{z})\right]
\, . \label{S0intermsofa} 
\end{align}
The temporal-Fourier transformation of the field $S^+_0(u,z,\bar{z})$ defined in \eqref{temporal fourier} is then found to be (suppressing the $(z,\bar{z})$ dependence):
\begin{align}
\tilde{S}^+_0(\omega>0)=\lim\limits_{\delta\rightarrow0^+}i\left[a_+^{\text{out}}-{a_+^{\text{in}}}\right](\omega+i\delta)\hspace{2.5mm};\hspace{2.5mm}
\tilde{S}^+_0(\omega<0)=\lim\limits_{\delta\rightarrow0^+}i\left[{a_-^{\text{in}}}^\dagger-{a_-^{\text{out}}}^\dagger\right](-\omega-i\delta) 
 \, . \label{67} 
\end{align}
Now, starting from \eqref{61}$_{u\to \infty}$ in position space at $\mathscr{I}$, then using the mode-expansions \eqref{66prime}-\eqref{66}, and the following delta-function convention:
\begin{align*}
    \int\limits_0^\infty d\omega\text{ }\delta(\omega)f(\omega)= \int\limits_{-\infty}^\infty d\omega\text{ }\delta(\omega) \theta(\omega)  f(\omega)  =\frac{1}{2}f(0^+) \, , 
\end{align*} one can obtain
\begin{align}
S^+_0(u\rightarrow\infty,z,\bar{z})&=\lim\limits_{\omega\rightarrow0^+}\lim\limits_{\delta\rightarrow0^+}\frac{\omega}{2}\left[\left\{a_+^{\text{out}}-a_+^{\text{in}}\right\}(\omega+i\delta,z,\bar{z})-\left\{{a_-^{\text{in}}}^\dagger-{a_-^{\text{out}}}^\dagger\right\}(\omega-i\delta,z,\bar{z})\right] \nonumber \\
&=\lim\limits_{\omega\rightarrow0^+}\lim\limits_{\delta\rightarrow0^+}\frac{\omega}{2}\left[a_+^{\text{out}}(\omega+i\delta,z,\bar{z})-{a_-^{\text{in}}}^\dagger(\omega-i\delta,z,\bar{z})\right]
\end{align} where the last equality holds inside of the correlators (compare with \eqref{leading and subleading soft operator}). The insertion of the field $iS^+_0(u\rightarrow\infty,z,\bar{z})$ in the Carrollian CFT correlators directly gives rise to the leading soft graviton theorem in the bulk \eqref{leading soft them position} expressed for Carrollian amplitudes. One can further see that the relation \eqref{63} leads to the following expressions for the objects $S^+_{k(e)}(u\rightarrow\infty,z,\bar{z})$ in terms of the (soft) graviton creation and annihilation operators: 
\begin{align}
S^+_{k(e)}(u\rightarrow\infty) &=\lim\limits_{\omega\rightarrow0^+}\lim\limits_{\delta\rightarrow0^+}\frac{1}{k!}\left(i\partial_\omega\right)^k\cdot\frac{\omega}{2}\left[\left\{a_+^{\text{out}}-{a_+^{\text{in}}}\right\}(\omega+i\delta)+(-)^k\left\{{a_-^{\text{out}}}^\dagger-{a_-^{\text{in}}}^\dagger\right\}(\omega-i\delta)\right] \nonumber\\
&=\lim\limits_{\omega\rightarrow0^+}\lim\limits_{\delta\rightarrow0^+}\frac{1}{k!}\left(i\partial_\omega\right)^k\cdot\frac{\omega}{2}\left[a_+^{\text{out}}(\omega+i\delta)+(-)^{k+1}{a_-^{\text{in}}}^\dagger(\omega-i\delta) \right] 
\end{align} where the last equality holds inside of the correlators. This is consistent with the earlier observation that the primary `Ward identity' $i\langle S_{k(e)}^+(\infty,z,\bar{z})X\rangle$ describes the bulk tree-level (projected, for $k\geq3$) (sub)$^k$-leading soft graviton theorem.

\subsection{More on the collinear limit}
\label{sec:More on the collinear limit}

We shall now proceed to physically interpret the LHS of the (bulk) null-momentum space `OPE' \eqref{48} in light of the momentum space representation \eqref{67} (suppressing the $i\delta$-prescription for brevity), stemming from the correspondence \eqref{61}, of the Carrollian CFT primary field $S^+_0$. We recall that the RHS of the `OPE' \eqref{48} has been recognized to be describing the leading order holomorphic collinear splitting \eqref{86} (with $\omega,\omega_p>0$), involving at least one positive helicity (in the `outgoing convention') graviton, of the 4D Einstein-Yang-Mills amplitudes.

Choosing $\epsilon=\epsilon_p=1$ in \eqref{48} and using $\tilde{S}^+_0(\omega>0)$ \eqref{67}, we get
\begin{align*}
\left[a_+^{\text{out}}-{a_+^{\text{in}}}\right](\omega,z,\bar{z})\tilde{\Phi}_{s_p}(\omega_p,z_p,\bar{z}_p)\sim -\frac{\bar{z}-\bar{z}_p}{{z-z_p}} \text{ }\frac{\omega_p}{\omega} 
\left(\frac{\omega_p+\omega}{\omega_p}\right)^{{s}_p}\text{ }\tilde{\Phi}_{s_p}\left(\omega_p+\omega,z_p,\frac{\omega_p\bar{z}_p+\omega\bar{z}}{\omega_p+\omega}\right) \, . 
\end{align*}
Since, the contribution of the ${a_+^{\text{in}}}$ operator to this `OPE' vanishes, we are left with
\begin{align}
a_+^{\text{out}}(\omega,z,\bar{z})\tilde{\Phi}_{s_p}(\omega_p,z_p,\bar{z}_p)\sim -\frac{\bar{z}-\bar{z}_p}{{z-z_p}} \text{ }\frac{\omega_p}{\omega} 
\left(\frac{\omega_p+\omega}{\omega_p}\right)^{{s}_p}\text{ }\tilde{\Phi}_{s_p}\left(\omega_p+\omega,z_p,\frac{\omega_p\bar{z}_p+\omega\bar{z}}{\omega_p+\omega}\right)\label{88}
\end{align} 
that consistently describes the following holomorphic collinear splitting involving an outgoing positive helicity graviton in the 4D Einstein-Yang-Mills theory
\begin{align}
{\Phi}^{*\text{out}}_{s_p}\left(\omega_p+\omega,z_p,\frac{\omega_p\bar{z}_p+\omega\bar{z}}{\omega_p+\omega}\right)\longrightarrow {a_+^{\text{out}}}(\omega,z,\bar{z}){\Phi}^{\text{out}}_{s_p}(\omega_p,z_p,\bar{z}_p) \, .\label{89}
\end{align}

Similarly, choosing $\epsilon=-\epsilon_p=1$ in \eqref{48} leads to (in the `all outgoing' convention):
\begin{align}
a_+^{\text{out}}(\omega,z,\bar{z})\tilde{\Phi}_{s_p}(-\omega_p,z_p,\bar{z}_p)\sim\frac{\bar{z}-\bar{z}_p}{{z-z_p}} \text{ }\frac{\omega_p}{\omega} 
\left(\frac{\omega_p-\omega}{\omega_p}\right)^{{s}_p}\text{ }\tilde{\Phi}_{s_p}\left(-(\omega_p-\omega),z_p,\frac{\omega_p\bar{z}_p-\omega\bar{z}}{\omega_p-\omega}\right)\label{90}
\end{align}
by virtue of the vanishing contribution from ${a_+^{\text{in}}}$, which describes the following two 4D Einstein-Yang-Mills collinear splittings by crossing symmetry:
\begin{align}
&{\Phi}^{\text{in}\dagger}_{(-s_p)}(\omega_p,z_p,\bar{z}_p)\longrightarrow {a_+^{\text{out}}}(\omega,z,\bar{z}){\Phi}^{*\text{in}\dagger}_{(-s_p)}\left(\omega_p-\omega,z_p,\frac{\omega_p\bar{z}_p-\omega\bar{z}}{\omega_p-\omega}\right)\hspace{10mm}(\omega_p>\omega) \, , \label{91}\\
&{\Phi}^{*\text{out}}_{s_p}\left(\omega-\omega_p,z_p,\frac{\omega\bar{z}-\omega_p\bar{z}_p}{\omega-\omega_p}\right){\Phi}^{\text{in}\dagger}_{(-s_p)}(\omega_p,z_p,\bar{z}_p)\longrightarrow {a_+^{\text{out}}}(\omega,z,\bar{z})\hspace{10mm}(\omega>\omega_p) \, . \nonumber
\end{align} 
Thus, we see that $\tilde{S}^+_0(\omega>0,z,\bar{z})$ given by \eqref{67} naturally corresponds to the outgoing positive helicity graviton.

On the other hand, choosing $\epsilon=\epsilon_p=-1$ in \eqref{48}, using $\tilde{S}^+_0(\omega<0)$ \eqref{67}, and remembering that ${a_+^{\text{out}\dagger}}$ does not contribute, one finds (with $a_+\equiv a_+^{\text{out}}$ and ${a_-^{\text{in}\dagger}}(\omega)=a_+(-\omega)$ for $\text{Re}(\omega)>0$, by crossing) that
\begin{align*}
a_+(-\omega,z,\bar{z})\tilde{\Phi}_{s_p}(-\omega_p,z_p,\bar{z}_p)\sim -\frac{\bar{z}-\bar{z}_p}{{z-z_p}} \text{ }\frac{\omega_p}{\omega} 
\left(\frac{\omega_p+\omega}{\omega_p}\right)^{{s}_p}\text{ }\tilde{\Phi}_{s_p}\left(-(\omega_p+\omega),z_p,\frac{\omega_p\bar{z}_p+\omega\bar{z}}{\omega_p+\omega}\right) \, ,
\end{align*}
where all the energies and helicities are denoted in the `all outgoing' convention in the last line, that describes the following collinear splitting involving an incoming negative helicity graviton in the 4D Einstein-Yang-Mills theory:
\begin{align*}
{a_-^{\text{in}\dagger}}(\omega,z,\bar{z}){\Phi}^{\text{in}\dagger}_{(-s_p)}(\omega_p,z_p,\bar{z}_p)\longrightarrow\Phi^{*\text{in}\dagger}_{(-s_p)}\left(\omega_p+\omega,z_p,\frac{\omega_p\bar{z}_p+\omega\bar{z}}{\omega_p+\omega}\right) \, .
\end{align*}
This is the CPT transformed version of the splitting scenario \eqref{89}, but with the same splitting coefficient \eqref{88}. Similarly, the CPT transformations of the collinear splittings \eqref{91} lead to:
\begin{align*}
&{a_-^{\text{in}\dagger}}(\omega,z,\bar{z}){\Phi}^{*\text{out}}_{s_p}\left(\omega_p-\omega,z_p,\frac{\omega_p\bar{z}_p-\omega\bar{z}}{\omega_p-\omega}\right)\longrightarrow{\Phi}^{\text{out}}_{s_p}(\omega_p,z_p,\bar{z}_p)\hspace{10mm}(\omega_p>\omega) \, , \\
&{a_-^{\text{in}\dagger}}(\omega,z,\bar{z})\longrightarrow{\Phi}^{*\text{in}\dagger}_{(-s_p)}\left(\omega-\omega_p,z_p,\frac{\omega\bar{z}-\omega_p\bar{z}_p}{\omega-\omega_p}\right){\Phi}^{\text{out}}_{s_p}(\omega_p,z_p,\bar{z}_p)\hspace{10mm}(\omega>\omega_p)
\end{align*}
described by the same splitting coefficient \eqref{90} (in the `all outgoing' convention):
\begin{align*}
a_+(-\omega,z,\bar{z})\tilde{\Phi}_{s_p}(\omega_p,z_p,\bar{z}_p)\sim\frac{\bar{z}-\bar{z}_p}{{z-z_p}} \text{ }\frac{\omega_p}{\omega} 
\left(\frac{\omega_p-\omega}{\omega_p}\right)^{{s}_p}\text{ }\tilde{\Phi}_{s_p}\left(\omega_p-\omega,z_p,\frac{\omega_p\bar{z}_p-\omega\bar{z}}{\omega_p-\omega}\right)
\end{align*}
(with $\omega>0$) which can be directly obtained by putting $\epsilon=-\epsilon_p=-1$ in \eqref{48} and using $\tilde{S}^+_0(\omega<0)$ \eqref{67}. Hence, $\tilde{S}^+_0(\omega<0)$ \eqref{67} corresponds to the incoming negative helicity graviton.

Thus, the momentum space `OPE' \eqref{48} and its associated splitting scenario \eqref{86} contain the six Einstein-Yang-Mills holomorphic collinear splittings: \eqref{89}, \eqref{91} and their CPT transformations, collectively given by, in the `all out' convention,
with $\Omega_i\equiv\epsilon_i\omega_i$:
\begin{align}
&{\Phi}^*_{s_p}\left(\Omega_p+\Omega,z_p,\frac{\Omega_p\bar{z}_p+\Omega\bar{z}}{\Omega_p+\Omega}\right)\longrightarrow h_{+2}(\Omega,z,\bar{z}){\Phi}_{s_p}(\Omega_p,z_p,\bar{z}_p) \nonumber\\
\lim\limits_{\delta\rightarrow0^+}& a_{+}(\Omega+i\delta,z,\bar{z}){\Phi}_{s_p}(\Omega_p,z_p,\bar{z}_p)\label{113}\\
&\sim\lim\limits_{\delta\rightarrow0^+}-\frac{\bar{z}-\bar{z}_p}{{z-z_p}} \text{ }\frac{\Omega_p-i\delta}{\Omega+i\delta}\left(\frac{\Omega_p+\Omega}{\Omega_p-i\delta}\right)^{{s}_p}{\Phi}_{s_p}\left(\Omega_p+\Omega,z_p,\frac{\left(\Omega_p-i\delta\right)\bar{z}_p+\left(\Omega+i\delta\right)\bar{z}}{\Omega_p+\Omega}\right) \, . \nonumber
\end{align}
The above discussion also shows that the bulk CPT invariance and crossing relations naturally justifies the following regularization prescription for the binomial expansion in \eqref{92} outside its circle of convergence:
\begin{align}
\lim\limits_{\delta\rightarrow0^+}\sum_{n=0}^\infty\frac{(-2\bar{h}_p)_{n}}{n!}\left(\frac{\Omega+i\delta}{\Omega_p+\Omega}\right)^{n-1}=\lim\limits_{\delta\rightarrow0^+}\frac{\Omega_p-i\delta}{\Omega+i\delta}\left(\frac{\Omega_p+\Omega}{\Omega_p-i\delta}\right)^{1-2\bar{h}_p}\label{114}
\end{align}
in the context of the 4D bulk tree-level massless amplitude calculation in the `all outgoing' convention\footnote{Alternatively, since the amplitudes are actually distributions due to the presence of the momentum-conserving delta-functions, \eqref{114} may be thought to be valid in a distributional sense. Explicit checks with bulk amplitudes in Appendix \ref{finite u check} confirm this interpretation.} such that the `OPE' \eqref{48} is expressed as (with $1-2\bar{h}_p=s_p$):
\begin{align}
&-i\tilde{S}^+_0(\Omega,z,\bar{z}){\Phi}_{s_p}(\Omega_p,z_p,\bar{z}_p)\nonumber\\
\sim&\lim\limits_{\delta\rightarrow0^+}-\frac{\bar{z}-\bar{z}_p}{{z-z_p}} \text{ }\frac{\Omega_p-i\delta}{\Omega+i\delta}\left(\frac{\Omega_p+\Omega}{\Omega_p-i\delta}\right)^{{s}_p}{\Phi}_{s_p}\left(\Omega_p+\Omega,z_p,\frac{\left(\Omega_p-i\delta\right)\bar{z}_p+\left(\Omega+i\delta\right)\bar{z}}{\Omega_p+\Omega}\right)
\label{fullOmega}
\end{align}
which is valid for any $\Omega,\Omega_p\in\mathbb{R}$ (the variables $\Omega$ and $\Omega_p$ are interpreted as (bilateral) Fourier conjugates to $u$ and $u_p$). Moreover, the validity of the soft graviton expansion \eqref{41} includes both $\Omega\rightarrow0^\pm$:
\begin{align*}
\lim\limits_{\Omega\rightarrow0}i\left\langle \tilde{S}_0^+(\Omega,z,\bar{z}) \tilde{X}_{\text{out}}\tilde{X}_{\text{in}}\right\rangle=
\lim\limits_{\Omega\rightarrow0}\lim\limits_{\delta\rightarrow0^+}\left[\frac{F^{(0)}}{\Omega+i\delta}+\sum\limits_{k=0}^\infty{F^{(k+1)}}{\left(\Omega+i\delta\right)^{k}}\right] \left\langle\tilde{X}_{\text{out}}\tilde{X}_{\text{in}}\right\rangle
\end{align*}
with the $\Omega\rightarrow0^-(0^+)$ limit denoting the incoming negative (outgoing positive) helicity soft graviton theorem. The above expansion can then be generated solely by an insertion of the operator $\lim\limits_{\Omega\rightarrow0}\lim\limits_{\delta\rightarrow0^+}-a_+(\Omega+i\delta,z,\bar{z})$ into the 4D $\mathcal{S}$-matrix. Hence, as a consequence of the CPT invariance and crossing symmetry of the 4D bulk gravity theory, the insertions of the operators $-i\tilde{S}^+_0(\Omega,z,\bar{z})$ and $\lim\limits_{\delta\rightarrow0^+}a_+(\Omega+i\delta,z,\bar{z})$ (with $\Omega\in\mathbb{R}$) into a bulk massless $\mathcal{S}$-matrix are equivalent to each other,  
leading to the following correspondence in the `all outgoing' convention: 
\begin{align}
\tilde{S}^+_0(\Omega,z,\bar{z})&=\lim\limits_{\delta\rightarrow0^+}ia_+(\Omega+i\delta,z,\bar{z})\hspace{4mm}(\text{for }\Omega\in\mathbb{R})\nonumber\\
\Longrightarrow\hspace{2mm}{{S}}^+_0(u,z,\bar{z})&=\lim\limits_{\delta\rightarrow0^+}i\int\limits_{-\infty}^\infty \frac{d\Omega}{2\pi}\text{ }e^{-i\Omega u}\text{ }a_+(\Omega+i\delta,z,\bar{z})\label{93}\\
&=\lim\limits_{\delta\rightarrow0^+}i\int\limits_{0}^\infty \frac{d\omega}{2\pi}\left[e^{-i\omega u}\text{ }a_+^{\text{out}}(\omega+i\delta,z,\bar{z})+e^{i\omega u}\text{ }{a_-^{\text{in}}}^\dagger(\omega-i\delta,z,\bar{z})\right]\nonumber \, .
\end{align}
Note that this is completely equivalent to the correspondence \eqref{S0intermsofa} at the level of bulk $\mathcal{S}$-matrix insertion, since ${a_+^{\text{in}}}$ and ${a_-^{\text{out}}}^\dagger$ do not contribute to the $\mathcal{S}$-matrix.
Thus, \eqref{93} establishes the direct correspondence between the conformal Carrollian field ${S}^+_0(u,z,\bar{z})$ and the 4D positive (in the `outgoing' sense i.e. both outgoing positive and incoming negative) helicity graviton. The correspondence in \eqref{93} is also consistent with the conventions chosen in Section 11 of \cite{Mason:2023mti}, where the Fourier integral is performed over all real values.

Furthermore, we note from the relation \eqref{102} (valid for $\Omega\in\mathbb{R}$) and the correspondence \eqref{93} that:
\begin{align}
&\tilde{{S}}^+_k(\Omega,z,\bar{z})=\lim\limits_{\delta\rightarrow0^+}\frac{\tilde{{S}}^+_0(\Omega,z,\bar{z})}{\left(-i\Omega+\delta\right)^k}=\lim\limits_{\delta\rightarrow0^+}\frac{ia_+(\Omega+i\delta,z,\bar{z})}{\left(-i\Omega+\delta\right)^k}\hspace{5mm}(\text{for }\Omega\in\mathbb{R})\nonumber\\
\Longrightarrow\hspace{5mm}&{S}^+_k(u,z,\bar{z})=\lim\limits_{\delta\rightarrow0^+}i\int\limits_{-\infty}^\infty \frac{d\Omega}{2\pi}\text{ }e^{-i\Omega u}\text{ }\frac{a_+(\Omega+i\delta,z,\bar{z})}{\left(-i\Omega+\delta\right)^k} \, .  \label{112}
\end{align}

With the temporal boundary condition for the field ${S}^+_k(u,z,\bar{z})$ unspecified, the above correspondence is formally generalized as (for $\Omega\in\mathbb{R}-\{0\}$):
\begin{align}
\tilde{{S}}^+_k(\Omega,z,\bar{z})=\frac{ia_+(\Omega,z,\bar{z})}{\left(-i\Omega\right)^k}
\hspace{2.5mm}\Longleftrightarrow\hspace{2.5mm}{S}^+_k(u,z,\bar{z})=\int\limits_{-\infty}^\infty \frac{d\Omega}{2\pi}\text{ }e^{-i\Omega u}\text{ }\frac{ia_+(\Omega,z,\bar{z})}{\left(-i\Omega\right)^k}  \label{128}
\end{align}
where the $\Omega$-integral is understood in the principal value sense.

In Appendix \ref{sec:Applications to bulk amplitudes}, we illustrate the definitions and the results presented in this section on explicit examples of massless scattering amplitudes.

\section{Discussion}
\label{sec:Discussion}

In this paper, using the Carrollian approach to flat space holography, we showed that the large $u$ limit of the Carrollian CFT stress tensor Ward identities correctly reproduces the leading and subleading soft graviton theorems in the bulk. This allowed us to explicitly express the Carrollian stress tensor in terms of the hard graviton operators, and thus the bulk radiative modes. The universality of the leading and subleading soft theorems reflects the presence of a stress tensor as part of the definition of the holographic Carrollian CFT. Moreover, we showed that the inclusion of an operator $S^+_2$ in the Carrollian CFT, from which a whole tower of $Lw_{1+\infty}$ currents automatically follows, implies an infinite hierarchy of projected soft graviton theorems for Einstein gravity in the bulk in the $u \to \infty$ limit. We also demonstrated that the $Lw_{1+\infty}$ currents coincide with an infinite tower of $\partial_u$-Carrollian ascendants of the hard graviton operator, and their OPEs at finite $u$ correctly encode the collinear limit of scattering amplitudes in the bulk. In particular, the fact that the stress tensor at finite $u$ encodes information about the bulk hard graviton is reminiscent of the AdS/CFT correspondence, where stress tensor correlation functions in the CFT coincide with bulk graviton correlators.

Let us emphasize that, in our approach, we began with the properties of the Carrollian CFT to deduce aspects of the bulk theory—opposite to the approach commonly used in the literature, which typically translates amplitude statements into the dual theory. This allows us to have more control over the properties the Carrollian CFT should satisfy, such as OPE associativity or constraints on the spectrum.

\medskip

In the following, we relate our results with previous analyses in the literature. 
\paragraph{Sourced Ward identities} In \cite{Donnay:2022aba,Donnay:2022wvx}, it was shown that sourced Ward identities for a Carrollian CFT stress tensor reproduce the leading and subleading soft graviton theorems. Our current analysis does not require the introduction of external sources but is nevertheless compatible with this previous result. Indeed, the stress tensor considered there was constructed from the non-radiative subsector of the gravitational solution space (inspired by results in 3D gravity \cite{Barnich:2006av,Fareghbal:2013ifa,Detournay:2014fva,Bagchi:2015wna,Ciambelli:2020eba,Campoleoni:2022wmf}) and encodes information on the Bondi mass and angular momentum aspects. This subsector is typically used to describe bound states. External sources with respect to this subsector were required to capture the radiation at null infinity (see also the discussion section of \cite{Donnay:2022wvx} for a summary). In the present work, we focus on the complementary subsector of 4D gravity—specifically, the `pure radiation' sector describing massless scattering. In particular, the `source operators' with respect to the non-radiative phase space considered in \cite{Donnay:2022wvx} are precisely the Carrollian primary operators for the massless scattering states considered here. Thus, the stress tensor constructed in this work should not be compared to the stress tensor considered in these earlier references, as they describe different subsectors. One of the important challenges of flat space holography is to understand the interplay between these two bulk subsectors--massless scattering and bound states--from a holographic perspective.

\paragraph{The large $u$ limit} In this work, we have shown that the large $u$ limit in the Carrollian CFT coincides with the soft limit in the bulk theory. This is physically intuitive, as the observation of soft/large wavelength particles at null infinity requires considering large time intervals at $\mathscr{I}$. This led us to the observation that, at finite time, the Carrollian CFT Ward identities encode not only the soft theorems but also the collinear factorization properties of bulk scattering amplitudes. Although similar in appearance, the approach used here differs from the one proposed very recently in \cite{Bagchi:2024gnn} (based on \cite{Dutta:2022vkg}). One key difference is that we consider the quantum Ward identities associated with the Carrollian CFT stress tensor to reproduce the soft theorems. By contrast, in \cite{Bagchi:2024gnn}, the stress tensor is obtained by solving classical Carrollian constraints/time evolution equations for the stress tensor (thereby omitting the contact terms arising from the insertion of Carrollian primaries in the correlators), and the time-independent integration functions on the celestial sphere are promoted as components of the Carrollian CFT stress tensor operator at $\mathscr{I}$. In fact, the stress tensor considered there is essentially the integrated BMS flux momenta discussed in \cite{Donnay:2021wrk}. These objects are non-local at $\mathscr{I}$ when expressed in terms of the asymptotic shear and naturally live on the celestial sphere. Indeed, they can be viewed as corners of the celestial diamonds discussed in \cite{Pasterski:2021fjn,Pasterski:2021dqe}. Therefore, the analysis in \cite{Bagchi:2024gnn} does not capture the finite $u$ information of the Carrollian CFT Ward identities—namely, the hard graviton operators—but only the soft subsector. Moreover, as discussed in Section \ref{sec:Comments on the symmetries} (and also in Appendix \ref{A}), the quantum symmetries of the holographic Carrollian CFT \eqref{algebra S0S1T} imply ${T^z}_{\bar{z}} = 0$ (together with ${T^{\bar{z}}}_{z} = 0$ if we include all the diffeomorphisms on the celestial sphere for superrotations). Thus, these stress tensor components do not encode further information about the hard graviton.

\medskip

Let us conclude with some perspectives. First, soft photon theorems for Carrollian amplitudes have been recently derived in \cite{Kim:2023qbl,Kraus:2024gso} from conformal Carrollian partition functions at $\mathscr{I}$. It would be interesting to explore whether the results obtained here can be directly derived from the Carrollian CFT partition function for gravity.

Additionally, it would be important to establish an independent definition of the Carrollian CFT that captures the features of scattering amplitudes in flat space. Preliminary steps have been made in this direction by taking Carrollian limits of holographic CFTs. For example, \cite{Alday:2024yyj} considers the limit at the level of holographic CFT correlators, while \cite{Bagchi:2024efs} examines the limit at the level of the Lagrangian. Furthermore, Carrollian CFT correlators have been computed from an intrinsic perspective in \cite{Cotler:2024xhb}. Based on the discussion in Section \ref{sec:Carrollian CFT stress tensor}, twistor theory appears to be the natural framework for formulating a holographic Carrollian CFT. An intriguing question is how the results obtained in this paper might emerge in the flat space limit of AdS/CFT limit. We leave these questions for future investigation.

\paragraph{Acknowledgements} We thank Arjun Bagchi, Shamik Banejee, Sourish Banerjee, Adam Kmec, Alok Laddha, Lionel Mason and Akshay Yelleshpur Srikant for useful discussions. We are grateful to the Erwin Schrödinger Institute in Vienna for invitation to the
workshop Carrollian Physics and Holography, where part of this project has been carried out. RR is supported by the Titchmarsh Research Fellowship at the Mathematical Institute and by the Walker Early Career Fellowship at Balliol College. This work was also supported by the Simons Collaboration on Celestial Holography. AS is supported by the Prime Minister's Research Fellowship, MHRD, India.

\appendix

\section{The 3D Carrollian CFT stress tensor}
\label{A}

To shed light on the intrinsic Carrollian derivation of the stress tensor Ward identities \eqref{eq:2}-\eqref{eq:5}, we discuss in this appendix on the classical aspects of the stress tensor in a generic 3D field theory on a flat Carrollian background (with topology $\mathbb{R}\times S^2$), with its action possessing the original BMS$_4$ ($\cong \text{Lorentz}\loplus \text{Supertranslations}$) algebra \cite{Bondi:1962px,Sachs:1962wk} as the kinematical symmetry. We shall closely follow the analogous discussion on the 2D Carrollian CFT stress tensor presented in \cite{Saha:2022gjw}.

\medskip

To set the stage, we first review the definitions of the relevant entities from a classical field theory. Suppose that under the infinitesimal transformation:
\begin{align}
x^{\mu}\rightarrow x^{\prime\mu}=x^{\mu}+\epsilon^a\zeta^{\mu}_{\hspace{1.5mm}a}(\mathbf{x})
\end{align}
a multi-component field transforms as a finite matrix representation as ($i$ denotes the matrix representation index):
\begin{align}
\Phi^i(\mathbf{x})\rightarrow{\tilde{\Phi}}^i(\mathbf{x^\prime})=\Phi^i(\mathbf{x})+\epsilon^a{(\mathcal{F}_a\cdot\Phi)}^i(\mathbf{x})\label{111} \, .
\end{align}
Defining the transformation generators as \cite{DiFrancesco:1997nk}:
\begin{align*}
\delta_{\bm{\epsilon}}\Phi^i(\mathbf{x})\equiv{\tilde{\Phi}}^i(\mathbf{x})-{{\Phi}}^i(\mathbf{x}):=-i\epsilon^aG_a(\mathbf{x})\Phi^i(\mathbf{x})\equiv -i\epsilon^aG_a\Phi^i(\mathbf{x})
\end{align*}
the generator $G_a(\mathbf{x})$ of the above transformation is found to be:
\begin{align}
-iG_a(\mathbf{x})\Phi^i(\mathbf{x})={(\mathcal{F}_a\cdot\Phi)}^i(\mathbf{x})-\zeta^{\mu}_{\hspace{1.5mm}a}(\mathbf{x})\partial_\mu\Phi^i(\mathbf{x}) 
  \, . \label{71}
\end{align}
The following action describing the classical theory of the fields $\mathbf{\Phi}$:
\begin{align}
S[\bm{\Phi}]=\int d^{d+1}\mathbf{x}\text{ }\mathcal{L}(\bm{\Phi},\partial_\mu\bm{\Phi})\label{72}
\end{align}
transforms on-shell under the above transformation as \cite{DiFrancesco:1997nk}:
\begin{align*}
S[\bm{\Phi}]\rightarrow S^\prime[\bm{\tilde{\Phi}}]=S[\bm{\Phi}]+\int d^{d+1}\mathbf{x}\text{ }\epsilon^a\partial_\mu j^\mu_{\hspace{1.5mm}a}(\mathbf{x})
\end{align*} 
with $j^\mu_{\hspace{1.5mm}a}(\mathbf{x})$ being the corresponding Noether current (summation over the repeated indices $i$ are implicit):
\begin{align}
j^\mu_{\hspace{1.5mm}a}(\mathbf{x})= {T_{(c)}}^\mu_{\hspace{1.5mm}\nu} \zeta^\nu_{\hspace{1.5mm}a}-{(\mathcal{F}_a\cdot\Phi)}^i\frac{\partial\mathcal{L}}{\partial(\partial_\mu\Phi^i)}  \, . \label{73}
\end{align}
Here, ${T_{(c)}}^\mu_{\hspace{1.5mm}\nu}$ is the canonical stress tensor which is the (on-shell) conserved Noether current associated to the global spacetime translation symmetry. More generally, if the transformation induced by the generator \eqref{71} is a symmetry of the action \eqref{72}, the corresponding Noether current \eqref{73} is on-shell conserved:
\begin{align*}
\partial_\mu j^\mu_{\hspace{1.5mm}a}=0 \, .
\end{align*}  

\medskip

Turning to the 3D Carrollian CFT, we now note that the infinitesimal version of the original BMS$_4$ transformations is given by:
\begin{align}
z\rightarrow z+\epsilon \mathcal{Y}(z)\hspace{2.5mm};\hspace{2.5mm}\bar{z}\rightarrow \bar{z}+\bar{\epsilon}\bar{\mathcal{Y}}(\bar{z})\hspace{2.5mm};\hspace{2.5mm}u\rightarrow u\left(1+\frac{\epsilon}{2}\mathcal{Y}^\prime(z)+\frac{\bar{\epsilon}}{2}\bar{\mathcal{Y}}^\prime(\bar{z})\right)+\epsilon^u\mathcal{T}(z,\bar{z})
\end{align}
with $\mathcal{Y}(z)$ and $\bar{\mathcal{Y}}(\bar{z})$ being at most quadratic polynomials of $z$ and $\bar{z}$ respectively, with the restrictions $\bar{z}=z^*$, $\bar{\epsilon}=\epsilon^*$ and $\bar{\mathcal{Y}}(\bar{z})=\mathcal{Y}(z)^*$, while $\mathcal{T}(z,\bar{z})$ is an arbitrary function restricted to be real. We shall keep treating $z$ and $\bar{z}$ as independent variables only to impose the said restrictions at the end.

\medskip

As discussed in \cite{Saha:2023hsl}, the generator of the infinitesimal supertranslation $\mathcal{T}(z,\bar{z})$ induces the following transformation on a 3D classical conformal Carrollian primary multiplet $\Phi$:
\begin{align}
&-iG_{\mathcal{T}}{\Phi}(u,z,\bar{z})=-\left[\mathcal{T}\partial_u+\left(\partial_z\mathcal{T}\right)\bm{\xi}+\left(\partial_{\bar{z}}\mathcal{T}\right)\bar{\bm{\xi}}\right]\cdot{\Phi}(u,z,\bar{z})\label{68}
\end{align}
with $\bm{\xi}$ and $\bar{\bm{\xi}}$ furnishing a finite-dimensional indecomposable but reducible representation of the two 3D Carrollian boosts. If the field $\Phi$ transforms under the 3D Carrollian spin-boost irrep, it must have $\bm{\xi}\cdot\Phi=\bar{\bm{\xi}}\cdot\Phi=0$. On the other hand, a Lorentz tensor (or $\text{SL}(2,\mathbb{C})$ (quasi-)primary) field $\Phi$ undergoes the following infinitesimal Lorentz transformations \cite{Saha:2023hsl}: 
\begin{align}
&-iG_{\mathcal{Y}}{\Phi}(u,z,\bar{z})=-\left[\mathcal{Y}(z)\partial_z+\mathcal{Y}^\prime(z)\left(\mathbf{h}+\frac{u}{2}\partial_u\right)+\mathcal{Y}^{\prime\prime}(z)\frac{u}{2}\bm{\xi}\right]\cdot{\Phi}(u,z,\bar{z}) \, ,\label{69}\\
&-iG_{\bar{\mathcal{Y}}}{\Phi}(u,z,\bar{z})=-\left[\bar{\mathcal{Y}}(\bar{z})\partial_{\bar{z}}+\bar{\mathcal{Y}}^\prime(\bar{z})\left(\bar{\mathbf{h}}+\frac{u}{2}\partial_u\right)+\bar{\mathcal{Y}}^{\prime\prime}(\bar{z})\frac{u}{2}\bar{\bm{\xi}}\right]\cdot{\Phi}(u,z,\bar{z}) \, , \label{70prime}
\end{align}
with $\mathbf{h},\bar{\mathbf{h}}=\frac{\Delta\mathbf{I}\pm\mathbf{J}}{2}$. The 3D conformal Carrollian multiplets with covariant transformation properties \eqref{68}-\eqref{70prime} under the original BMS$_4$ group will be referred to as the BMS$_4$ primary fields. 

\medskip

We now consider a 3D classical field theory on flat Carrollian background with topology $\mathbb{R}\times S^2$, described by the following action \cite{Bagchi:2019clu}:
\begin{align}
S[\bm{\Phi}]=\int du\int\limits_{S^2} d^2\vec{x}\text{ }\mathcal{L}(\bm{\Phi},\partial_\mu\bm{\Phi})\label{107}
\end{align}
which we demand to be (original) BMS$_4$ invariant. Treating the celestial sphere $S^2$ as the Riemann sphere, we use the (mostly) flat metric on it. Following the relativistic case \cite{Mack:1969rr}, the `fundamental' fields appearing in the Lagrangian are assumed to be BMS$_4$ primaries.

\medskip

Then, the Noether current associated with the $\mathcal{T}(z,\bar{z})$ supertranslation is given below, from \eqref{73} and \eqref{68}, as:
\begin{align}
j^\mu_{\text{ST}}= {T_{(c)}}^\mu_{\hspace{1.5mm}u}\mathcal{T}+\left[(\partial_z\mathcal{T}){(\bm{\xi}\cdot\Phi)}^i+(\partial_{\bar{z}}\mathcal{T}){(\bar{\bm{\xi}}\cdot\Phi)}^i\right]\frac{\partial\mathcal{L}}{\partial(\partial_\mu\Phi^i)} \, .
\end{align}
Since this supertranslation is demanded to be a symmetry of the Carrollian CFT action, the above Noether current is on-shell conserved. Its conservation law leads to the following condition on the stress tensor as well as on the Lagrangian itself since $\mathcal{T}(z,\bar{z})$ is arbitrary:
\begin{align}
\partial_{\mu}j^\mu_{\text{ST}}=0\hspace{2.5mm}\text{(on-shell)}\text{ }\Longrightarrow\text{ } &{T_{(c)}}^z_{\hspace{1.5mm}u}+\partial_{\mu}\left({(\bm{\xi}\cdot\Phi)}^i\frac{\partial\mathcal{L}}{\partial(\partial_\mu\Phi^i)}\right)=0\label{76} \, ,  \\
&{T_{(c)}}^{\bar{z}}_{\hspace{1.5mm}u}+\partial_{\mu}\left({(\bar{\bm{\xi}}\cdot\Phi)}^i\frac{\partial\mathcal{L}}{\partial(\partial_\mu\Phi^i)}\right)=0\label{75}   \, , \\
{({\bm{\xi}}\cdot\Phi)}^i\frac{\partial\mathcal{L}}{\partial(\partial_z\Phi^i)}=&{(\bar{\bm{\xi}}\cdot\Phi)}^i\frac{\partial\mathcal{L}}{\partial(\partial_{\bar{z}}\Phi^i)}={(\bar{\bm{\xi}}\cdot\Phi)}^i\frac{\partial\mathcal{L}}{\partial(\partial_z\Phi^i)}+{({\bm{\xi}}\cdot\Phi)}^i\frac{\partial\mathcal{L}}{\partial(\partial_{\bar{z}}\Phi^i)}=0 
 \, . \label{74}
\end{align}
The constraints \eqref{74} are hence necessary for a 3D Carrollian CFT Lagrangian to satisfy (on-shell) if the action is to be supertranslation invariant. The on-shell conservation of the stress tensor ${T_{(c)}}^\mu_{\hspace{1.5mm}u}$, together with the on-shell conditions \eqref{76}-\eqref{74} then implies:  
\begin{align}
\partial_u {T_{(c)}}^u_{\hspace{1.5mm}u}-\partial_u\partial_{z}\left({(\bm{\xi}\cdot\Phi)}^i\frac{\partial\mathcal{L}}{\partial(\partial_u\Phi^i)}\right)-\partial_u\partial_{\bar{z}}\left({(\bar{\bm{\xi}}\cdot\Phi)}^i\frac{\partial\mathcal{L}}{\partial(\partial_u\Phi^i)}\right)=0\hspace{5mm}\text{(on-shell)}\, .\label{77}
\end{align}
Thus, the stress tensor of a 3D Carrollian CFT can be Belinfante-improved (off-shell), as the consequence of the conditions \eqref{76}, \eqref{75} and \eqref{77} stemming from the supertranslation invariance, as below (${T}^\mu_{\hspace{1.5mm}\nu}$ is the improved stress tensor):
\begin{align}
{T}^z_{\hspace{1.5mm}u}={T}^{\bar{z}}_{\hspace{1.5mm}u}=0\hspace{2mm},\hspace{2mm}&{T}^u_{\hspace{1.5mm}u}={T_{(c)}}^u_{\hspace{1.5mm}u}-\partial_z\left({(\bm{\xi}\cdot\Phi)}^i\frac{\partial\mathcal{L}}{\partial(\partial_u\Phi^i)}\right)-\partial_{\bar{z}}\left({(\bar{\bm{\xi}}\cdot\Phi)}^i\frac{\partial\mathcal{L}}{\partial(\partial_u\Phi^i)}\right)\hspace{2.5mm}\text{(off-shell)}\label{84}\\
\Longrightarrow\hspace{2.5mm}&\partial_u{T}^u_{\hspace{1.5mm}u}=0\hspace{2.5mm}\text{(on-shell)}\nonumber
\end{align}
such that ${T}^\mu_{\hspace{1.5mm}u}$ remains on-shell conserved. It is the improvement ${T}^z_{\hspace{1.5mm}u}={T}^{\bar{z}}_{\hspace{1.5mm}u}=0$ that allows us to re-express the conserved supertranslation currents $j^\mu_{\text{ST}}$ into the following simple form:
\begin{align}
j^\mu_{\text{ST}}={T}^\mu_{\hspace{1.5mm}u}\cdot \mathcal{T}(z,\bar{z})\hspace{5mm}\text{(off-shell)} \, .
\end{align} 

\medskip

We now investigate the consequences of the $\text{SL}(2,\mathbb{C})$ (Lorentz) invariance of the 3D Carrollian CFT action. It turns out that, for this purpose, we can safely treat $z$ and $\bar{z}$ as independent variables without altering any results. From \eqref{73} and \eqref{69}, the Noether current corresponding to the holomorphic Lorentz $\mathcal{Y}(z)$ invariance is found to be (with $\mathcal{Y}^{\prime\prime\prime}(z)=0$):
\begin{align}
j^\mu_{\text{HL}}= \mathcal{Y}(z){T_{(c)}}^\mu_{\hspace{1.5mm}z}+\frac{u}{2} \mathcal{Y}^\prime(z){T_{(c)}}^\mu_{\hspace{1.5mm}u}+\mathcal{Y}^\prime(z){(\mathbf{h}\cdot\Phi)}^i\frac{\partial\mathcal{L}}{\partial(\partial_\mu\Phi^i)}+\frac{u}{2}\mathcal{Y}^{\prime\prime}(z){(\bm{\xi}\cdot\Phi)}^i\frac{\partial\mathcal{L}}{\partial(\partial_\mu\Phi^i)} \, .
\end{align}
Conservation of this current implies the following on-shell conditions:
\begin{align}
\partial_{\mu}j^\mu_{\text{HL}}=0\hspace{2.5mm}\text{(on-shell)}\text{ }\Longrightarrow\text{ }&{T_{(c)}}^z_{\hspace{1.5mm}z}+\frac{1}{2}{T_{(c)}}^u_{\hspace{1.5mm}u}+\partial_{\mu}\left({(\mathbf{h}\cdot\Phi)}^i\frac{\partial\mathcal{L}}{\partial(\partial_\mu\Phi^i)}\right)=0 \, , \label{80}\\
&{(\mathbf{h}\cdot\Phi)}^i\frac{\partial\mathcal{L}}{\partial(\partial_z\Phi^i)}+\frac{1}{2}{(\bm{\xi}\cdot\Phi)}^i\frac{\partial\mathcal{L}}{\partial(\partial_u\Phi^i)}=0 \, , \label{78}
\end{align}
along with the known relation \eqref{76}. Similarly, from the anti-holomorphic Lorentz invariance, we obtain the `complex-conjugates' of the above conditions:
\begin{align}
\partial_{\mu}j^\mu_{\text{AL}}=0\hspace{2.5mm}\text{(on-shell)}\text{ }\Longrightarrow\text{ }&{T_{(c)}}^{\bar{z}}_{\hspace{1.5mm}\bar{z}}+\frac{1}{2}{T_{(c)}}^u_{\hspace{1.5mm}u}+\partial_{\mu}\left({(\bar{\mathbf{h}}\cdot\Phi)}^i\frac{\partial\mathcal{L}}{\partial(\partial_\mu\Phi^i)}\right)=0 \, , \label{81}\\
&{(\bar{\mathbf{h}}\cdot\Phi)}^i\frac{\partial\mathcal{L}}{\partial(\partial_{\bar{z}}\Phi^i)}+\frac{1}{2}{(\bar{\bm{\xi}}\cdot\Phi)}^i\frac{\partial\mathcal{L}}{\partial(\partial_u\Phi^i)}=0 \, . \label{79}
\end{align}
The restriction \eqref{78} and \eqref{79} are the necessary on-shell conditions for the Lorentz invariance of the action.

\medskip

Now, using the improvement \eqref{77} and the restrictions \eqref{78} and \eqref{79}, we reach from the on-shell relation \eqref{80} to:
\begin{align*}
{T_{(c)}}^z_{\hspace{1.5mm}z}+\frac{1}{2}T^u_{\hspace{1.5mm}u}+\partial_{u}\left({(\mathbf{h}\cdot\Phi)}^i\frac{\partial\mathcal{L}}{\partial(\partial_u\Phi^i)}\right)+\partial_{\bar{z}}\left({(\mathbf{J}\cdot\Phi)}^i\frac{\partial\mathcal{L}}{\partial(\partial_{\bar{z}}\Phi^i)}\right)=0 \, . 
\end{align*}
Thus, one can Belinfante-improve the ${T_{(c)}}^z_{\hspace{1.5mm}z}$ component as below: 
\begin{align}
T^z_{\hspace{1.5mm}z}={T_{(c)}}^z_{\hspace{1.5mm}z}+\partial_{u}\left({(\mathbf{h}\cdot\Phi)}^i\frac{\partial\mathcal{L}}{\partial(\partial_u\Phi^i)}\right)+\partial_{\bar{z}}\left({(\mathbf{J}\cdot\Phi)}^i\frac{\partial\mathcal{L}}{\partial(\partial_{\bar{z}}\Phi^i)}\right)\hspace{5mm}\text{(off-shell)}\label{82}
\end{align}
leading to the following on-shell relation:
\begin{align*}
T^z_{\hspace{1.5mm}z}+\frac{1}{2}T^u_{\hspace{1.5mm}u}=0 \, .
\end{align*}
Similarly, starting from \eqref{81} and improving the ${T_{(c)}}^{\bar{z}}_{\hspace{1.5mm}\bar{z}}$ component as:
\begin{align}
T^{\bar{z}}_{\hspace{1.5mm}\bar{z}}={T_{(c)}}^{\bar{z}}_{\hspace{1.5mm}\bar{z}}+\partial_{u}\left({(\bar{\mathbf{h}}\cdot\Phi)}^i\frac{\partial\mathcal{L}}{\partial(\partial_u\Phi^i)}\right)-\partial_{{z}}\left({(\mathbf{J}\cdot\Phi)}^i\frac{\partial\mathcal{L}}{\partial(\partial_{{z}}\Phi^i)}\right)\hspace{5mm}\text{(off-shell)}
\end{align}
we are finally led to the on-shell symmetric, trace-less nature of the improved stress tensor:
\begin{align}
T^z_{\hspace{1.5mm}z}=T^{\bar{z}}_{\hspace{1.5mm}\bar{z}}=-\frac{1}{2}T^u_{\hspace{1.5mm}u}\hspace{5mm}\Longrightarrow\hspace{5mm}T^\mu_{\hspace{1.5mm}\mu}=0\hspace{5mm}\text{(on-shell)} \, .  \label{83}
\end{align}
It follows from \eqref{82} and the conservation of ${T_{(c)}}^{\mu}_{\hspace{1.5mm}{z}}$ that the components ${T_{(c)}}^{u}_{\hspace{1.5mm}{z}}$ and ${T_{(c)}}^{\bar{z}}_{\hspace{1.5mm}{z}}$ can be (off-shell) improved as below:
\begin{align}
T^{u}_{\hspace{1.5mm}{z}}={T_{(c)}}^{u}_{\hspace{1.5mm}{z}}-\partial_{z}\left({({\mathbf{h}}\cdot\Phi)}^i\frac{\partial\mathcal{L}}{\partial(\partial_u\Phi^i)}\right)\hspace{5mm};\hspace{5mm}
T^{\bar{z}}_{\hspace{1.5mm}{z}}={T_{(c)}}^{\bar{z}}_{\hspace{1.5mm}{z}}-\partial_{z}\left({({\mathbf{J}}\cdot\Phi)}^i\frac{\partial\mathcal{L}}{\partial(\partial_{\bar{z}}\Phi^i)}\right)\label{110}
\end{align}
and similarly, for ${T_{(c)}}^{u}_{\hspace{1.5mm}{\bar{z}}}$ and ${T_{(c)}}^{{z}}_{\hspace{1.5mm}\bar{z}}$, one has:
\begin{align}
T^{u}_{\hspace{1.5mm}\bar{z}}={T_{(c)}}^{u}_{\hspace{1.5mm}\bar{z}}-\partial_{\bar{z}}\left({(\bar{\mathbf{h}}\cdot\Phi)}^i\frac{\partial\mathcal{L}}{\partial(\partial_u\Phi^i)}\right)\hspace{5mm};\hspace{5mm}
T^{{z}}_{\hspace{1.5mm}\bar{z}}={T_{(c)}}^{{z}}_{\hspace{1.5mm}\bar{z}}+\partial_{\bar{z}}\left({({\mathbf{J}}\cdot\Phi)}^i\frac{\partial\mathcal{L}}{\partial(\partial_{{z}}\Phi^i)}\right)\label{109}
\end{align}
such that both $T^{\mu}_{\hspace{1.5mm}{z}}$ and $T^{\mu}_{\hspace{1.5mm}\bar{z}}$ remain conserved on-shell.

\medskip

Thus, we have shown, as the consequences of the Lorentz and supertranslation invariance, that the stress tensor in a 3D classical Carrollian CFT can be improved to become (on-shell) symmetric in spatial indices and trace-less \eqref{83} along with the (off-shell) vanishing of the components $T^{i}_{\hspace{1.5mm}{u}}$ \eqref{84}. These conclusions follow even when the assumption that the fields appearing in the Lagrangian transform as BMS$_4$ primary fields is given up; no knowledge of the specific BMS$_4$ transformation properties of the `fundamental' fields is required. Due to those symmetry properties of the stress tensor, the conserved Noether current associated with any original BMS$_4$ symmetry-transformation $x^{\mu}\rightarrow x^{\mu}+\epsilon^a\zeta^{\mu}_{\hspace{1.5mm}a}(\mathbf{x})$ can be expressed in the following simple form: 
\begin{align}
j^\mu_{\hspace{1.5mm}a}={T}^\mu_{\hspace{1.5mm}\nu}\zeta^\nu_{\hspace{1.5mm}a}\hspace{5mm}\text{(off-shell)}\, .\label{85}
\end{align}   

\medskip

Intriguingly, no additional condition or restriction on the Lagrangians other than \eqref{76}-\eqref{74} and \eqref{80}-\eqref{79} arises if we demand invariance of the action under arbitrary superrotations contained in the extended BMS$_4$ group (even when $z$ and $\bar{z}$ are kept completely independent):
\begin{align*}
&z\rightarrow z+\epsilon z^{n+1}\hspace{2.5mm};\hspace{2.5mm}\bar{z}\rightarrow \bar{z}\hspace{2.5mm};\hspace{2.5mm}u\rightarrow u+\epsilon\frac{u}{2}(n+1)z^n\hspace{5mm}\text{(holomorphic superrotation)}\\
&z\rightarrow z\hspace{2.5mm};\hspace{2.5mm}\bar{z}\rightarrow \bar{z}+\epsilon \bar{z}^{n+1}\hspace{2.5mm};\hspace{2.5mm}u\rightarrow u+\epsilon\frac{u}{2}(n+1)\bar{z}^n\hspace{5mm}\text{(anti-holomorphic superrotation)}
\end{align*}
where $n\in\mathbb{Z}$. Thus, the classical $\text{SL}(2,\mathbb{C})$ Lorentz and supertranslation symmetries of the 3D Carrollian CFT action \textit{automatically} lead to its invariance under an infinite number of arbitrary superrotations, implying the extended BMS$_4$ symmetry of the action. This symmetry-enhancement of the classical action of 3D Carrollian CFT from the original (global) BMS$_4$ to the extended (local) BMS$_4$ or, more precisely, from the global $\text{SL}(2,\mathbb{C})$ (Lorentz) part thereof to the local superrotations, is perfectly analogous to the enhancement of the global conformal symmetry to the local infinite-dimensional one in a 2D CFT. By virtue of the properties \eqref{84} and \eqref{83} of the improved stress tensor, each of the conserved Noether currents corresponding to the extended BMS$_4$ symmetry can be (off-shell) expressed in the form \eqref{85}.

\medskip

On the other hand, the modes contained in the object $S^+_{1e}(u,z,\bar{z})$ generate the following infinitesimal Carrollian diffeomorphisms on the Hilbert space \cite{Banerjee:2020zlg}:
\begin{align}
&z\rightarrow z\hspace{2.5mm};\hspace{2.5mm}\bar{z}\rightarrow \bar{z}+\epsilon z^n\bar{z}^{q+1}\hspace{2.5mm};\hspace{2.5mm}u\rightarrow u+\epsilon\frac{u}{2}(q+1)z^n\bar{z}^q\hspace{5mm}\text{(with $q=0,\pm1$ and $n\in\mathbb{Z}$)}\label{106}
\end{align}
under which a 3D conformal Carrollian primary multiplet $\Phi$ transforms as \cite{Saha:2023hsl}:
\begin{align}
&\left[j^{(-)}_{n}\text{ },\text{ }\Phi(\mathbf{x}_p)\right]=-iz_p^{n}\partial_{\bar{z}_p}\Phi(\mathbf{x}_p)\, ,\nonumber\\
&\left[j^{(0)}_{n}\text{ },\text{ }\Phi(\mathbf{x}_p)\right]=-i\left[z^n_p\left(\bar{z}_p\partial_{\bar{z}_p}+\frac{u_p}{2}\partial_{u_p}+\bar{\mathbf{h}}_p\right)+nz^{n-1}_p\frac{u_p}{2}\bm{\xi}_p\right]\Phi(\mathbf{x}_p) \, , \label{108}\\
&\left[j^{(+)}_{n}\text{ },\text{ }\Phi(\mathbf{x}_p)\right]=-i\left[z_p^n\left(\bar{z}^2_p\partial_{\bar{z}_p}+\bar{z}_pu_p\partial_{u_p}+2\bar{z}_p\bar{\mathbf{h}}_p+u_p\bar{{\bm{\xi}}}_p\right)+nz_p^{n-1}\bar{z}_p{u_p}\bm{\xi}_p\right]\Phi(\mathbf{x}_p) \, . \nonumber
\end{align}
Since the modes $j^{(a)}_{n}$ generate a symmetry at the level of the Carrollian CFT OPE, the action \eqref{107} should be invariant under the diffeomorphisms \eqref{106}. From \eqref{73} and \eqref{108}, the Noether currents associated to these diffeomorphisms are given as:
\begin{align}
j^\mu_{\text{Diff}}&=z^n\bar{z}^{q+1}{T_{(c)}}^\mu_{\hspace{1.5mm}\bar{z}}+\frac{u}{2}(q+1)z^n\bar{z}^q {T_{(c)}}^\mu_{\hspace{1.5mm}u}+(q+1)z^n\bar{z}^q{(\bar{\mathbf{h}}\cdot\Phi)}^i\frac{\partial\mathcal{L}}{\partial(\partial_\mu\Phi^i)}\nonumber\\
&\qquad\qquad+\frac{u}{2}(q+1)qz^n\bar{z}^{q-1}{(\bar{\bm{\xi}}\cdot\Phi)}^i\frac{\partial\mathcal{L}}{\partial(\partial_\mu\Phi^i)}+\frac{u}{2}(q+1)nz^{n-1}\bar{z}^{q}{({\bm{\xi}}\cdot\Phi)}^i\frac{\partial\mathcal{L}}{\partial(\partial_\mu\Phi^i)} \, .
\end{align}
The conservation of these diffeomorphism currents, together with the (original) BMS$_4$ invariance conditions \eqref{76}-\eqref{74} and \eqref{80}-\eqref{79}, impose the following additional (on-shell) restrictions on the Lagrangian:
\begin{align}
\partial_{\mu}j^\mu_{\text{Diff}}=0\hspace{2.5mm}\text{(on-shell)}\text{ }\Longrightarrow\text{ }{T_{(c)}}^{z}_{\hspace{1.5mm}\bar{z}}={({\mathbf{J}}\cdot\Phi)}^i\frac{\partial\mathcal{L}}{\partial(\partial_{{z}}\Phi^i)}=0
\end{align}
which keep even the improved component $T^{z}_{\hspace{1.5mm}\bar{z}}$ fixed at $0$ off-shell, as seen from \eqref{109}. In fact, if we extend the transformations \eqref{106} to include any arbitrary diffeomorphisms on the sphere \cite{Campiglia:2014yka,Campiglia:2015yka,Compere:2018ylh} as the symmetries of the action \eqref{107}, we also recover the `conjugate' restrictions in a similar way:
\begin{align}
{T_{(c)}}^{\bar{z}}_{\hspace{1.5mm}{z}}={({\mathbf{J}}\cdot\Phi)}^i\frac{\partial\mathcal{L}}{\partial(\partial_{\bar{z}}\Phi^i)}=0
\end{align}
which persist upon improvement: $T^{\bar{z}}_{\hspace{1.5mm}{z}}=0$ (off-shell), from \eqref{110}. The Noether currents corresponding to the Carrollian diffeomorphisms can then all be expressed in the form \eqref{85}.

\medskip

Assuming only the classical (original) BMS$_4$ invariance of the action \eqref{107}, we now sketch the derivation of the stress tensor Ward identities in the corresponding QFT in the Carrollian path-integral formalism \cite{Chen:2023pqf}. In this formalism, an $n$-point correlator (covariant time-ordered) is defined as (suppressing the field tensor indices):
\begin{align}
\langle X\rangle\equiv\langle\Phi_1(\mathbf{x_1})\Phi_2(\mathbf{x_2})...\Phi_n(\mathbf{x_n})\rangle:=\frac{\int[\mathcal{D}\bm{\Phi}]\text{ }\Phi_1(\mathbf{x_1})\Phi_2(\mathbf{x_2})...\Phi_n(\mathbf{x_n})\text{ }e^{iS[\bm{\Phi}]}}{\int[\mathcal{D}\bm{{\Phi}}]\text{ }e^{iS[\bm{{\Phi}}]}} \, .
\end{align}
A field transformation, e.g. \eqref{111}, will be a (local) symmetry of the QFT if, for any $X$:
\begin{align}
\langle X\rangle={\langle X\rangle}^\prime\equiv\langle\tilde{\Phi}_1(\mathbf{x_1})\tilde{\Phi}_2(\mathbf{x_2})...\tilde{\Phi}_n(\mathbf{x_n})\rangle:=\frac{\int[\mathcal{D}\bm{\tilde{\Phi}}]\text{ }\tilde{\Phi}_1(\mathbf{x_1})\tilde{\Phi}_2(\mathbf{x_2})...\tilde{\Phi}_n(\mathbf{x_n})\text{ }e^{iS^\prime[\bm{\tilde{\Phi}}]}}{\int[\mathcal{D}\bm{\tilde{\Phi}}]\text{ }e^{iS^\prime[\bm{\tilde{\Phi}}]}} \, .
\end{align} 
Under the assumption of invariance of the path-integral measure: $[\mathcal{D}\bm{\Phi}]=[\mathcal{D}\bm{\tilde{\Phi}}]$, this symmetry condition, at the 1st order in $\bm\epsilon$, leads to the Ward identity \cite{DiFrancesco:1997nk}:
\begin{align}
-\delta_{\bm{\epsilon}}\langle X\rangle\equiv\sum_{i=1}^n\langle\Phi_1(\mathbf{x_1}))...(i\epsilon^aG_a\Phi_i(\mathbf{x_i}))...\Phi_n(\mathbf{x_n})\rangle=i\int du\text{ } d^{2}\vec{x}\text{ }\epsilon^a\langle\partial_\mu j^\mu_{\hspace{1.5mm}a}(\mathbf{x})\left(X-\langle X\rangle\right)\rangle \, .
\end{align}
In the 3D Carrollian CFTs, the Ward identity takes the following differential form when $X$ is a string of conformal Carrollian primary fields, up to derivatives of temporal contact-terms:
\begin{align}
\partial_\mu\langle j^\mu_{\hspace{1.5mm}a}(\mathbf{x})(X-\langle X\rangle)\rangle=\sum\limits_{i=1}^n\text{ }\delta(u-u_i)\delta^2(\vec{x}-\vec{x_i}){{(G_a)}_i\langle X\rangle} \, . 
\end{align}
To reach the stress tensor Ward identities \eqref{eq:2}-\eqref{eq:5} from here, all we need is the form \eqref{85} of the Noether currents $j^\mu_{\hspace{1.5mm}a}$ and the fact that $\partial_\mu\langle j^\mu_{\hspace{1.5mm}a}(\mathbf{x})\rangle=0$. Had $X$ been a string of arbitrary conformal Carrollian non-primary fields, derivatives of the spatial delta-function would have also appeared on the RHS of these Ward identities.

\medskip

Now, the Carrollian diffeomorphisms \eqref{106} are local symmetries of the 3D quantum Carrollian CFT. As a consequence, we have the following trivial Ward identity: $\langle T^z_{\hspace{1.5mm}\bar{z}}(\mathbf{x})X\rangle=0$, which can be taken to be exact. Thus, the component $T^z_{\hspace{1.5mm}\bar{z}}$ vanishes both at the classical as well as the quantum level. If the Lagrangian is hermitian, the `conjugate' component $T^{\bar{z}}_{\hspace{1.5mm}{z}}$ also has to vanish.

\section{OPE formulae}
\label{sec:OPE formulae}

\subsection*{OPE formulae for \texorpdfstring{$S_0^+(u,z,\bar{z})$}{S0}}
\label{OPEformulaS0}

We now discuss the Carrollian OPE between $S_0^+(u,z,\bar{z})$ and a general Carrollian operator (primary or descendent). These formulae will be useful when extending the analysis to $Lw_{1+\infty}$ in Section \ref{s3.1}. 

We will treat the coordinates $z$ and $\bar{z}$
as independent variables so that any term like $\frac{\left(\bar{z}-\bar{z}_p\right)^r}{\left(z-z_p\right)^s}$ with $r\geq0,s\geq1$ has (meromorphic) pole singularity. Together with this, the form of the Ward identity \eqref{3} is suggestive of the below decomposition of $S^+_0$ inside the correlator, following \cite{Banerjee:2020zlg}:
\begin{align}
&S^+_0(u,z,\bar{z})=\bar{z}P_{-1}(u,z,\bar{z})-P_0(u,z,\bar{z})\label{4}\\
\text{with}\hspace{5mm}&\langle P_{-1}(u,z,\bar{z})X\rangle=-i\sum\limits_{p=1}^n\text{ }\theta(u-u_p)\left(\frac{\partial_{u_p}}{z-z_p}+\frac{\bm{\xi}_p}{{(z-z_p)}^2}\right)\langle X\rangle\nonumber\\
\text{and}\hspace{5mm}&\langle P_0(u,z,\bar{z})X\rangle=-i\sum\limits_{p=1}^n\text{ }\theta(u-u_p)\left(\frac{\bar{z}_p\partial_{u_p}+\bar{{\bm{\xi}}}_p}{z-z_p}+\frac{\bar{z}_p\bm{\xi}_p}{{(z-z_p)}^2}\right)\langle X\rangle\nonumber\\
\Longrightarrow\hspace{5mm}&\langle\bar{\partial} P_i(u,z,\bar{z})X\rangle=-i\sum\limits_{p=1}^n\text{ }\theta(u-u_p)\text{ }\left[\text{contact terms on $S^2$}\right] \, . \nonumber
\end{align}
These relations are reminiscent of the holomorphic Ka\v c-Moody like Ward identities in a 2D Euclidean CFT. Both $P_i$ and $S^+_0$ clearly have the same holomorphic weight $h=\frac{3}{2}$. For later use, we introduce the mode decomposition 
\begin{equation}
    P_i = \sum_{n\in \mathbb{Z}} P_{n,i} z^{-n-2}\, ,\qquad i = 0,-1
\end{equation} where the modes have been re-labelled such that $n$ takes integer values.

Below, we note the OPE of $S^+_0$ with a general conformal Carrollian (not necessarily primary) multiplet $\Phi$ (that is mutually local with $S^+_0$) in the $j\epsilon$-form \cite{Saha:2023hsl}: 
\begin{align}
S^+_0(u,z,\bar{z})\Phi(\mathbf{x}_p)\sim\lim\limits_{\epsilon\rightarrow0^+}&-i\left[(\bar{z}-\bar{z}_p)\left(\sum\limits_{n\geq0}^J\frac{\left(P_{n,-1}\Phi\right)}{{(\Delta\tilde{z}_p)}^{n+2}}+\frac{\partial_{u_p}\Phi}{(\Delta\tilde{z}_p)}\right)-\sum\limits_{n\geq-1}^K\frac{\left(P_{n,0}\Phi\right)}{{(\Delta\tilde{z}_p)}^{n+2}}\right](\mathbf{x}_p)\label{5}\\
\Longrightarrow\hspace{5mm}&\bar{\partial}^2 S^+_0(u,z,\bar{z})\Phi(\mathbf{x}_p)\sim0\label{6}\\
\text{and}\hspace{5mm}&\partial_u S^+_0(u,z,\bar{z})\Phi(\mathbf{x}_p)\sim0\label{7}
\end{align}
where (especially on the first line) $\sim$ denotes `modulo terms holomorphic (regular) in $\Delta\tilde{z}_p$', and $K$, $J$ are integers depending on the particular field. This OPE is clearly a Laurent expansion in the holomorphic variable $z$ (or $\tilde{z}=z-j\epsilon u$) but is regular in $\bar{z}$. The relations \eqref{6} and \eqref{7} are, on the other hand, exact as OPE statements since all the contact terms in the RHS vanish in the OPE limit.

While denoting the residues of the higher order poles in the OPE \eqref{5}, we have followed the `normal-ordering' notation introduced in section 6.5 of \cite{DiFrancesco:1997nk}. For a conformal Carrollian primary multiplet $\Phi$ as defined in \eqref{70prime}, we specially have:
\begin{align}
\left(P_{n+1,-1}\Phi\right)=0=\left(P_{n,0}\Phi\right)\text{\hspace{5mm} for $n\geq0$} \, . \label{124}
\end{align}
The OPE involving a 3D conformal Carrollian descendant multiplet can be obtained following a standard 2D CFT like method, starting from the OPE of its parent primary field. The higher order residues are then readily extracted by comparison with the general OPE \eqref{5}. A 3D conformal Carrollian quasi-primary (i.e. merely an $\mathfrak{iso}(3,1)$ covariant) multiplet needs to only satisfy $\left(P_{0,0}\Phi\right)=0$. 

\subsection*{OPE formulae for \texorpdfstring{$S_1^+(u,z,\bar{z})$}{S0}}

Since $z$ and $\bar{z}$ are treated independently, looking at the form of \eqref{10} we decompose $S^+_{1e}$ inside a correlator as below:
\begin{align}
&S^+_{1e}(u,z,\bar{z})=j^{(+)}_e(u,z,\bar{z})-2\bar{z}j^{(0)}_e(u,z,\bar{z})+\bar{z}^2j^{(-)}_e(u,z,\bar{z})\label{18}\\
\Longrightarrow\hspace{5mm}&\langle\bar{\partial} j^a_e(u,z,\bar{z})X\rangle=-i\sum\limits_{p=1}^n\text{ }\theta(u-u_p)\text{ }\left[\text{contact terms on $S^2$}\right] \, . \nonumber
\end{align}
Like $S^+_1$, all three $j^a_e$ have holomorphic weight $h=1$, so their Ward identities are practically the same as the holomorphic Ka\v c-Moody current Ward identities in a usual 2D Euclidean CFT.

Finally, the general OPE in the $j\epsilon$-form of the $S^+_1$ field was found in \cite{Saha:2023hsl} to be:
\begin{align}
&\left\{S_1^+-({u-u_p})S^+_0\right\}(u,z,\bar{z})\Phi(\mathbf{x}_p) \label{14} \\
&\qquad\qquad\sim\lim\limits_{\epsilon\rightarrow0^+}-i\left[(\bar{z}-\bar{z}_p)^2\left(\sum\limits_{n\geq1}^L\frac{\left(j^{(-)}_{n}\Phi\right)}{{(\Delta\tilde{z}_p)}^{n+1}}+\frac{\partial_{\bar{z}_p}\Phi}{\Delta\tilde{z}_p}\right)+\sum\limits_{n\geq0}^M\frac{\left(j^{(+)}_{n}\Phi\right)}{{(\Delta\tilde{z}_p)}^{n+1}}\right.\nonumber\\
&\qquad\qquad\qquad\qquad\qquad\qquad\qquad\qquad\left.-2(\bar{z}-\bar{z}_p)\left(\sum\limits_{n\geq1}^N\frac{\left(j^{(0)}_{n}\Phi\right)}{{(\Delta\tilde{z}_p)}^{n+1}}+\frac{\bar{h}_p\Phi}{\Delta\tilde{z}_p}\right)\right](\mathbf{x}_p)  \nonumber \\
&\hspace{9mm}\Longrightarrow\hspace{5mm}\bar{\partial}^3 S^+_1(u,z,\bar{z})\Phi(\mathbf{x}_p)\sim0\label{15}\\
&\hspace{9mm}\text{and}\hspace{5mm}\left(\partial_u S^+_1-S^+_0\right)(u,z,\bar{z})\Phi(\mathbf{x}_p)\sim0 \, .  \label{16}
\end{align}
Again, the last two relations above are exact in the OPE limit. A conformal Carrollian primary multiplet $\Phi$ must satisfy:
\begin{align}
\left(j^{(-)}_{n}\Phi\right)=\left(j^{(0)}_{n}\Phi\right)=\left(j^{(+)}_{n-1}\Phi\right)=0\text{\hspace{5mm} for $n\geq1$}\label{125}
\end{align}
whereas for an $\text{SL}(2,\mathbb{C})$ or Lorentz covariant quasi-primary, we only have $\left(j^{(+)}_{0}\Phi\right)=0$.

The corresponding correlators and OPEs involving the $S^-_1$ field (replacing $S^+_1$) can be found by simply complex conjugating $(z\rightarrow\bar{z},\bar{h}\rightarrow h,\bm{\xi}\rightarrow\bm{\bar{\xi}})$ the above mentioned equations. But it was shown in \cite{Saha:2023hsl}, following \cite{Banerjee:2022wht}, that the fields $S^+_0$ and $S^-_1$ are not mutually local. This is so because while $\left(\bar{\partial}^2 S^+_0\right)S^-_1\sim0$ respects \eqref{6}, $S^-_1\bar{\partial}^2 S^+_0$ contains anti-meromorphic poles, thus violating the OPE commutativity property. In contrast, the OPE of $S^+_0$ with $S^+_1$ does not suffer from this problem; hence, $S^+_0$ and $S^+_1$ can be simultaneously treated as local fields.

\section{Holomorphic superrotations generator}
\label{sec:Full Lorentz symmetries}

An alternative way to convert the $S^2$ contact terms in the Ward identity \eqref{a13} (or, its conjugate version)
into pole singularities is to extract from the RHS a $\bar{\partial}$ (or, $\partial$) -derivative. By inverting these derivatives, the conformal Carrollian fields $T$ and $\bar{T}$ are defined as:
\begin{align}
&T(u,z,\bar{z})=\int\limits_{S^2} d^2{r}^\prime\frac{T^u_{\hspace{1.5mm}z}(u,\vec{x}^\prime)}{z-z^\prime}+\pi\int\limits_{-\infty}^udu^\prime T^{\bar{z}}_{\hspace{1.5mm}z}(u^\prime,\vec{x})\hspace{2.5mm}\Longrightarrow\hspace{2.5mm}2\bar{\partial}T=\partial^3S^-_1 \, ,\\
&\bar{T}(u,z,\bar{z})=\int\limits_{S^2} d^2{r}^\prime\frac{T^u_{\hspace{1.5mm}\bar{z}}(u,\vec{x}^\prime)}{\bar{z}-\bar{z}^\prime}+\pi\int\limits_{-\infty}^udu^\prime T^{{z}}_{\hspace{1.5mm}\bar{z}}(u^\prime,\vec{x})\hspace{2.5mm}\Longrightarrow\hspace{2.5mm}2{\partial}\bar{T}=\bar{\partial}^3S^+_1 \, .
\end{align}
The fields $T$ and $\bar{T}$ respectively have dimensions $(\Delta,s)=(2,\pm2)$. Since $S^\pm_1$ have $(\Delta,s)=(0,\pm2)$, the above relations directly imply that $(S^+_1,\bar{T})$ and $(S^-_1,T)$ are two shadow pairs (on $S^2$). Hence, if we choose $S^+_1$ as a local field, $\bar{T}$ must be treated as its non-local shadow.

\medskip

With a string $X$ of mutually local primaries, the $T$ Ward identity was found, up to temporal contact terms, from the superrotation Ward identity to be \cite{Saha:2023hsl}:
\begin{align}
&i\langle T(u,z,\bar{z})X\rangle
=\sum\limits_{p=1}^n\text{ }\theta(u-u_p)\left[\frac{h_p}{(z-z_p)^2}+\frac{\partial_{z_p}}{z-z_p}\right.\nonumber\\
&\left.\hspace{46mm}-\frac{u-u_p}{2}\left\{\frac{\partial_{u_p}}{(z-z_p)^2}+\frac{2\bm{\xi}_p}{{(z-z_p)}^3}+\pi\bar{\bm{\xi}}_p\partial\delta^2(\vec{x}-\vec{x}_p)\right\}\right]\langle X\rangle\label{19}\\
&\text{with\hspace{2.5mm}}\langle \bar{\partial} T(u,z,\bar{z})X\rangle=-i\sum\limits_{p=1}^n\theta(u-u_p)\left[\text{contact terms on $S^2$}\right]\nonumber\\
&\text{and\hspace{2.5mm}}\partial_u\langle T(u,z,\bar{z})X\rangle-\frac{1}{2}\langle \partial\bar{\partial} S^+_0(u,z,\bar{z})X\rangle=[\text{temporal contact terms}] \, .  \nonumber
\end{align} 
The last relation inspires the following decomposition of the $T$ field inside the correlator \cite{Saha:2023hsl}:
\begin{align}
&T(u,z,\bar{z})=\frac{u}{2}\partial\bar{\partial} S^+_0(u,z,\bar{z})+T_e(u,z,\bar{z})\\
\Longrightarrow\hspace{2.5mm}&{\partial}_u\langle T_{e}(u,z,\bar{z}) X\rangle=[\text{temporal contact terms}]\nonumber
\end{align}
where the object (not a Carrollian field) $T_e$ contains the modes generating the holomorphic superrotations:
\begin{align}
z\rightarrow z^\prime=\mathcal{Y}(z)\hspace{2.5mm};\hspace{2.5mm}\bar{z}\rightarrow \bar{z}^\prime=\bar{z}\hspace{2.5mm};\hspace{2.5mm}u\rightarrow u^\prime=u\left(\frac{d\mathcal{Y}}{dz}\right)^{\frac{1}{2}}\label{17}
\end{align}
with $\mathcal{Y}(z)$ being a meromorphic function. The $\langle T_{e}(u,z,\bar{z}) X\rangle$ `Ward identity' is the 3D Carrollian CFT incarnation of the 2D celestial conformal stress tensor Ward identity \cite{Kapec:2016jld}. Consistently, $T_e$ is the $S^2$ shadow transformation of $S^-_{1e}$, from whose `Ward identity' we can obtain the subleading negative-helicity energetically soft graviton theorem \cite{Cachazo:2014fwa,Kapec:2014opa}.

\medskip

We note down the generic OPE involving the $T$ field in the $j\epsilon$-form \cite{Saha:2023hsl} below:
\begin{align}
&\left(T-\frac{u-u_p}{2}\text{ }\partial\bar{\partial}S^+_0\right)(u,z,\bar{z})\Phi(\mathbf{x}_p)\sim\lim\limits_{\epsilon\rightarrow0^+}-i\left[\sum\limits_{n\geq1}^P\frac{\left(L_{n}\Phi\right)}{{(\Delta\tilde{z}_p)}^{n+2}}+\frac{h_p\Phi}{(\Delta\tilde{z}_p)^2}+\frac{\partial_{z_p}\Phi}{\Delta\tilde{z}_p}\right](\mathbf{x}_p)\label{8}\\
\Longrightarrow\hspace{5mm}&\bar{\partial} T(u,z,\bar{z})\Phi(\mathbf{x}_p)\sim0\hspace{5mm}\text{and}\hspace{5mm}\left(\partial_u T-\frac{1}{2}\partial\bar{\partial} S^+_0\right)(u,z,\bar{z})\Phi(\mathbf{x}_p)\sim0\hspace{10mm}\text{(OPE exact)} \, .
\end{align}
For a conformal Carrollian primary,
\begin{align}
  \left(L_{n}\Phi\right)=0 \hspace{5mm}\text{for $n\geq1$} \, ,\label{126} 
\end{align}
whereas an $\text{SL}(2,\mathbb{C})$ Lorentz quasi-primary needs to satisfy only $\left(L_{1}\Phi\right)=0$.

\medskip

Without referring to the transformation properties of the conformal Carrollian primaries at all, we shall define the field (multiplet) $\Phi$ as a primary under the symmetry-algebra \eqref{algebra S0S1T} if it satisfies \eqref{124}, \eqref{125}, and \eqref{126}, or equivalently, transforms as \eqref{68} under the wedge supertranslations $\mathcal{T}_{\wedge}(z,\bar{z})$ with $\bar{\partial}^2\mathcal{T}_{\wedge}=0$, as \eqref{69} under the holomorphic superrotations $\mathcal{Y}(z)$, and as \eqref{108} under the wedge diffeomorphisms $\bar{V}(z,\bar{z})$ with $\bar{\partial}^3\bar{V}=0$.     

\medskip

It is worth mentioning that the globally defined (on $S^2$) subset of the transformations \eqref{9}, \eqref{13}, and \eqref{17} consists of the ten $\text{ISL}(2,\mathbb{C})$ Poincar\'e transformations. The above-described extraction of the various powers of $\partial$ or $\bar\partial$ from the 3D Carrollian CFT stress tensor Ward identities is precisely the minimum amount of extraction needed to construct the Carrollian fields that collectively contain all the Poincar\'e generators. In fact, we have not extracted a $\bar{\partial}$ from the $T$ Ward identity \eqref{19} even though it has a contact term (on $S^2$) singularity, since the field $T$ already contains the modes generating the three holomorphic Lorentz transformations. Furthermore, as explained in \cite{Saha:2023hsl}, we can treat the three Carrollian fields $S^+_0$, $S^+_1$, and $T$ as mutually local, $\text{SL}(2,\mathbb{C})$ quasi-primary fields (descending from the Identity field).

\section{Central charges}
\label{sec:Central charge}

In this appendix, we shall obtain the central charges of the 3D conformal Carrollian field theory by calculating the two-point correlators between the stress-tensor components, using the following correspondences: 
\begin{align}
T^u_{\hspace{1.5mm}u}(u,z,\bar{z})&= -2\bar{\partial}^2\int\limits_{-\infty}^udu^\prime\text{ }\partial_{u^\prime}\left[C_{zz}(u^\prime,z,\bar{z})+D_{zz}(u^\prime,z,\bar{z})\right]\\
T^u_{\hspace{1.5mm}\bar{z}}(u,z,\bar{z})&=-\bar{\partial}^3\int\limits_{-\infty}^udu^\prime\text{ }(u-u^\prime)\text{ }\partial_{u^\prime}\left[C_{zz}(u^\prime,z,\bar{z})+D_{zz}(u^\prime,z,\bar{z})\right]\\
T^u_{\hspace{1.5mm}{z}}(u,z,\bar{z})&=-{\partial}^3\int\limits_{-\infty}^udu^\prime\text{ }(u-u^\prime)\text{ }\partial_{u^\prime}\left[C_{\bar{z}\bar{z}}(u^\prime,z,\bar{z})+D_{\bar{z}\bar{z}}(u^\prime,z,\bar{z})\right]
\end{align}
One of the crucial ingredients is the bulk null momentum space mode expansions \eqref{66} so that we have:
\begin{align}
&\partial_{u}\left[C_{zz}(u,z,\bar{z})+D_{zz}(u,z,\bar{z})\right]\nonumber\\
&\qquad\qquad\qquad=-\frac{1}{4\pi^2}\int\limits_{0}^\infty {d\omega}\text{ }\omega\left[e^{-i\omega u}\left\{a_+^{\text{out}}-a_+^{\text{in}}\right\}+e^{i\omega u}\left\{{a_-^{\text{out}}}^\dagger-{a_-^{\text{in}}}^\dagger\right\}\right](\omega,z,\bar{z})
\end{align}
while the other being the graviton mode commutation relation given below \cite{Donnay:2022wvx}:
\begin{align}
\left[a_{\alpha}^{\text{out}}\left(\omega_1,z_1,\bar{z}_1\right),{a_{\alpha^\prime}^{\text{out}}}^\dagger\left(\omega_2,z_2,\bar{z}_2\right)\right]=\frac{16\pi^3}{\omega_1}\delta\left(\omega_1-\omega_2\right)\delta^{(2)}\left(z_1-z_2\right)\delta_{\alpha\alpha^\prime}\label{119}
\end{align}
and similarly, for the incoming modes. We also recall that the operators $a_{\alpha}^{\text{out}}$ and $a_{\alpha}^{\text{in}}$ are related via the unitary $S$-operator as below \cite{Caron-Huot:2023ikn}:
\begin{align}
a_{\alpha}^{\text{out}}=S^\dagger a_{\alpha}^{\text{in}}S\hspace{5mm}\text{with}\hspace{5mm}a_{\alpha}^{\text{in}}\equiv a_{\alpha}
\end{align}
and that the vacuum has the following properties:
\begin{align}
S|0\rangle=|0\rangle\hspace{5mm}\text{and}\hspace{5mm}S a_{\alpha}^\dagger|0\rangle=a_{\alpha}^\dagger|0\rangle\label{120}
\end{align}

\medskip
 
By virtue of \eqref{119}-\eqref{120}, we then have:
\begin{align*}
&\left\langle0\left|\partial_{u_1}\left[C_{zz}(u_1,z_1,\bar{z}_1)+D_{zz}(u_1,z_1,\bar{z}_1)\right]\partial_{u_2}\left[C_{zz}(u_2,z_2,\bar{z}_2)+D_{zz}(u_2,z_2,\bar{z}_2)\right]\right|0\right\rangle\\
=&\int\limits_{0}^\infty \frac{d\omega_1}{4\pi^2}\int\limits_{0}^\infty \frac{d\omega_2}{4\pi^2}\text{ }\omega_1\omega_2\left\langle0\left|\left[e^{-i\omega_1 u_1}\left\{S^\dagger a_+S-a_+\right\}+e^{i\omega_1 u_1}\left\{S^\dagger a_-^\dagger S-a_-^\dagger\right\}\right](\omega_1,z_1,\bar{z}_1)\right.\right.\\
&\left.\left.\qquad\qquad\qquad\qquad\times\left[e^{-i\omega_2 u_2}\left\{S^\dagger a_+S-a_+\right\}+e^{i\omega_2 u_2}\left\{S^\dagger a_-^\dagger S-a_-^\dagger\right\}\right](\omega_2,z_2,\bar{z}_2)\right|0\right\rangle\text{ }=\text{ }0
\end{align*}
which implies that the each of the two-point correlators $\left\langle T^u_{\hspace{1.5mm}\mu}\left(\mathbf{x}_1\right)T^u_{\hspace{1.5mm}\nu}\left(\mathbf{x}_2\right)\right\rangle$ with $\mu,\nu\in\left\{u,\bar{z}\right\}$ vanishes. The $\left\langle T^u_{\hspace{1.5mm}z}\left(\mathbf{x}_1\right)T^u_{\hspace{1.5mm}z}\left(\mathbf{x}_2\right)\right\rangle$ correlator is similarly $0$. The above calculation hence confirms that all the $\left\langle S^+_k\left(\mathbf{x}_1\right)S^+_l\left(\mathbf{x}_2\right)\right\rangle$ and the $\left\langle T\left(\mathbf{x}_1\right)T\left(\mathbf{x}_2\right)\right\rangle$ correlators vanish.

\medskip

The following contribution is also zero similarly:
\begin{align*}
&\left\langle0\left|\partial_{u_1}\left[C_{zz}(u_1,z_1,\bar{z}_1)+D_{zz}(u_1,z_1,\bar{z}_1)\right]\partial_{u_2}\left[C_{\bar{z}\bar{z}}(u_2,z_2,\bar{z}_2)+D_{\bar{z}\bar{z}}(u_2,z_2,\bar{z}_2)\right]\right|0\right\rangle\\
=&\int\limits_{0}^\infty \frac{d\omega_1}{4\pi^2}\int\limits_{0}^\infty \frac{d\omega_2}{4\pi^2}\text{ }\omega_1\omega_2\left\langle0\left|\left[e^{-i\omega_1 u_1}\left\{S^\dagger a_+S-a_+\right\}+e^{i\omega_1 u_1}\left\{S^\dagger a_-^\dagger S-a_-^\dagger\right\}\right](\omega_1,z_1,\bar{z}_1)\right.\right.\\
&\left.\left.\qquad\qquad\qquad\qquad\times\left[e^{-i\omega_2 u_2}\left\{S^\dagger a_-S-a_-\right\}+e^{i\omega_2 u_2}\left\{S^\dagger a_+^\dagger S-a_+^\dagger\right\}\right](\omega_2,z_2,\bar{z}_2)\right|0\right\rangle=0
\end{align*}
which makes the $\left\langle T^u_{\hspace{1.5mm}\mu}\left(\mathbf{x}_1\right)T^u_{\hspace{1.5mm}z}\left(\mathbf{x}_2\right)\right\rangle$ correlators with $\mu\in\{u,\bar{z}\}$ and each of the $\left\langle S^+_k\left(\mathbf{x}_1\right)T\left(\mathbf{x}_2\right)\right\rangle$ correlators vanish.

Hence, we conclude that the above stress-tensor correlators contain no central charge.

\section{Infinite tower of currents}
\label{sec:Infinite tower of currents}

We shall first sketch the outlines of the derivation, detailed in \cite{Saha:2023abr}, of the singular parts of the mutual OPEs between the fields $S^+_2$, $S^+_0$ and $S^+_1$. The knowledge of the singular terms of the mutual OPEs between the latter two fields was crucial in this endeavour, that were completely determined in \cite{Saha:2023hsl}, by demanding that the OPE commutativity property holds, after making the appropriate ansatz according to the general forms \eqref{5}, and \eqref{14} and truncating those ansatz by the assumptions that:
\begin{enumerate}
\item no time-independent local field in the theory possesses negative Carrollian scaling dimension $\Delta<0$; moreover, the time-independent local field with $\Delta=0=s$ is uniquely the Identity operator, 
\item the fields $S^+_0$ and $S^+_1$ are Lorentz quasi-primaries (i.e. \textit{a priori} covariant under $\text{SL}(2,\mathbb{C})$ only).
\end{enumerate}
The singular parts of the mutual OPEs (in the $j\epsilon$-form) between $S^+_0$ and $S^+_1$ were obtained as \cite{Saha:2023hsl}:
\begin{align}
&S_0^+(\mathbf{x})S^+_0(\mathbf{x}_p)\sim0\hspace{5mm};\hspace{5mm} S_0^+(\mathbf{x})S^+_1(\mathbf{x}_p)\sim\lim\limits_{\epsilon\rightarrow0^+}-i\frac{\bar{z}-\bar{z}_p}{{(\Delta\tilde{z}_p)}}\text{ }S^+_0(\mathbf{x}_p) \, ,\nonumber\\
&S_1^+(\mathbf{x})S^+_0(\mathbf{x}_p)\sim\lim\limits_{\epsilon\rightarrow0^+}-i\left[\frac{(\bar{z}-\bar{z}_p)^2}{{(\Delta\tilde{z}_p)}}\text{ }\partial_{\bar{z}_p} S^+_0+\frac{\bar{z}-\bar{z}_p}{{(\Delta\tilde{z}_p)}}\text{ }S^+_0\right](\mathbf{x}_p)  \, , \\
&\left\{S_1^+-(u-u_p)S^+_0\right\}(\mathbf{x})S^+_1(\mathbf{x}_p)\sim\lim\limits_{\epsilon\rightarrow0^+}-i\left[\frac{(\bar{z}-\bar{z}_p)^2}{{(\Delta\tilde{z}_p)}}\text{ }\partial_{\bar{z}_p} S^+_1+\frac{\bar{z}-\bar{z}_p}{{(\Delta\tilde{z}_p)}}\text{ }2S^+_1\right](\mathbf{x}_p) \, . \nonumber
\end{align}
Then the general form of the $S^+_0$ OPE \eqref{5}, the relation \eqref{20}, the $S^+_0S^+_1$ OPE, and the first assumption together completely fix the singular part of the $S^+_0S^+_2$ OPE to be \cite{Saha:2023abr}:
\begin{align}
S_0^+(\mathbf{x})S^+_2(\mathbf{x}_p)\sim\lim\limits_{\epsilon\rightarrow0^+}-i\frac{\bar{z}-\bar{z}_p}{{(\Delta\tilde{z}_p)}}\text{ }S^+_1(\mathbf{x}_p) \, . \label{22}
\end{align}
The singular terms of the $S^+_2S^+_0$ OPE can be readily obtained from the above using the OPE (bosonic) commutativity property, and the restrictions \eqref{6}, \eqref{7}, \eqref{15}, and \eqref{16}. Similarly, starting from an ansatz respecting the general form of the $S^+_1$ OPE \eqref{14}, the singular structure of the $S^+_1S^+_2$ OPE was found to be:
\begin{align}
S_1^+(\mathbf{x})S^+_2(\mathbf{x}_p)\sim\lim\limits_{\epsilon\rightarrow0^+}-i\left[\frac{(\bar{z}-\bar{z}_p)^2}{{(\Delta\tilde{z}_p)}}\text{ }\bar{\partial}_{p} S^+_2+\frac{\bar{z}-\bar{z}_p}{{(\Delta\tilde{z}_p)}}\text{ }3S^+_2+(u-u_p)\frac{\bar{z}-\bar{z}_p}{{(\Delta\tilde{z}_p)}}\text{ }S^+_1\right](\mathbf{x}_p) \, .
\end{align}
The $S^+_2S^+_1$ OPE can then be deduced using the OPE commutativity property with the following restriction on $S^+_2$, analogous to \eqref{6} and \eqref{15}, in mind:
\begin{align}
\bar{\partial}^4 S^+_2(u,z,\bar{z})\Phi(\mathbf{x}_p)\sim0 \hspace{15mm}\text{(OPE exact)}\label{23}
\end{align}
which was inferred in \cite{Saha:2023abr} by observing from \eqref{22} that:
\begin{align*}
S_0^+\bar{\partial}^{3+N} S^+_2\sim0\hspace{2.5mm}{(N\in\mathbb{N})}\hspace{5mm}\Longrightarrow\hspace{5mm}\left(\bar{\partial}^{3+N} S^+_2\right)S^+_0\sim0\hspace{10mm}\text{(OPE exact)}
\end{align*}
and then demanding that the conformal Carrollian OPEs must be associative, just as in a 2D Euclidean CFT \cite{Zamolodchikov:1989mz}. The condition \eqref{23} readily implies the following decomposition of the object $S^+_{2e}$ (see \eqref{21}) inside a correlator:
\begin{align}
&S^+_{2e}(u,z,\bar{z})=-k^{(+2)}_e(u,z,\bar{z})+3\bar{z}k^{(+1)}_e(u,z,\bar{z})-3\bar{z}^2k^{(0)}_e(u,z,\bar{z})+\bar{z}^3k^{(-1)}_e(u,z,\bar{z})\label{24}\\
\text{such that}\hspace{2.5mm}&\langle\bar{\partial} k^a_e(u,z,\bar{z})X\rangle=-i\sum\limits_{p=1}^n\text{ }\theta(u-u_p)\text{ }\left[\text{contact terms on $S^2$}\right]\nonumber
\end{align}
just analogous to \eqref{4} and \eqref{18}. The holomorphic weight of the field $S^+_2$ as well as all the four $k^a_e$ is $h=\frac{1}{2}$. For future reference, we note down the $S^+_2S^+_1$ OPE below:
\begin{align}
S_2^+(\mathbf{x})S^+_1(\mathbf{x}_p)&\sim\lim\limits_{\epsilon\rightarrow0^+}-i\left[\frac{\left(\bar{z}-\bar{z}_p\right)^3}{{(\Delta\tilde{z}_p)}}\text{ }\frac{1}{2}\bar{\partial}^2_p S^+_2+\frac{\left(\bar{z}-\bar{z}_p\right)^2}{{(\Delta\tilde{z}_p)}}\text{ }2\bar{\partial}_p S^+_2+\frac{\bar{z}-\bar{z}_p}{{(\Delta\tilde{z}_p)}}\text{ }3S^+_2\right.\nonumber\\
&\left.+(u-u_p)\left\{\frac{(\bar{z}-\bar{z}_p)^2}{{(\Delta\tilde{z}_p)}}\text{ }\bar{\partial}_{p} S^+_1+\frac{\bar{z}-\bar{z}_p}{{(\Delta\tilde{z}_p)}}\text{ }2S^+_1\right\}+\frac{1}{2}(u-u_p)^2\text{ }\frac{\bar{z}-\bar{z}_p}{{(\Delta\tilde{z}_p)}}\text{ }S^+_0\right](\mathbf{x}_p) \, . \label{25}
\end{align}

\medskip

To find the (singular part of the) $S^+_2S^+_2$ OPE, we first make the following ansatz consistent with the decomposition \eqref{21} and the restriction \eqref{24} \cite{Saha:2023abr}:
\begin{align}
S_2^+(\mathbf{x})S^+_2(\mathbf{x}_p)\sim\lim\limits_{\epsilon\rightarrow0^+}-i&\left[\sum_{r=0}^3\sum_{s\geq1}\frac{(\bar{z}-\bar{z}_p)^r}{{(\Delta\tilde{z}_p)}^s}\text{ }A_{r,s}+(u-u_p)\left\{\frac{(\bar{z}-\bar{z}_p)^2}{{(\Delta\tilde{z}_p)}}\text{ }\bar{\partial}_{p} S^+_2+\frac{\bar{z}-\bar{z}_p}{{(\Delta\tilde{z}_p)}}\text{ }3S^+_2\right\}\right.\nonumber\\
&\left.\hspace{36mm}+\frac{1}{2}(u-u_p)^2\text{ }\frac{\bar{z}-\bar{z}_p}{{(\Delta\tilde{z}_p)}}\text{ }S^+_1\right](\mathbf{x}_p)\label{49}
\end{align}
with $A_{r,s}(\mathbf{x}_p)$ being some yet-to-be-determined fields that are mutually local with $S^+_2,S^+_1,S^+_0$. Now, the singular part of the $S_2^+(\mathbf{x})\partial_{u_p}S^+_2(\mathbf{x}_p)$ OPE obtained from this ansatz must be in complete agreement with that of the $S_2^+(\mathbf{x})S^+_1(\mathbf{x}_p)$ OPE \eqref{25} because of the relation \eqref{20}. The non-trivial constraints, exact in the OPE limit, coming entirely from the $\mathcal{O}\left((u-u_p)^0\right)$ terms are listed below (where $\dot{}$ represents time-derivative):
\begin{align}
&\dot{A}_{r,s\geq2}\sim0\hspace{5mm},\hspace{5mm}\dot{A}_{0,1}\sim0\nonumber\\
&\dot{A}_{3,1}\sim\frac{1}{2}\bar{\partial}^2_p S^+_2\hspace{5mm},\hspace{5mm}\dot{A}_{2,1}\sim3\bar{\partial}_p S^+_2\hspace{5mm},\hspace{5mm}\dot{A}_{1,1}\sim6S^+_2 \, .  \label{26} 
\end{align}
Thus all the local fields ${A}_{r,s\geq2}$ and ${A}_{0,1}$ are time-independent (in the OPE limit) and have $\Delta<0$; all of them are hence set to zero by the first assumption. More interestingly, the constraints \eqref{26} imply that, for the consistency of the $S^+_2S^+_2$ OPE, there must \textit{automatically} exist a local field $S^+_3$ in the 3D Carrollian CFT, that, in the OPE limit, obeys \cite{Saha:2023abr}:
\begin{align} \label{next spin}
\left(\partial_u S^+_3-S^+_2\right)(u,z,\bar{z})\Phi(\mathbf{x}_p)\sim0 \, .
\end{align}
The dimensions of such a field $S^+_3$ are $(\Delta,s)=(-2,2)$. Consistently with \eqref{21}, the $S^+_3$ field can be decomposed inside a correlator as:
\begin{align}
&S^+_3(u,z,\bar{z})=S^+_{3e}(u,z,\bar{z})+uS^+_{2e}(u,z,\bar{z})+\frac{u^2}{2}S^+_{1e}(u,z,\bar{z})+\frac{u^3}{3!}S^+_0(u,z,\bar{z})\\
\Longrightarrow\hspace{5mm}&{\partial_u}\langle S^+_{3e}(u,z,\bar{z})X\rangle=\left[\text{temporal contact terms}\right] \, .   \nonumber
\end{align}
The automatic presence of the local field $S^+_3$ may be interpreted as the non-closure of the algebra of the modes contained in the object $S^+_{2e}$. Finally, the $S^+_2S^+_2$ OPE is written down below:
\begin{align}
S_2^+(\mathbf{x})S^+_2(\mathbf{x}_p)&\sim\lim\limits_{\epsilon\rightarrow0^+}-i\left[\frac{(\bar{z}-\bar{z}_p)^3}{{(\Delta\tilde{z}_p)}}\text{ }\frac{1}{2}\bar{\partial}^2_pS^+_3+\frac{(\bar{z}-\bar{z}_p)^2}{{(\Delta\tilde{z}_p)}}\text{ }3\bar{\partial}_pS^+_3+\frac{\bar{z}-\bar{z}_p}{{(\Delta\tilde{z}_p)}}\text{ }6S^+_3\right.\nonumber\\
&\left.+(u-u_p)\left\{\frac{(\bar{z}-\bar{z}_p)^2}{{(\Delta\tilde{z}_p)}}\text{ }\bar{\partial}_{p} S^+_2+\frac{\bar{z}-\bar{z}_p}{{(\Delta\tilde{z}_p)}}\text{ }3S^+_2\right\}+\frac{1}{2}(u-u_p)^2\text{ }\frac{\bar{z}-\bar{z}_p}{{(\Delta\tilde{z}_p)}}\text{ }S^+_1\right](\mathbf{x}_p) \, .
\end{align}

\medskip

Again, from the $S^+_0S^+_3$ OPE together with the restriction \eqref{23} (and OPE associativity), it was shown in \cite{Saha:2023abr} that:
\begin{align} \label{interm1}
\bar{\partial}^5 S^+_3(u,z,\bar{z})\Phi(\mathbf{x}_p)\sim0\hspace{15mm}\text{(OPE exact)}
\end{align}
revealing that an $S^+_3\Phi$ OPE must be a quartic polynomial in $(\bar{z}-\bar{z}_p)$. Hence, analogous to \eqref{24}, $S^+_{3e}$ can be decomposed into five objects $l^a_e(u,z,\bar{z})$ whose correlators are both holomorphic and time-independent in the OPE limit. The $S^+_2S^+_3$ OPE was derived in \cite{Saha:2023abr} using the knowledge of the $S^+_2S^+_2$ OPE in exactly the same way (as reviewed above) the latter was obtained. The derivation revealed that for the $S^+_2S^+_3$ OPE to be consistent, a local field $S^+_4$ with dimensions $(\Delta,s)=(-3,2)$ must \textit{automatically} be present in the theory, that satisfies:
\begin{align} \label{interm2}
&\left(\partial_u S^+_4-S^+_3\right)(u,z,\bar{z})\Phi(\mathbf{x}_p)\sim0\hspace{15mm}\text{(OPE exact)}\\
\Longrightarrow\hspace{5mm}&\text{ }\bar{\partial}^6 S^+_4(u,z,\bar{z})\Phi(\mathbf{x}_p)\sim0 
 \, . \nonumber 
\end{align}
Recognising the above pattern, the \textit{automatic} existence of a tower of mutually local Carrollian fields $S^+_{k+1}$ ($k\geq2$) satisfying the following conditions, just to render the $S^+_2S^+_k$ OPEs consistent, was recursively realised in \cite{Saha:2023abr}:
\begin{align*}
\left(\partial_t S^+_{k+1}-S^+_k\right)(t,z,\bar{z})\Phi(\mathbf{x}_p)\sim0\hspace{5mm}
\Longrightarrow\hspace{5mm}\bar{\partial}^{k+2} S^+_k(t,z,\bar{z})\Phi(\mathbf{x}_p)\sim0\hspace{10mm}\text{(OPE exact)} \, .
\end{align*}
It is to be reemphasized that we only needed to postulate the presence of a field $S^+_2$ obeying \eqref{20} in the 3D CarrCFT to begin with. Recursing the steps to derive the $S^+_2S^+_2$ OPE \eqref{49}, the general $S^+_2S^+_k$ OPE was obtained in \cite{Saha:2023abr} as:
\begin{align}
&iS_2^+(\mathbf{x})S^+_k(\mathbf{x}_p) \label{35}  \\
&\sim\lim\limits_{\epsilon\rightarrow0^+}\left[\frac{(\bar{z}-\bar{z}_p)^3}{{(\Delta\tilde{z}_p)}}\text{ }\frac{1}{2}\bar{\partial}_p^2S^+_{k+1}+\frac{(\bar{z}-\bar{z}_p)^2}{{(\Delta\tilde{z}_p)}}(k+1)\bar{\partial}_pS^+_{k+1}+\frac{\bar{z}-\bar{z}_p}{{(\Delta\tilde{z}_p)}}\text{ }\frac{1}{2}(k+1)(k+2)S^+_{k+1}\right.\nonumber\\
&\quad\left.+(u-u_p)\left\{\frac{(\bar{z}-\bar{z}_p)^2}{{(\Delta\tilde{z}_p)}}\text{ }\bar{\partial}_{p} S^+_k+\frac{\bar{z}-\bar{z}_p}{{(\Delta\tilde{z}_p)}}\text{ }(k+1)S^+_k\right\}+\frac{1}{2}(u-u_p)^2\text{ }\frac{\bar{z}-\bar{z}_p}{{(\Delta\tilde{z}_p)}}\text{ }S^+_{k-1}\right](\mathbf{x}_p) \, . \nonumber
\end{align}

\section{OPE formulae for \texorpdfstring{$S_2^+(u,z,\bar{z})$}{S2} and subsubleading soft theorem}
\label{sec:S2 formulae}

We first discuss how the Ward identity of the $S^+_2$ field encodes the positive helicity subsubleading energetically soft graviton theorem \cite{Cachazo:2014fwa}. The starting point in \cite{Saha:2023abr} was to postulate that the field $S^+_2$ with a conformal Carrollian primary $\Phi$ of a certain type, with $\left(\bar{{\bm{\xi}}}\cdot \Phi\right)=0=\left(\bm{\xi}\cdot \Phi\right)$ and dimensions $(h,\bar{h})$, has the following OPE:
\begin{align}
S_2^+(\mathbf{x})\Phi(\mathbf{x}_p)&\sim\lim\limits_{\epsilon\rightarrow0^+}-i\left[\frac{(\bar{z}-\bar{z}_p)^3}{{(\Delta\tilde{z}_p)}}\text{ }\frac{1}{2}\bar{\partial}_p^2\Phi_1-\frac{(\bar{z}-\bar{z}_p)^2}{{(\Delta\tilde{z}_p)}}\text{ }2\bar{h}\bar{\partial}_p\Phi_1+\frac{\bar{z}-\bar{z}_p}{{(\Delta\tilde{z}_p)}}\text{ }\frac{1}{2}(2\bar{h})(2\bar{h}-1)\Phi_1\right.\nonumber\\
&\left.+(u-u_p)\left\{\frac{(\bar{z}-\bar{z}_p)^2}{{(\Delta\tilde{z}_p)}}\text{ }\bar{\partial}_{p} \Phi-\frac{\bar{z}-\bar{z}_p}{{(\Delta\tilde{z}_p)}}\text{ }2\bar{h}\Phi\right\}+\frac{1}{2}(u-u_p)^2\text{ }\frac{\bar{z}-\bar{z}_p}{{(\Delta\tilde{z}_p)}}\text{ }\dot{\Phi}\right](\mathbf{x}_p)\label{36}
\end{align} 
with the unique local Carrollian field $\Phi_1$ exactly satisfying $\dot{\Phi}_1\sim\Phi$ in the OPE limit. This was motivated\footnote{Unlike its celestial counterpart (see however \cite{Mitra:2024ugt}), the conformal Carrollian weight $\Delta$ is expected to be discrete. The reason is that a 3D Carrollian CFT primary corresponding to a 4D bulk mass-less scattering field must possess $\Delta=1$, and hence, all its descendants have integer Carrollian weights.} by the OPEs \eqref{35} which are valid for an infinite number of conformal Carrollian primaries $S^+_k$, each having $\left(\bar{{\bm{\xi}}}\cdot S^+_k\right)=0=\left(\bm{\xi}\cdot S^+_k\right)$ and dimensions $(h,\bar{h})=\left(\frac{3-k}{2},-\frac{k+1}{2}\right)$. The OPE of the field $\Phi$ may not completely fix that of the field $\Phi_1$; rather, it may happen that:
\begin{align*}
\Phi_1(u,z,\bar{z})\sim\int\limits_{-\infty}^u du^\prime\text{ }\Phi(u^\prime,z,\bar{z})+\text{(terms with temporal step-function factor)}
\end{align*}
as are the cases for the fields $S^+_k$. We shall end this appendix with a $\text{Witt}\loplus 
Lw_{1+\infty}$ symmetry-based partial derivation of the above OPE. 

\medskip

Decomposing the $S^+_2$ field according to \eqref{29} and \eqref{30}, using the mode-expansion \eqref{37}, and applying the 3D Carrollian CFT OPE $\longleftrightarrow$ commutator prescription \cite{Saha:2023hsl} on the $S^+_2$ OPE \eqref{36}, it was found that the modes contained in the object $S^+_{2e}$ induce the following infinitesimal transformations on a Carrollian CFT primary $\Phi(\mathbf{x}_p)$ \cite{Saha:2023abr}:
\begin{align}
&i\left[H^2_{-\frac{3}{2};n}\text{ },\text{ }\Phi\right]=3{z}_p^{n-\frac{1}{2}}\bar{\partial}_p^2\Phi_1(\mathbf{x}_p) 
\, ,  \nonumber\\
&i\left[H^2_{-\frac{1}{2};n}\text{ },\text{ }\Phi\right]={z}_p^{n-\frac{1}{2}}\left[3\bar{z}_p\bar{\partial}_p^2\Phi_1+4\bar{h}\bar{\partial}_p\Phi_1+2u_p\bar{\partial}_p\Phi\right](\mathbf{x}_p)   \, ,\label{54}\\
&i\left[H^2_{\frac{1}{2};n}\text{ },\text{ }\Phi\right]={z}_p^{n-\frac{1}{2}}\left[3\bar{z}^2_p\bar{\partial}_p^2\Phi_1+8\bar{h}\bar{z}_p\bar{\partial}_p\Phi_1+2\bar{h}(2\bar{h}-1)\Phi_1+4\bar{z}_p u_p\bar{\partial}_p\Phi+4\bar{h}u_p\Phi+u_p^2\dot{\Phi}\right](\mathbf{x}_p)   \, ,\nonumber\\
&i\left[H^2_{\frac{3}{2};n}\text{ },\text{ }\Phi\right]=3{z}_p^{n-\frac{1}{2}}\left[\bar{z}^3_p\bar{\partial}_p^2\Phi_1+4\bar{h}\bar{z}^2_p\bar{\partial}_p\Phi_1+2\bar{h}(2\bar{h}-1)\bar{z}_p\Phi_1+2\bar{z}^2_p u_p\bar{\partial}_p\Phi+4\bar{h}u_p\bar{z}_p\Phi+u_p^2\bar{z}_p\dot{\Phi}\right](\mathbf{x}_p) \, , \nonumber
\end{align} 
with the four modes $H^2_{a;\frac{1}{2}}$ generating four global transformations. Thus, the 3D Carrollian CFT correlators must be invariant under these four global transformations, besides the ten Poincar\'e ones, if the field $S^+_2$ exists in the theory. Interestingly, the above transformations generated by $S^+_2$ are non-local in time (but local in space) since they act on the primary field with time-integration.

\medskip

Demanding that the $\langle S_2^+(u,z,\bar{z})X\rangle$ correlator, where $X$ is a string of the above-mentioned special Carrollian CFT primaries all with $\left(\bar{{\bm{\xi}}}\cdot \Phi\right)=0=\left(\bm{\xi}\cdot \Phi\right)$, be finite whenever $z\neq z_p$ (especially at $z=\infty$) and recalling that the field $S^+_2$ has $h>0$, the $S^+_2$ Ward identity corresponding to the OPE \eqref{36} is uniquely given by (up to temporal contact terms): 
\begin{align}
i\langle S_2^+(u,z,\bar{z}) X\rangle=\sum\limits_{p=1}^n\theta&(u-u_p)\left[\frac{(\bar{z}-\bar{z}_p)^3}{{z-z_p}}\text{ }\frac{1}{2}\bar{\partial}_p^2\partial_{u_p}^{-1}-\frac{(\bar{z}-\bar{z}_p)^2}{{z-z_p}}\text{ }2\bar{h}_p\bar{\partial}_p\partial_{u_p}^{-1}+\frac{\bar{z}-\bar{z}_p}{{z-z_p}}\text{ }\bar{h}_p(2\bar{h}_p-1)\partial_{u_p}^{-1}\right.\nonumber\\
+&\left.(u-u_p)\left\{\frac{(\bar{z}-\bar{z}_p)^2}{z-{z}_p}\bar{\partial}_{p}-2\bar{h}_p\frac{\bar{z}-\bar{z}_p}{z-{z}_p}\right\}+\frac{(u-u_p)^2}{2}\text{ }\frac{\bar{z}-\bar{z}_p}{z-{z}_p}\partial_{u_p}\right]\langle X\rangle 
  \, . \label{60}
\end{align}
With the anti-derivative operator $\partial_{u_p}^{-1}$ changing $(\Delta,s)\rightarrow(\Delta-1,s)$, the $\langle S_{2e}^+(u,z,\bar{z}) X\rangle$ part of this Ward identity is the 3D conformal Carrollian analogue of the 2D celestial positive-helicity subsubleading conformally soft graviton Ward identity \cite{Pate:2019lpp}. Consequently, the $\langle S_{2e}^+(u,z,\bar{z}) X\rangle$ `Ward identity' in the explicit $u\rightarrow\infty$ limit, with each primary in $X$ having $\Delta_p=1$, is equivalent to the subsubleading energetically soft graviton theorem. The $S_2^+$ Ward identity corresponding to the postulated $S^+_2\Phi$ OPE \eqref{36} thus uniquely encodes the subsubleading soft graviton theorem as desired, providing an indirect \textit{a posteriori} justification for the said postulate.   

\medskip

We now partially derive the (singular part of the) $S^+_2\Phi$ OPE \eqref{36}, postulated in \cite{Saha:2023abr}, using a $\text{Witt}\loplus 
Lw_{1+\infty}$ symmetry argument. We begin with the following generalization of the ansatz \eqref{49} for a conformal Carrollian primary $\Phi$ with $\left(\bar{{\bm{\xi}}}\cdot \Phi\right)=0=\left(\bm{\xi}\cdot \Phi\right)$ and dimensions $(h,\bar{h})$:
\begin{align}
S_2^+(\mathbf{x})\Phi(\mathbf{x}_p)\sim\lim\limits_{\epsilon\rightarrow0^+}-i&\left[\sum_{d=0}^3\sum_{s\geq1}\frac{(\bar{z}-\bar{z}_p)^d}{{(\Delta\tilde{z}_p)}^s}\text{ }B_{d,s}+(u-u_p)\left\{\frac{(\bar{z}-\bar{z}_p)^2}{{(\Delta\tilde{z}_p)}}\text{ }\bar{\partial}_{p} \Phi-\frac{\bar{z}-\bar{z}_p}{{(\Delta\tilde{z}_p)}}\text{ }2\bar{h}\Phi\right\}\right.\nonumber\\
&\left.\hspace{36mm}+\frac{1}{2}(u-t_p)^2\text{ }\frac{\bar{z}-\bar{z}_p}{{(\Delta\tilde{z}_p)}}\text{ }\dot{\Phi}\right](\mathbf{x}_p)\label{56}
\end{align}
with the goal to determine the local fields $B_{d,s}(\mathbf{x}_p)$ using symmetry principles. Using the 3D Carrollian CFT OPE $\longleftrightarrow$ commutator prescription \cite{Saha:2023hsl} on this OPE-ansatz, we deduce that the modes $H^2_{-\frac{3}{2};n}$ with $n\geq\frac{1}{2}$ (cf. the decomposition \eqref{30} and the mode-expansion \eqref{37} for $k=2$), contained in the object $S^+_{2e}$, infinitesimally transform the field $\Phi$ as below:
\begin{align}
    i\left[H^2_{-\frac{3}{2};n}\text{ },\text{ }\Phi(\mathbf{x}_p)\right]=6{z}_p^{n-\frac{1}{2}}\sum_{s=0}^{n-\frac{1}{2}}\frac{\left(n+\frac{1}{2}-s\right)_s}{z_p^s\cdot s!}\text{ } B_{s}(\mathbf{x}_p)\label{50}
\end{align}
with $B_s\equiv B_{3,s}$. Now, the following part of the algebra \eqref{34}:
\begin{align}
i\left[H^0_{a;n}\text{ },\text{ }H^2_{-\frac{3}{2};m}\right]=\left[3a+\frac{3}{2}\right]H^{1}_{a-\frac{3}{2};n+m}\hspace{5mm};\hspace{5mm} i\left[H^1_{a;n}\text{ },\text{ }H^2_{-\frac{3}{2};m}\right]=\left[3a+3\right]H^{2}_{a-\frac{3}{2};n+m}\label{51}
\end{align} 
together with the commutator below: 
\begin{align*}
    i\left[L_n\text{ },\text{ }H^2_{-\frac{3}{2};m}\right]=\left[-\frac{n}{2}-m\right]H^{2}_{-\frac{3}{2};n+m}
\end{align*}
suggests that it may be possible to impose some constraints on the local fields $B_{3,s}$ using the Jacobi identity involving the primary field $\Phi(\mathbf{x}_p)$, the mode $H^2_{-\frac{3}{2};m\geq\frac{1}{2}}$ and one of the five modes $\left\{H^0_{-\frac{1}{2};n\geq-\frac{1}{2}},H^0_{\frac{1}{2};n\geq-\frac{1}{2}},H^1_{-1;n\geq0},H^1_{0;n\geq0},L_{n\geq-1}\right\}$. Furthermore, we can completely fix the fields $B_{d,s}$ for $d=\{2,1,0\}$ in terms of the fields $B_{3,s}\equiv B_s$ and $\Phi$ (and their descendants/ascendants under the `universal' subalgebra of \eqref{34}) just by the use of the Jacobi identities involving $\Phi(\mathbf{x}_p)$, $H^1_{1;n\geq0}$ and $H^2_{a;m\geq\frac{1}{2}}$ for $a=\left\{-\frac{3}{2},-\frac{1}{2},\frac{1}{2}\right\}$ consecutively.

\medskip

We now consider the Jacobi identity involving $\Phi(\mathbf{x}_p)$, $H^2_{-\frac{3}{2};m\geq\frac{1}{2}}$ and one of $H^0_{\pm\frac{1}{2};n\geq-\frac{1}{2}}$. One of the necessary commutators, $\left[H^0_{a;n}\text{ },\text{ }B_s(\mathbf{x}_p)\right]$ can be obtained from the $S^+_0B_s$ OPE (in the general form) \eqref{5} to be (with $n\geq-\frac{1}{2}$):
\begin{align}
    i\left[H^0_{a;n}\text{ },\text{ }B_s(\mathbf{x}_p)\right]=\bar{z}_p^{a-\frac{1}{2}}{z}_p^{n-\frac{1}{2}}\sum_{r=-1}^{n-\frac{1}{2}}\frac{\left(n+\frac{1}{2}-r\right)_{r+1}}{z_p^r\cdot (r+1)!}\left[\bar{z}_p\left(P_{r,-1}B_s\right)+\left(a+\frac{1}{2}\right)\left(P_{r,0}B_s\right)\right](\mathbf{x}_p)
\end{align}
(recall that $\left(P_{-1,-1}B_s\right)\equiv\dot{B}_s$). Using this, along with the commutators \eqref{50}, \eqref{51} and 
\begin{align*}
i\left[H^0_{a;n}\text{ },\text{ }\Phi(\mathbf{x}_p)\right]=\bar{z}_p^{a+\frac{1}{2}}{z}_p^{n-\frac{1}{2}}\dot{\Phi}(\mathbf{x}_p)    
\end{align*}
we extract the following constraints on $B_s$ from the said Jacobi identities:
\begin{align}
    &\left(P_{r,-1}B_s\right)=0\hspace{2.5mm}\text{ for }s\geq0,r\geq0\hspace{5mm};\hspace{5mm}
    \left(P_{r,0}B_s\right)=0\hspace{2.5mm}\text{ for }s\geq1,r\geq-1 \, ,\nonumber\\
&\left(P_{-1,0}B_0\right)=\bar{\partial}_p\Phi\hspace{5mm};\hspace{5mm}\left(P_{r,0}B_0\right)=0=\left(P_{r,-1}B_0\right)\hspace{2.5mm}\text{ for }r\geq0 \, . \label{59}
\end{align}
Obeying these constraints, we can readily write the $S^+_0B_s$ OPEs as below:
\begin{align}
&S^+_0(u,z,\bar{z})B_0(\mathbf{x}_p)\sim\lim\limits_{\epsilon\rightarrow0^+}-i\left[(\bar{z}-\bar{z}_p)\text{ }\frac{\partial_{u_p}B_0}{\Delta\tilde{z}_p}-\frac{\bar{\partial}_p\Phi}{{\Delta\tilde{z}_p}}\right](\mathbf{x}_p)\label{52} \, ,\\
&S^+_0(u,z,\bar{z})B_{s}(\mathbf{x}_p)\sim\lim\limits_{\epsilon\rightarrow0^+}-i(\bar{z}-\bar{z}_p)\text{ }\frac{\partial_{u_p}B_s(\mathbf{x}_p)}{\Delta\tilde{z}_p}\hspace{5mm}\text{ for }s\geq1 \, .  \nonumber    
\end{align}

\medskip

Next, replacing $H^0_{\pm\frac{1}{2};n\geq-\frac{1}{2}}$ in the above Jacobi identity first with $H^1_{-1;n\geq0}$, then with $H^1_{0;n\geq0}$ and finally with $L_{n\geq0}$, we similarly obtain the following constraints (cf. the OPEs \eqref{14} and \eqref{8}):
\begin{align}
&\left(j^{(-)}_{r}B_s\right)=0\hspace{1.5mm};\hspace{1.5mm}\left(j^{(0)}_{r}B_s\right)=\frac{3}{2}B_{r+s}\hspace{1.5mm};\hspace{1.5mm}\left(L_{r}B_s\right)=-\frac{1}{2}(r+2s+1)B_{r+s}\hspace{3mm}\text{all for }s\geq0,r\geq1\nonumber\\
\Longrightarrow\hspace{2.5mm}&\left(j^{(-)}_{s}B_0\right)=0\hspace{2.5mm};\hspace{2.5mm}B_s=\frac{2}{3}\left(j^{(0)}_{s}B_0\right)=-\frac{2}{s+1}\left(L_{s}B_0\right)\hspace{5mm}\text{all for $s\geq1$}\label{53}
\end{align}
implying that once we get to know the $S^+_1B_0$ or the $TB_0$ OPE, all the fields $B_{s\geq1}$ can be directly read off from the coefficients of the singular terms in the said OPEs.

\medskip

We observe that the simplest choice for the field $B_0(\mathbf{x}_p)$ such that the OPE \eqref{52} is satisfied is $\frac{1}{2}\bar{\partial}^2_p\Phi_1(\mathbf{x}_p)$ with $\dot{\Phi}_1\sim\Phi$ exactly (in the OPE limit) and the $S^+_0\Phi_1$ OPE being:
\begin{align}
S^+_0(u,z,\bar{z})\Phi_1(\mathbf{x}_p)\sim\lim\limits_{\epsilon\rightarrow0^+}-i(\bar{z}-\bar{z}_p)\text{ }\frac{\Phi(\mathbf{x}_p)}{\Delta\tilde{z}_p}   \, . \label{58} 
\end{align}
Clearly, the $S^+_2S^+_k$ OPEs \eqref{35} conform to this choice. Remaining compatible with the constraints 
\eqref{53}, we can further choose the field $\Phi_1$ to be a conformal Carrollian primary satisfying the following OPEs:
\begin{align}
    &S_1^+(\mathbf{x})\Phi_1(\mathbf{x}_p)\sim\lim\limits_{\epsilon\rightarrow0^+}-i\left[\frac{(\bar{z}-\bar{z}_p)^2}{{(\Delta\tilde{z}_p)}}\text{ }\bar{\partial}_{p} \Phi_1-\frac{\bar{z}-\bar{z}_p}{{(\Delta\tilde{z}_p)}}\text{ }(2\bar{h}-1)\Phi_1+(u-u_p)\text{ }\frac{\bar{z}-\bar{z}_p}{{(\Delta\tilde{z}_p)}}\text{ }{\Phi}\right](\mathbf{x}_p) \, ,\\
&T(u,z,\bar{z})\Phi_1(\mathbf{x}_p)\sim\lim\limits_{\epsilon\rightarrow0^+}-i\left[\frac{\left(h-\frac{1}{2}\right)\Phi_1}{(\Delta\tilde{z}_p)^2}+\frac{\partial_{z_p}\Phi_1}{\Delta\tilde{z}_p}-\frac{u-u_p}{2}\frac{\Phi}{(\Delta\tilde{z}_p)^2}\right](\mathbf{x}_p) \, ,
\end{align}
rendering all the fields $B_{s\geq1}$ to be 0. With this simplest choice of the field $B_0$, the commutator \eqref{50} simply reduces to:
\begin{align}
    i\left[H^2_{-\frac{3}{2};n}\text{ },\text{ }\Phi(\mathbf{x}_p)\right]=3{z}_p^{n-\frac{1}{2}}\bar{\partial}^2_p\Phi_1(\mathbf{x}_p) \, . \label{55}
\end{align}

\medskip

With the explicit commutator \eqref{55} at hand, one can now obtain the $\left[H^2_{-\frac{1}{2};n}\text{ },\text{ }\Phi(\mathbf{x}_p)\right]$ one, as mentioned earlier, from the Jacobi Identity involving $\Phi$, $H^2_{-\frac{3}{2};n}$ and $H^1_{1;0}$ to be:
\begin{align}
    i\left[H^2_{-\frac{1}{2};n}\text{ },\text{ }\Phi(\mathbf{x}_p)\right]={z}_p^{n-\frac{1}{2}}\left[3\bar{z}_p\bar{\partial}_p^2\Phi_1+4\bar{h}\bar{\partial}_p\Phi_1+2u_p\bar{\partial}_p\Phi\right](\mathbf{x}_p) \, . \label{57}
\end{align}
On the other hand, using the 3D Carrollian CFT OPE $\longleftrightarrow$ commutator map on the OPE-ansatz \eqref{56}, we get (for $n\geq\frac{1}{2}$):
\begin{align*}
    i\left[H^2_{-\frac{1}{2};n}\text{ },\text{ }\Phi(\mathbf{x}_p)\right]=2{z}_p^{n-\frac{1}{2}}\left[u_p\bar{\partial}_p\Phi+\sum_{s=0}^{n-\frac{1}{2}}\frac{\left(n+\frac{1}{2}-s\right)_s}{z_p^s\cdot s!}\left(3\bar{z}_p B_{3,s}-B_{2,s}\right)\right](\mathbf{x}_p)
\end{align*}
which, upon comparison with \eqref{57}, implies that $B_{2,0}=-2\bar{h}\bar{\partial}_p\Phi$ and $B_{2,s\geq1}=0$. 

\medskip

Having recognized the commutators \eqref{55} and \eqref{57} to be the first two of the transformations \eqref{54}, we can find the remaining two by replacing $H^2_{-\frac{3}{2};n}$, first with $H^2_{-\frac{1}{2};n}$ and next with $H^2_{\frac{1}{2};n}$ in the above-mentioned Jacobi identity. Following the derivation of the fields $B_{2,s}$, one can figure out that $B_{1,0}=\bar{h}(2\bar{h}-1)\Phi_1$ and $B_{1,s\geq1}=0$ from the third commutator in \eqref{54} and finally, from the last one, $B_{0,s\geq0}=0$. Substituting the expressions obtained above for the fields $B_{d,s}$ back into the ansatz \eqref{56}, we find the singular part of the $S^+_2\Phi$ OPE to be exactly \eqref{36}.

\medskip

It is important to emphasize that we have not uniquely determined the $\left[H^2_{a;n}\text{ },\text{ }\Phi(\mathbf{x}_p)\right]$ commutators to be \eqref{54} from the constraints \eqref{59} and \eqref{53} arising from the symmetry arguments. Rather, the commutators \eqref{54} follow from symmetry once we make the following simplest possible choice for the field $B_{3,0}$, compatible with the said constraints: $B_{3,0}=\frac{1}{2}\bar{\partial}^2\Phi_1$ with $\Phi_1$ being a Carrollian CFT primary satisfying the OPE \eqref{58}. Thus, the transformation properties of the Carrollian CFT primary $\Phi$ under the action of the modes of the `universal' generator fields $S^+_0,S^+_1,T$ can not uniquely fix the transformations induced by the modes $H^2_{a;n}$ contained in the object $S^+_{2e}$. This is consistent with the fact that the field $S^+_2$ generates four new global spacetime transformations independent of the ten Poincar\'e ones (induced by the three `universal' generators).

\medskip

As shown in \cite{Saha:2023abr}, the $\left\langle S^+_2X\right\rangle$ Ward identity \eqref{60} corresponding to the $S^+_2\Phi$ OPE \eqref{36} encodes the 4D bulk  soft graviton theorem up to the subsubleading order when each of the Carrollian CFT primaries in $X$ possesses $\left(\bar{{\bm{\xi}}}\cdot \Phi\right)=0=\left(\bm{\xi}\cdot \Phi\right)$ and scaling dimension $\Delta=1$. But, since the $S^+_2\Phi$ OPE \eqref{36} is not unique from the symmetry perspective, so is the corresponding Ward identity \eqref{60}. This translates into the following additional criterion for a Carrollian CFT primary $\Phi$ if it is to correspond to a 4D bulk  massless scattering field: under the action of the global generator $H^2_{-\frac{3}{2};\frac{1}{2}}$, it must transform as:
\begin{align}
  i\left[H^2_{-\frac{3}{2};\frac{1}{2}}\text{ },\text{ }\Phi(\mathbf{x}_p)\right]=3\bar{\partial}^2_p\Phi_1(\mathbf{x}_p) 
\end{align}
as seen from the commutator \eqref{55}, with $\Phi_1$ being a conformal Carrollian primary (with $\Delta=0$) satisfying the OPE \eqref{58}. In other words, the choice $\frac{1}{2}\bar{\partial}^2\Phi_1$ for the field $B_{3,0}$ is fortunately both the simplest and the one relevant for flat space holography. 

\medskip

Once we have found out how the field $\Phi$ transforms under the action of the modes $H^2_{a;n}$, be it \eqref{54} or any other possibilities respecting the constraints \eqref{59} and \eqref{53}, we can recursively infer what transformations the modes $H^{k\geq3}_{a;n}$ induce on $\Phi$ just by using the Jacobi identities involving $\Phi(\mathbf{x})$, $H^2_{a;n}$ and $H^{k-1}_{b;m}$, starting with $k=3$, because of the following part of the algebra \eqref{34}:
\begin{align*}
    i\left[H^2_{a;n}\text{ },\text{ }H^{k-1}_{b;m}\right]=\left[ak-3b\right]H^{k}_{a+b;n+m} \, .
\end{align*}
For example, if the $\left[H^2_{a;n}\text{ },\text{ }\Phi(\mathbf{x})\right]$ commutators are indeed given by \eqref{54}, all the $\left[H^{k\geq3}_{a;n}\text{ },\text{ }\Phi(\mathbf{x})\right]$ ones can be shown to be arising from the $S^+_{k\geq3}\Phi$ OPEs \eqref{38}. All the fields $\Phi_{r\geq0}$ (cf. \eqref{44}) in these OPEs are Carrollian CFT primaries satisfying the following OPEs:
\begin{align}
&S^+_0(u,z,\bar{z})\Phi_r(\mathbf{x}_p)\sim\lim\limits_{\epsilon\rightarrow0^+}-i(\bar{z}-\bar{z}_p)\text{ }\frac{\Phi_{r-1}(\mathbf{x}_p)}{\Delta\tilde{z}_p} \, ,\nonumber\\
    &S_1^+(\mathbf{x})\Phi_r(\mathbf{x}_p)\sim\lim\limits_{\epsilon\rightarrow0^+}-i\left[\frac{(\bar{z}-\bar{z}_p)^2}{{(\Delta\tilde{z}_p)}}\text{ }\bar{\partial}_{p} \Phi_r-\frac{\bar{z}-\bar{z}_p}{{(\Delta\tilde{z}_p)}}\text{ }(2\bar{h}-r)\Phi_r+(u-u_p)\text{ }\frac{\bar{z}-\bar{z}_p}{{(\Delta\tilde{z}_p)}}\text{ }{\Phi_{r-1}}\right](\mathbf{x}_p) \, , \nonumber\\
&T(u,z,\bar{z})\Phi_r(\mathbf{x}_p)\sim\lim\limits_{\epsilon\rightarrow0^+}-i\left[\frac{\left(h-\frac{r}{2}\right)\Phi_r}{(\Delta\tilde{z}_p)^2}+\frac{\partial_{z_p}\Phi_r}{\Delta\tilde{z}_p}-\frac{u-u_p}{2}\frac{\Phi_{r-1}}{(\Delta\tilde{z}_p)^2}\right](\mathbf{x}_p) \, .
\end{align}
Similarly as the $S^+_2\Phi$ OPE \eqref{36}, the $S^+_{k\geq3}\Phi$ OPEs given by \eqref{38} are of the form relevant for the 3D Carrollian holographic description of the 4D bulk  physics of massless scattering.

\section{Verification from bulk amplitudes}
\label{sec:Applications to bulk amplitudes}

In this Appendix, we illustrate the results obtained in the core of the paper by performing various consistency checks on massless scattering amplitudes.

\subsection{Soft graviton theorems}\label{G.1}
\subsubsection*{The leading case}
We shall now use the correspondence \eqref{S0intermsofa} or equivalently, the correspondence \eqref{93} to verify the $\langle S^+_0(u,z,\bar{z})X\rangle$ Ward identity \eqref{3} at $u\rightarrow\infty$ directly from the correspondence between the 3D Carrollian CFT position space correlators and the 4D bulk null-momentum space $\mathcal{S}$-matrices \eqref{eq:AtoC}. Only in the Carrollian $u\rightarrow\infty$ limit which translates into the energetically soft $\omega\rightarrow0$ limit, the Carrollian CFT correlators on both sides of \eqref{3} can simultaneously correspond to bulk massless $\mathcal{S}$-matrices. For a finite $u$ (that implies finite $\omega$), however, energy conservation prevents both sides of \eqref{3} from simultaneously corresponding to an $\mathcal{S}$-matrix.

We consider the case where $X$ only contains three Carrollian CFT primaries corresponding to one positive helicity graviton and two negative helicity gravitons (in the `all outgoing' convention). Then the Carrollian CFT correlator in the LHS of \eqref{3} corresponds to the following bulk tree-level Einstein-Yang-Mills four-graviton (MHV) amplitude (with the $S^+_0$ field corresponding to the graviton $1^{+2}$) \cite{Banerjee:2019prz,Mason:2023mti}:
\begin{align}
\mathcal{A}_4(1^{+2},2^{-2},3^{-2},4^{+2})=4\epsilon_1\epsilon_2\epsilon_3\epsilon_4\frac{\omega_2\omega_3\omega_4}{\omega_1}\frac{z_{23}^4\bar{z}_{14}^4}{z_{12}\bar{z}_{12}z_{13}\bar{z}_{13}z_{14}\bar{z}_{14}}\delta^{(4)}(\sum_{i=1}^4\epsilon_i\omega_iq_i)\label{98}
\end{align}
with the four-particle delta function simplified as (with $z=\frac{z_{12}z_{34}}{z_{13}z_{24}}$) \cite{Pate:2019mfs}:
\begin{align}
\delta^{(4)}\left(\sum_{i=1}^4\epsilon_i\omega_iq_i\right)=\frac{1}{4\omega_4}\text{ }&\delta\left(\omega_1+\epsilon_1\epsilon_4\omega_4\frac{z_{34}\bar{z}_{24}}{z_{13}\bar{z}_{12}}\right)\text{ }\delta\left(\omega_2-\epsilon_2\epsilon_4\omega_4\frac{z_{34}\bar{z}_{14}}{z_{23}\bar{z}_{12}}\right)\nonumber\\
\times\text{ }&\delta\left(\omega_3+\epsilon_3\epsilon_4\omega_4\frac{z_{24}\bar{z}_{14}}{z_{23}\bar{z}_{13}}\right)\text{ }\delta\left(z_{12}z_{34}\bar{z}_{24}\bar{z}_{13}-z_{24}z_{13}\bar{z}_{12}\bar{z}_{34}\right) \, . \label{95}
\end{align}
Using the map between the 3D Carrollian CFT correlators and the 4D  massless amplitudes \eqref{eq:AtoC} and the correspondence \eqref{S0intermsofa}, the correlator $\langle {S}^+_0(u,z,\bar{z})X\rangle$ corresponding to the above amplitude is found to be (with $\mathcal{A}_4\left(1^{+2},2^{-2},3^{-2},4^{+2}\right)\equiv\mathcal{A}_4\left(\epsilon_1\omega_1,\left\{\epsilon_p\omega_p\right\}\right)$ with $p\in\{2,3,4\}$):
\begin{align}
&\langle {S}^+_0(u_1,z_1,\bar{z}_1)X\rangle\nonumber\\
=&\lim\limits_{\delta\rightarrow0^+}i\int\limits_{0}^\infty \frac{d\omega_1}{2\pi}\ldots\int\limits_{0}^\infty \frac{d\omega_4}{2\pi}\text{ }e^{-i\sum\limits_{r=2}^4\epsilon_i\omega_iu_i}\left[e^{-i\omega_1u_1}\mathcal{A}_4\left(\omega_1+i\delta,\left\{\epsilon_p\omega_p\right\}\right)\right.\nonumber\\
&\hspace{75mm}\left.+\text{ }e^{i\omega_1u_1}\mathcal{A}_4\left(-\omega_1+i\delta,\left\{\epsilon_p\omega_p\right\}\right)\right]\nonumber\\
=&\lim\limits_{\delta\rightarrow0^+}i\int\limits_{-\infty}^\infty \frac{d\Omega_1}{2\pi}\int\limits_{0}^\infty \frac{d\omega_2}{2\pi}\int\limits_{0}^\infty \frac{d\omega_3}{2\pi}\int\limits_{0}^\infty \frac{d\omega_4}{2\pi}\text{ }e^{-i\left[\Omega_1u_1+\sum\limits_{i=2}^4\epsilon_i\omega_iu_i\right]}\text{ }\mathcal{A}_4\left(\Omega_1+i\delta,\left\{\epsilon_p\omega_p\right\}\right) 
  \label{94}
\end{align}
which explicitly shows the equivalence between the correspondences \eqref{S0intermsofa} and \eqref{93}. The $\Omega_1$-delta-function then needs to be analytically continued; this is found to be:
\begin{align}
&\delta\left(\Omega_1+\epsilon_4\omega_4\frac{z_{34}\bar{z}_{24}}{z_{13}\bar{z}_{12}}\right)=\lim\limits_{\mu\rightarrow0^+}\frac{1}{2\pi i}\left[\frac{1}{\Omega_1+\epsilon_4\omega_4\frac{z_{34}\bar{z}_{24}}{z_{13}\bar{z}_{12}}-i\mu}-\frac{1}{\Omega_1+\epsilon_4\omega_4\frac{z_{34}\bar{z}_{24}}{z_{13}\bar{z}_{12}}+i\mu}\right]\nonumber\\
&\xrightarrow[]{\Omega_1\rightarrow\left(\Omega_1+i\delta\right)}\lim\limits_{\mu\rightarrow0^+}\frac{1}{2\pi i}\left[\frac{1}{\Omega_1+i\delta+\epsilon_4\omega_4\frac{z_{34}\bar{z}_{24}}{z_{13}\bar{z}_{12}}-i\mu}-\frac{1}{\Omega_1+i\delta+\epsilon_4\omega_4\frac{z_{34}\bar{z}_{24}}{z_{13}\bar{z}_{12}}+i\mu}\right]
\end{align}
Now, as said before, we shall be imposing the limit $u_1\rightarrow\infty$. Let us then look into the $\Omega_1$-integral first (with $\alpha\equiv \epsilon_4\omega_4\frac{z_{34}\bar{z}_{24}}{z_{13}\bar{z}_{12}}\in\mathbb{R}$):
\begin{align*}
&\lim\limits_{\mu\rightarrow0^+}\lim\limits_{u_1\rightarrow\infty}\lim\limits_{\delta\rightarrow0^+}i\int\limits_{-\infty}^\infty \frac{d\Omega_1}{2\pi}\text{ }\frac{e^{-i\Omega_1u_1}}{\Omega_1+i\delta}\text{ }\frac{1}{2\pi i}\left[\frac{1}{\Omega_1+\alpha-i(\mu-\delta)}-\frac{1}{\Omega_1+\alpha+i(\mu+\delta)}\right]\\
=&\lim\limits_{\mu\rightarrow0^+}\lim\limits_{u_1\rightarrow\infty}\lim\limits_{\delta\rightarrow0^+}\left[\frac{e^{-\delta u_1}}{2\pi i}\left\{\frac{1}{\alpha-i\mu}-\frac{1}{\alpha+i\mu}\right\}+\frac{e^{-(\delta+\mu) u_1}}{2\pi i}\frac{e^{i\alpha u_1}}{\alpha+i\mu}\right]\\
=&\lim\limits_{\mu\rightarrow0^+}\frac{1}{2\pi i}\left\{\frac{1}{\alpha-i\mu}-\frac{1}{\alpha+i\mu}\right\}+0=\delta\left(\alpha\right) \, .
\end{align*} 
Similarly, it can be verified that the $\Omega_1$-integral yields a vanishing result when $u_1\rightarrow-\infty$, in agreement with the retarded initial condition \eqref{eq:6}.

Next, for the simplifications of the various delta-functions involving $z_{ij}$ and $\bar{z}_{ij}$ appearing in \eqref{94}, we shall closely follow \cite{Pate:2019mfs}. Assuming $z_{ij}\neq0$ and $\bar{z}_{12}\neq0$, we first rewrite $\delta\left(\omega_4\frac{z_{34}\bar{z}_{24}}{z_{13}\bar{z}_{12}}\right)$ as:
\begin{align}
\delta\left(\omega_4\frac{z_{34}\bar{z}_{24}}{z_{13}\bar{z}_{12}}\right)=\text{sgn}\left(z_{13}\bar{z}_{12}z_{34}\right)\frac{z_{13}\bar{z}_{12}}{\omega_4 z_{34}}\delta\left(\bar{z}_{24}\right) \, . \label{96}
\end{align}
On the support of the delta function $\delta\left(\bar{z}_{24}\right)$, we then simplify the following factor appearing in \eqref{95} as:
\begin{align}
\delta\left(z_{12}z_{34}\bar{z}_{24}\bar{z}_{13}-z_{24}z_{13}\bar{z}_{12}\bar{z}_{34}\right)=\delta\left(z_{24}z_{13}\bar{z}_{12}\bar{z}_{34}\right)=\frac{\text{sgn}\left(z_{24}z_{13}\bar{z}_{12}\right)}{z_{24}z_{13}\bar{z}_{12}}\delta\left(\bar{z}_{34}\right)   \, . \label{97}
\end{align}
We now perform the three remaining integrals in \eqref{94} using the amplitude \eqref{98} and the four-particle delta function \eqref{95} and the simplifications \eqref{96}-\eqref{97} to find the value of the LHS of \eqref{3}$\big\vert_{u\rightarrow\infty}$ as:
\begin{align}
\langle S^+_0(\infty,z_1,\bar{z}_1)X\rangle&=\int\limits_{0}^\infty \frac{d\omega_2}{2\pi}\int\limits_{0}^\infty \frac{d\omega_3}{2\pi}\int\limits_{0}^\infty \frac{d\omega_4}{2\pi}\text{ }e^{-i\sum\limits_{i=2}^4\epsilon_i\omega_iu_i}\text{ }\epsilon_2\epsilon_3\epsilon_4\frac{\omega_2\omega_3}{\omega_4}\frac{z_{23}^4\bar{z}_{14}}{z_{12}z_{13}z_{14}}\frac{\text{sgn}\left(z_{24}z_{34}\right)}{z_{24}z_{34}}\nonumber\\
&\qquad\qquad\qquad\times\text{ }\delta\left(\bar{z}_{24}\right)\delta\left(\bar{z}_{34}\right)\delta\left(\omega_2-\epsilon_2\epsilon_4\omega_4\frac{z_{34}}{z_{23}}\right)
\delta\left(\omega_3-\epsilon_3\epsilon_4\omega_4\frac{z_{42}}{z_{23}}\right)\nonumber\\
&=\int\limits_{0}^\infty \frac{d\omega_4}{(2\pi)^3}\text{ }e^{-i\epsilon_4\omega_4\left[u_4+u_2\frac{z_{34}}{z_{23}}+u_3\frac{z_{42}}{z_{23}}\right]}\epsilon_4\omega_4\frac{z_{23}^2\bar{z}_{14}}{z_{12}z_{13}z_{14}}{\text{sgn}\left(z_{42}z_{34}\right)}\delta\left(\bar{z}_{24}\right)\delta\left(\bar{z}_{34}\right)\nonumber\\
&\qquad\qquad\qquad\times\text{ }\Theta\left(\epsilon_2\epsilon_4\frac{z_{34}}{z_{23}}\right)\Theta\left(\epsilon_3\epsilon_4\frac{z_{42}}{z_{23}}\right)\nonumber\\
&=-\frac{\epsilon_2\epsilon_3\epsilon_4}{(2\pi)^3}\text{ }\frac{\Theta\left(\epsilon_2\epsilon_4\frac{z_{34}}{z_{23}}\right)\Theta\left(\epsilon_3\epsilon_4\frac{z_{42}}{z_{23}}\right)}{{\left[u_4z_{23}+u_2{z_{34}}+u_3{z_{42}}\right]}^{2}}\text{ }\frac{z_{23}^4\bar{z}_{14}}{z_{12}z_{13}z_{14}}\text{ }\delta\left(\bar{z}_{24}\right)\delta\left(\bar{z}_{34}\right) \, . \label{99}
\end{align}

To evaluate the RHS of \eqref{3}, we first note down the following tree-level Einstein-Yang-Mills three-graviton (MHV) amplitude \cite{Puhm:2019zbl}:
\begin{align}
\mathcal{A}_3(2^{-2},3^{-2},4^{+2})=4\frac{\omega_2^2\omega_3^2}{\omega_4^2}\frac{z_{23}^6}{z_{34}^2z_{42}^2}\delta^{(4)}(\sum_{i=2}^4\epsilon_i\omega_iq_i)
\end{align}
with the three-particle delta function simplified below \cite{Pate:2019mfs}:
\begin{align}
\delta^{(4)}\left(\sum_{i=2}^4\epsilon_i\omega_iq_i\right)=\frac{\text{sgn}\left(z_{23}z_{24}\right)}{4\omega_2^2z_{24}z_{23}}\text{ }\delta\left(\omega_3-\epsilon_2\epsilon_3\omega_2\frac{z_{42}}{z_{34}}\right)\text{ }\delta\left(\omega_4-\epsilon_2\epsilon_4\omega_2\frac{z_{23}}{z_{34}}\right)\text{ }\delta\left(\bar{z}_{34}\right)\delta\left(\bar{z}_{24}\right) \, .
\end{align}
The corresponding Carrollian correlator $\langle X\rangle$ is then obtained to be:
\begin{align}
\langle X\rangle&=\int\limits_{0}^\infty \frac{d\omega_2}{2\pi}\int\limits_{0}^\infty \frac{d\omega_3}{2\pi}\int\limits_{0}^\infty \frac{d\omega_4}{2\pi}\text{ }e^{-i\sum\limits_{i=2}^4\epsilon_i\omega_iu_i}\text{ }\mathcal{A}_3(2^{-2},3^{-2},4^{+2})\nonumber\\
&=\int\limits_{0}^\infty \frac{d\omega_2}{(2\pi)^3}\text{ }e^{-i\epsilon_2\omega_2\left[u_2+u_3\frac{z_{42}}{z_{34}}+u_4\frac{z_{23}}{z_{34}}\right]}\text{ }\delta\left(\bar{z}_{24}\right)\delta\left(\bar{z}_{34}\right)\frac{\text{sgn}\left(z_{23}z_{24}\right)}{z_{24}z_{23}}\frac{z_{23}^4}{z_{34}^2}\text{ }\Theta\left(\epsilon_2\epsilon_4\frac{z_{23}}{z_{34}}\right)\Theta\left(\epsilon_2\epsilon_3\frac{z_{42}}{z_{34}}\right)\nonumber\\
&=\frac{i\epsilon_2\epsilon_3\epsilon_4}{(2\pi)^3}\text{ }\frac{\Theta\left(\epsilon_2\epsilon_4\frac{z_{34}}{z_{23}}\right)\Theta\left(\epsilon_3\epsilon_4\frac{z_{42}}{z_{23}}\right)}{{\left[u_4z_{23}+u_2{z_{34}}+u_3{z_{42}}\right]}}\text{ }\frac{z_{23}^3}{z_{24}z_{34}}\text{ }\delta\left(\bar{z}_{24}\right)\delta\left(\bar{z}_{34}\right) \, .
\end{align}
Using this expression for $\langle X\rangle$, we now proceed to calculate the RHS of the Ward identity \eqref{3}$_{u_1\rightarrow\infty}$ (with $\bm{\xi}_p=\bar{{\bm{\xi}}}_p=0$) to find:
\begin{align*}
-i\sum\limits_{p=2}^4\frac{\bar{z}_1-\bar{z}_p}{z_1-z_p}\partial_{u_p}\langle X\rangle&=-\frac{\epsilon_2\epsilon_3\epsilon_4}{(2\pi)^3}\text{ }\frac{\Theta\left(\epsilon_2\epsilon_4\frac{z_{34}}{z_{23}}\right)\Theta\left(\epsilon_3\epsilon_4\frac{z_{42}}{z_{23}}\right)}{{\left[u_4z_{23}+u_2{z_{34}}+u_3{z_{42}}\right]}^2}\text{ }\frac{z_{23}^3\bar{z}_{14}}{z_{24}z_{34}}\left[\frac{z_{34}}{z_{12}}+\frac{z_{42}}{z_{13}}+\frac{z_{23}}{z_{14}}\right]\delta\left(\bar{z}_{24}\right)\delta\left(\bar{z}_{34}\right)\\
&=-\frac{\epsilon_2\epsilon_3\epsilon_4}{(2\pi)^3}\text{ }\frac{\Theta\left(\epsilon_2\epsilon_4\frac{z_{34}}{z_{23}}\right)\Theta\left(\epsilon_3\epsilon_4\frac{z_{42}}{z_{23}}\right)}{{\left[u_4z_{23}+u_2{z_{34}}+u_3{z_{42}}\right]}^{2}}\text{ }\frac{z_{23}^4\bar{z}_{14}}{z_{12}z_{13}z_{14}}\text{ }\delta\left(\bar{z}_{24}\right)\delta\left(\bar{z}_{34}\right)
\end{align*} 
which matches exactly with the LHS of \eqref{3}$_{u_1\rightarrow\infty}$ given by \eqref{99}.

We have similarly verified the $\langle S^-_0(u_1\rightarrow\infty,z_1,\bar{z}_1)X\rangle$ Ward identity for the case where $X$ only contains three Carrollian CFT primaries corresponding to two positive helicity gluons and one negative helicity gluon (and the $S^-_0$ field of course corresponds to the graviton $1^{-2}$) in the `all outgoing' convention. The four-point tree-level Einstein-Yang-Mills amplitude relevant for the LHS is \cite{Stieberger:2024shv}:
\begin{align}
\mathcal{A}_4(1^{-2},2^{-1},3^{+1},4^{+1})=2\epsilon_1\epsilon_2\epsilon_3\epsilon_4\frac{\omega_4^2}{\omega_1}\frac{z_{12}z_{14}}{z_{13}}\frac{\bar{z}_{34}^3\bar{z}_{42}}{\bar{z}_{12}\bar{z}_{13}\bar{z}_{14}\bar{z}_{23}}\delta^{(4)}(\sum_{i=1}^4\epsilon_i\omega_iq_i)
\end{align}
while the tree-level Einstein-Yang-Mills three-gluon (anti-MHV) amplitude corresponding to the Carrollian CFT correlator $\langle X\rangle$ is \cite{Pasterski:2017ylz}:
\begin{align}
\mathcal{A}_3(2^{-1},3^{+1},4^{+1})=2\frac{\omega_3\omega_4}{\omega_2}\frac{\bar{z}_{34}^3}{\bar{z}_{23}\bar{z}_{42}}\delta^{(4)}(\sum_{i=2}^4\epsilon_i\omega_iq_i) \, .
\end{align}
Both sides of the $\langle S^-_0(u_1\rightarrow\infty,z_1,\bar{z}_1)X\rangle$ Ward identity are then found to be equal to:
\begin{align*}
\langle S^-_0(u_1\rightarrow\infty,z_1,\bar{z}_1)X\rangle&=-i\sum\limits_{p=2}^4\frac{z_1-z_p}{\bar{z}_1-\bar{z}_p}\partial_{u_p}\langle X\rangle\\
&=-\frac{i}{2(2\pi)^3}\text{ }\frac{\Theta\left(\epsilon_2\epsilon_4\frac{\bar{z}_{34}}{\bar{z}_{23}}\right)\Theta\left(\epsilon_3\epsilon_4\frac{\bar{z}_{42}}{\bar{z}_{23}}\right)}{{\left[u_4\bar{z}_{23}+u_2{\bar{z}_{34}}+u_3{\bar{z}_{42}}\right]}}\text{ }\frac{z_{12}\bar{z}_{34}^2}{\bar{z}_{12}\bar{z}_{13}\bar{z}_{14}}\text{ }\delta\left({z}_{24}\right)\delta\left({z}_{34}\right) \, .
\end{align*} 
Intriguingly, while the correlator $\langle X\rangle$ corresponding to the tree-level three-gluon Einstein-Yang-Mills amplitude is divergent, each one of the correlators $\partial_{u_p}\langle X\rangle$, and hence the RHS above, are convergent \cite{Mason:2023mti}.

Having checked the validity of the Ward identity \eqref{3}$_{u\rightarrow\infty}$ with the above explicit Einstein-Yang-Mills examples, we now provide a proof of the same using a general 4D  massless tree-level amplitude irrespective of the bulk theory. The main ingredient is the tree-level universality of the leading \cite{Weinberg:1965nx} and the subleading soft graviton theorems \cite{Cachazo:2014fwa} which (for positive helicity in the `outgoing' convention) state that:
\begin{align}
&\lim\limits_{\Omega\rightarrow0}\lim\limits_{\delta\rightarrow0^+}\mathcal{A}_{n+1}\left(h_{+2}(\Omega+i\delta,z,\bar{z}),\left\{\tilde{\Phi}_{s_p}\left(\epsilon_p\omega_p,z_p,\bar{z}_p\right)\right\}_{1\leq p\leq n}\right)=\lim\limits_{\Omega\rightarrow0}\lim\limits_{\delta\rightarrow0^+}-\left[\sum\limits_{q=1}^n\frac{\bar{z}-\bar{z}_q}{z-z_q}\frac{\epsilon_q\omega_q}{\Omega+i\delta}\right.\nonumber\\
&\left.+\sum\limits_{q=1}^n\frac{{\left(\bar{z}-\bar{z}_q\right)}^2\bar{\partial}_q+\left(\bar{z}-\bar{z}_q\right)\left(s_q+\omega_q\partial_{\omega_q}\right)}{z-z_q}+\mathcal{O}\left(\Omega+i\delta\right)\right]\mathcal{A}_{n}\left(\left\{\tilde{\Phi}_{s_p}\left(\epsilon_p\omega_p,z_p,\bar{z}_p\right)\right\}_{1\leq p\leq n}\right) \, . \label{101}
\end{align} 
By virtue of the correspondence \eqref{93}, and the map between the 3D Carrollian CFT correlators and the 4D massless amplitudes \eqref{eq:AtoC}, the LHS of \eqref{3} reads:
\begin{align}
\langle S^+_0(u,z,\bar{z})X\rangle=\lim\limits_{\delta\rightarrow0^+} i&\int\limits_{-\infty}^\infty \frac{d\Omega}{2\pi}\int\limits_{0}^\infty \frac{d\omega_1}{2\pi}\ldots\int\limits_{0}^\infty \frac{d\omega_n}{2\pi}\left[e^{-i\Omega u-i\sum\limits_{r=1}^n\epsilon_r\omega_ru_r}\right.\nonumber\\
&\left.\times\text{ }\mathcal{A}_{n+1}\left(h_{+2}(\Omega+i\delta,z,\bar{z}),\left\{\tilde{\Phi}_{s_p}\left(\epsilon_p\omega_p,z_p,\bar{z}_p\right)\right\}_{1\leq p\leq n}\right)\right] \, . \label{100}
\end{align}
Since the massless amplitude $\mathcal{A}_{n+1}$ is expected to die off at $\left|\Omega\right|\rightarrow\infty$, the $\Omega$-integrals in the two terms at the RHS for the case when $u\rightarrow\infty$ can be combined into a sum of counter-clockwise complex contour-integrals (over $\zeta$):
\begin{align*}
&\lim\limits_{u\rightarrow\infty}\lim\limits_{\delta\rightarrow0^+}i\int\limits_{-\infty}^\infty \frac{d\Omega}{2\pi}\text{ }e^{-i\Omega u}\mathcal{A}_{n+1}\left(\Omega+i\delta\right)\\
=&-\lim\limits_{u\rightarrow\infty}\lim\limits_{\delta\rightarrow0^+}i\oint\limits_{-i\delta} \frac{d\zeta}{2\pi}\text{ }e^{-i\zeta u}\mathcal{A}_{n+1}\left(\zeta+i\delta\right)-\sum_{\Omega^*\neq0}\text{ }\lim\limits_{u\rightarrow\infty}\lim\limits_{\delta\rightarrow0^+}i\oint\limits_{\Omega^*-i\delta} \frac{d\zeta}{2\pi}\text{ }e^{-i\zeta u}\mathcal{A}_{n+1}\left(\zeta+i\delta\right) 
\end{align*} 
with $\Omega^*=\Omega^*\left(\{\epsilon_p\omega_p\},\left\{\left|z_{pq}\right|^2\right\},\left\{\left|z-z_{p}\right|^2\right\}\right)\in\mathbb{R}$. The singularity of $\mathcal{A}_{n+1}\left(\zeta+i\delta\right)$ at $\zeta=-i\delta$ is the leading soft graviton (simple) pole while the $\zeta=\Omega^*-i\delta$ (with $\Omega^*\neq0$) poles may correspond to the multi-collinear singularities. Now, we have:
\begin{align*}
\lim\limits_{u\rightarrow\infty}\lim\limits_{\delta\rightarrow0^+}\oint\limits_{\Omega^*-i\delta} \frac{d\zeta}{2\pi}\text{ }e^{-i\zeta u}\mathcal{A}_{n+1}\left(\zeta+i\delta\right)\propto e^{-i\Omega^* u} 
\end{align*} 
which at large $u$ oscillates violently for $\Omega^*\neq0$; upon performing the remaining $\left\{\omega_p\right\}$-integrals in \eqref{100}, each of the $\Omega^*\neq0$ terms are expected to produce vanishing result by the Riemann-Lebesgue lemma. Hence, in the $u\rightarrow\infty$ limit, we can safely ignore these $\Omega^*\neq0$ terms to write just:  \begin{align}
\lim\limits_{u\rightarrow\infty}\lim\limits_{\delta\rightarrow0^+}\int\limits_{-\infty}^\infty \frac{d\Omega}{2\pi}\text{ }e^{-i\Omega u}\mathcal{A}_{n+1}\left(\Omega+i\delta\right)=\lim\limits_{u\rightarrow\infty}\lim\limits_{\delta\rightarrow0^+}-\oint\limits_{-i\delta} \frac{d\zeta}{2\pi}\text{ }e^{-i\zeta u}\mathcal{A}_{n+1}\left(\zeta+i\delta\right) \, . \label{103}
\end{align}
Using this and the knowledge of the singular behavior of $\mathcal{A}_{n+1}\left(\zeta+i0^+\right)$ near $\zeta=-i0^+$ given by \eqref{101}, we evaluate the $\Omega$-integral in \eqref{100}$_{u\rightarrow\infty}$ to obtain: 
\begin{align}
\langle S^+_0(u\rightarrow\infty,z,\bar{z})X\rangle&=\lim\limits_{u\rightarrow\infty}\lim\limits_{\delta\rightarrow0^+}\int\limits_{0}^\infty \frac{d\omega_1}{2\pi}\ldots\int\limits_{0}^\infty \frac{d\omega_n}{2\pi}\text{ }e^{-i\sum\limits_{r=1}^n\epsilon_r\omega_ru_r}\oint\limits_{-i\delta}\frac{i d\zeta}{2\pi}\text{ }e^{-i\zeta u}\nonumber\\
&\qquad\qquad\qquad\times\text{ }\sum\limits_{q=1}^n\frac{\bar{z}-\bar{z}_q}{z-z_q}\frac{\epsilon_q\omega_q}{\zeta+i\delta}\mathcal{A}_{n}\left(\left\{\tilde{\Phi}_{s_p}\left(\epsilon_p\omega_p,z_p,\bar{z}_p\right)\right\}_{1\leq p\leq n}\right)\nonumber\\
&=-\int\limits_{0}^\infty \frac{d\omega_1}{2\pi}\ldots\int\limits_{0}^\infty \frac{d\omega_n}{2\pi}\text{ }e^{-i\sum\limits_{r=1}^n\epsilon_r\omega_ru_r}\sum\limits_{q=1}^n\frac{\bar{z}-\bar{z}_q}{z-z_q}\text{ }{\epsilon_q\omega_q}\text{ }\mathcal{A}_{n}\left(\left\{\tilde{\Phi}_p\right\}\right)\nonumber\\
&=-i\sum\limits_{q=1}^n\frac{\bar{z}-\bar{z}_q}{z-z_q}\text{ }\partial_{u_q}\left[\int\limits_{0}^\infty \frac{d\omega_1}{2\pi}\ldots\int\limits_{0}^\infty \frac{d\omega_n}{2\pi}\text{ }e^{-i\sum\limits_{r=1}^n\epsilon_r\omega_ru_r}\text{ }\mathcal{A}_{n}\left(\left\{\tilde{\Phi}_p\right\}\right)\right]\nonumber\\
&=-i\sum\limits_{q=1}^n\frac{\bar{z}-\bar{z}_q}{z-z_q}\text{ }\partial_{u_q}\langle X\rangle
\end{align}
which is exactly the Ward identity \eqref{3} at $u\rightarrow\infty$ (with $\bm{\xi}_p=\bar{{\bm{\xi}}}_p=0$). Furthermore, the $\Omega$-integral in \eqref{100}$_{u\rightarrow-\infty}$ automatically vanishes, in consistency with the retarded initial condition.

\subsubsection*{The subleading case}

Following the above discussion, we shall now proceed to derive the $\langle S^+_1(u,z,\bar{z})X\rangle$ Ward identity \eqref{10} in the $u\rightarrow\infty$ limit from general considerations of the bulk physics. Using \eqref{112}$_{k=1}$, the LHS of \eqref{10} is given by:
\begin{align}
\langle S^+_1(u,z,\bar{z})X\rangle=-&\lim\limits_{\delta\rightarrow0^+}\int\limits_{-\infty}^\infty \frac{d\Omega}{2\pi}\int\limits_{0}^\infty \frac{d\omega_1}{2\pi}\ldots\int\limits_{0}^\infty \frac{d\omega_n}{2\pi}\left[e^{-i\sum\limits_{r=1}^n\epsilon_r\omega_ru_r}\text{ }\frac{e^{-i\Omega u}}{\Omega+i\delta}\right.\nonumber\\
&\left.\times\text{ }\mathcal{A}_{n+1}\left(h_{+2}(\Omega+i\delta,z,\bar{z}),\left\{\tilde{\Phi}_{s_p}\left(\epsilon_p\omega_p,z_p,\bar{z}_p\right)\right\}_{1\leq p\leq n}\right)\right]
 \, . \label{104}
\end{align}
Similarly as \eqref{103}, one also finds that:
\begin{align*}
\lim\limits_{u\rightarrow\infty}\lim\limits_{\delta\rightarrow0^+}\int\limits_{-\infty}^\infty \frac{d\Omega}{2\pi}\text{ }\frac{e^{-i\Omega u}}{\Omega+i\delta}\mathcal{A}_{n+1}\left(\Omega+i\delta\right)=\lim\limits_{u\rightarrow\infty}\lim\limits_{\delta\rightarrow0^+}-\oint\limits_{-i\delta} \frac{d\zeta}{2\pi}\text{ }\frac{e^{-i\zeta u}}{\zeta+i\delta}\mathcal{A}_{n+1}\left(\zeta+i\delta\right)
\end{align*}
which, together with the Laurent expansion \eqref{101} of $\mathcal{A}_{n+1}\left(\zeta+i\delta\right)$ around $\zeta=-i\delta$, leads us from \eqref{104} to:
\begin{align}
&\langle S^+_1(u\rightarrow\infty,z,\bar{z})X\rangle=-\lim\limits_{u\rightarrow\infty}\lim\limits_{\delta\rightarrow0^+}\int\limits_{0}^\infty \frac{d\omega_1}{2\pi}\ldots\int\limits_{0}^\infty \frac{d\omega_n}{2\pi}\text{ }e^{-i\sum\limits_{r=1}^n\epsilon_r\omega_ru_r}\oint\limits_{-i\delta}\frac{d\zeta}{2\pi}\text{ }e^{-i\zeta u}\nonumber\\
&\qquad\times\left[\sum\limits_{q=1}^n\left\{\frac{\bar{z}-\bar{z}_q}{z-z_q}\frac{\epsilon_q\omega_q}{(\zeta+i\delta)^2}+\frac{{\left(\bar{z}-\bar{z}_q\right)}^2\bar{\partial}_q+\left(\bar{z}-\bar{z}_q\right)\left(s_q+\omega_q\partial_{\omega_q}\right)}{\left(z-z_q\right)(\zeta+i\delta)}\right\}\right]\mathcal{A}_{n}\left(\left\{\tilde{\Phi}_p\right\}_{1\leq p\leq n}\right)\nonumber\\
=&\lim\limits_{u\rightarrow\infty}-i\int\limits_{0}^\infty \frac{d\omega_1}{2\pi}\ldots\int\limits_{0}^\infty \frac{d\omega_n}{2\pi}\text{ }e^{-i\sum\limits_{r=1}^n\epsilon_r\omega_ru_r}\sum\limits_{q=1}^n\frac{{\left(\bar{z}-\bar{z}_q\right)}^2\bar{\partial}_q+\left(\bar{z}-\bar{z}_q\right)\left(s_q-iu\epsilon_q\omega_q+\omega_q\partial_{\omega_q}\right)}{z-z_q}\text{ }\mathcal{A}_{n}\nonumber\\
=&\lim\limits_{u\rightarrow\infty}-i\sum\limits_{q=1}^n\frac{{\left(\bar{z}-\bar{z}_q\right)}^2\bar{\partial}_q+\left(\bar{z}-\bar{z}_q\right)\left\{s_q-1+\left(u-u_q\right)\partial_{u_q}\right\}}{z-z_q}\int\limits_{0}^\infty \frac{d\omega_1}{2\pi}\ldots\int\limits_{0}^\infty \frac{d\omega_n}{2\pi}\text{ }e^{-i\sum\limits_{r=1}^n\epsilon_r\omega_ru_r}\text{ }\mathcal{A}_{n}\nonumber\\
=&\lim\limits_{u\rightarrow\infty}-i\sum\limits_{q=1}^n\left[\frac{{\left(\bar{z}-\bar{z}_q\right)}^2}{z-z_q}\bar{\partial}_q-\frac{\bar{z}-\bar{z}_q}{z-z_q}\left(1-s_q\right)+\left(u-u_q\right)\frac{\bar{z}-\bar{z}_q}{z-z_q}\partial_{u_q}\right]\langle X\rangle\label{105}
\end{align}
that\footnote{\label{extendedFourier}To go from the second to the third line, we have ignored a finite term, the one in the last line below:
\begin{align*}
\int\limits_{0}^\infty \frac{d\omega_1}{2\pi}\ldots\int\limits_{0}^\infty \frac{d\omega_n}{2\pi}\text{ }e^{-i\sum\limits_{r=1}^n\epsilon_r\omega_ru_r}\left(1+\omega_q\partial_{\omega_q}\right)\mathcal{A}_{n}=&-u_q\partial_{u_q}\langle X\rangle\\
&+\left(\prod_{r\neq q}\int\limits_{0}^\infty \frac{d\omega_r}{2\pi}\text{ }e^{-i\epsilon_r\omega_ru_r}\right)\left[e^{-i\epsilon_q\omega_qu_q}{\omega_q}\mathcal{A}_{n}\left(\left\{\omega_i\right\}\right)\right]^{\omega_q=\infty}_{\omega_q=0}
\end{align*}
While the $\omega_q\rightarrow\infty$ contribution can be thrown away by regulating $u_q$ as $u_q\rightarrow u_q-i\epsilon_q\delta$ with $\delta\rightarrow0^+$ \cite{Banerjee:2019prz}, the term $\lim\limits_{\omega_q\rightarrow0}\omega_q\mathcal{A}_n\sim\mathcal{A}_{n-1}$ is finite when the $q$-th particle is a graviton or gauge boson. Omitting this term appears unjustified since, after all, our focus is on the 4D  soft theorems; moreover, we have kept all the other $\mathcal{O}\left(u^0\right)$ terms in \eqref{105}! In contrast, no such `boundary' contribution arises if the range of $\epsilon_q\omega_q\mapsto\Omega_q$ integral is expanded from one half to the whole of $\mathbb{R}-\{0\}$ and the following regulator is introduced: $u_q\rightarrow u_q-i\delta\text{sgn}\left(\Omega_q\right)$. (Note that we have already done this for the $S^+_0$ and $S^+_1$ fields via the correspondence \eqref{93}. In other words, the following quantity does not suffer from the above problem:
\begin{align*}
\langle \mathscr{X}\rangle\left(\left\{u_p,z_p,\bar{z}_p\right\}\right)\equiv\int\limits_{-\infty}^\infty \frac{d\Omega_1}{2\pi}\ldots\int\limits_{-\infty}^\infty \frac{d\Omega_n}{2\pi}\text{ }e^{-i\sum\limits_{r=1}^n\Omega_ru_r}\mathcal{A}_{n}\left(\left\{\tilde{\Phi}_{s_p}\left(\Omega_p,z_p,\bar{z}_p\right)\right\}\right) \, .
\end{align*}
Alternatively, the said ambiguous term is expected to be originating from the non-Fourier soft mode appearing in the asymptotic mode-expansion of the graviton or gauge boson \cite{Kraus:2024gso,Jorstad:2024yzm}.} is recognized to be the RHS of \eqref{10}$_{u\rightarrow\infty}$ with $\bm{\xi}_p=\bar{{\bm{\xi}}}_p=0$ and $2\bar{h}_p=1-s_p$. 

\medskip

Using the negative helicity counterpart of \eqref{101}, one can similarly verify the $\langle S^-_1(u\rightarrow\infty,z,\bar{z})X\rangle$ Ward identity with $2{h}_p=1+s_p$. Thus, we directly recover the fact that the 3D Carrollian CFT primary corresponding to a 4D bulk massless scattering field must possess Carrollian scaling dimension $\Delta_p=1$ and transform as a 3D Carrollian spin-boost singlet (i.e. with $\bm{\xi}_p=\bar{{\bm{\xi}}}_p=0$) with the Carrollian spin identified with the bulk `outgoing' helicity $s_p$. 

\subsubsection*{Infinite tower of soft theorems}

We recall from section \ref{sec:Infinite tower of soft graviton theorems} that the Ward identities $\langle S^+_k(u,z,\bar{z})X\rangle$ \eqref{39} for $k\geq3$ are partially fixed by the $\text{Witt}\loplus Lw_{1+\infty}$ symmetry; the RHS of \eqref{39} is valid upto a polynomial of order $(k-3)$ in $z$. The part that is determined by symmetry can be obtained in the $u\rightarrow\infty$ limit from the all order soft graviton expansion of the `projected' tree-level amplitudes \cite{Hamada:2018vrw,Li:2018gnc} by following the above described procedure. This is enabled by \eqref{112} and the following generalization of \eqref{103}:
\begin{align*}
\lim\limits_{u\rightarrow\infty}\lim\limits_{\delta\rightarrow0^+}\int\limits_{-\infty}^\infty \frac{d\Omega}{2\pi}\text{ }\frac{e^{-i\Omega u}}{\left(\Omega+i\delta\right)^k}\mathcal{A}_{n+1}\left(\Omega+i\delta\right)=\lim\limits_{u\rightarrow\infty}\lim\limits_{\delta\rightarrow0^+}-\oint\limits_{-i\delta} \frac{d\zeta}{2\pi}\text{ }\frac{e^{-i\zeta u}}{(\zeta+i\delta)^k}\mathcal{A}_{n+1}\left(\zeta+i\delta\right) \, .
\end{align*}

\subsection{OPEs at finite \texorpdfstring{$u$}{u}}\label{finite u check}

To demonstrate that the correspondence \eqref{62} (or its extension \eqref{112}) for the ${S}^+_k(u,z,\bar{z})$ field is also compatible with the conformal Carrollian physics at finite $u$, we now verify, starting from the 4D bulk theory, the leading holomorphic singularity of the $S^+_kS^+_l$ OPE \eqref{31} that was derived in \cite{Saha:2023abr} using intrinsic Carrollian symmetry arguments. Following \cite{Mason:2023mti}, our strategy will be to apply the ($k(\in\mathbb{Z}_{\geq0})$-generalization \eqref{112} of the bilateral extension of the) Fourier transformation map \eqref{eq:AtoC} leading from the 4D  massless amplitudes to the 3D Carrollian CFT primary correlators \cite{Donnay:2022aba,Donnay:2022wvx}, on both sides of the 4D Einstein-Yang-Mills leading-order universal holomorphic collinear splitting \eqref{113} with $\Phi_{s_p}$ taken to be the positive helicity graviton $h_{+2}$ in the `all outgoing' convention:
\begin{align}
&\lim\limits_{\delta\rightarrow0^+} a_{+}(\Omega+i\delta,z,\bar{z})a_+(\Omega_p,z_p,\bar{z}_p)\label{116}\\
&\qquad\sim\lim\limits_{\delta\rightarrow0^+}-\frac{\bar{z}-\bar{z}_p}{{z-z_p}} \text{ }\frac{\Omega_p-i\delta}{\Omega+i\delta}\left(\frac{\Omega_p+\Omega}{\Omega_p-i\delta}\right)^2a_+\left(\Omega_p+\Omega,z_p,\frac{\left(\Omega_p-i\delta\right)\bar{z}_p+\left(\Omega+i\delta\right)\bar{z}}{\Omega_p+\Omega}\right)   \, .\nonumber 
\end{align}
One can safely use the said map on both sides of the collinear splitting \eqref{113} since, upon inserting the remaining operators, both sides can be separately interpreted as bulk $\mathcal{S}$-matrices with conserved four-momentum.
It is important to note that, as a function of (complexified) $\Omega$, the RHS of \eqref{116} has singularity only at $\Omega=-i\delta$ (but not at $\Omega=-\Omega_p$ or anywhere else). Proceeding with the above strategy, one finds at the RHS (with $u\neq u_p$):
\begin{align}
&\lim\limits_{\delta\rightarrow0^+}i^{k+l}\int\limits_{-\infty}^\infty \frac{d\Omega}{2\pi}\frac{e^{-i\Omega u}}{\left(\Omega+i\delta\right)^k}\int\limits_{-\infty}^{\infty}\frac{d\Omega_p}{2\pi}\frac{e^{-i\Omega_pu_p}}{\Omega_p^l}\frac{\bar{z}-\bar{z}_p}{{z-z_p}} \text{ }\frac{\Omega_p-i\delta}{\Omega+i\delta}\left(\frac{\Omega_p+\Omega}{\Omega_p-i\delta}\right)^2\nonumber\\
&\hspace{65mm}\times\text{ } a_+\left(\Omega_p+\Omega,z_p,\frac{(\Omega_p-i\delta)\bar{z}_p+(\Omega+i\delta)\bar{z}}{\Omega_p+\Omega}\right)\nonumber\\
=&\lim\limits_{\delta\rightarrow0^+} i^{k}\partial_{u_p}^{-l}\int\limits_{-\infty}^\infty \frac{d\Omega}{2\pi}\frac{e^{-i\Omega u}}{(\Omega+i\delta)^k}\int\limits_{-\infty}^{\infty}\frac{d\Omega_p}{2\pi}{e^{-i\Omega_pu_p}}\frac{\bar{z}-\bar{z}_p}{{z-z_p}}\frac{\Omega_p-i\delta}{\Omega+i\delta}\left(\frac{\Omega_p+\Omega}{\Omega_p-i\delta}\right)^2\sum_{m=0}^\infty \left(\frac{\Omega+i\delta}{\Omega_p+\Omega}\right)^{m}\nonumber\\
&\hspace{65mm}\times\frac{\left(\bar{z}-\bar{z}_p\right)^m}{m!}\text{ }\bar{\partial}_p^m a_+\left(\Omega_p+\Omega,z_p,\bar{z}_p\right)\nonumber\\
=&\lim\limits_{\delta\rightarrow0^+} i^{k}\partial_{u_p}^{-l}\int\limits_{-\infty}^\infty \frac{d\Omega}{2\pi}\frac{e^{-i\Omega\left(u-u_p\right)}}{(\Omega+i\delta)^k}\int\limits_{-\infty}^{\infty}\frac{d\Omega_p}{2\pi}{e^{-i\Omega_pu_p}}\frac{\bar{z}-\bar{z}_p}{{z-z_p}}\frac{\Omega_p^2}{(\Omega+i\delta)\left(\Omega_p-\Omega-i\delta\right)}\nonumber\\
&\hspace{50mm}\times\sum_{m=0}^\infty\left(\frac{\Omega+i\delta}{\Omega_p}\right)^{m}\frac{\left(\bar{z}-\bar{z}_p\right)^m}{m!}\text{ }\bar{\partial}_p^m a_+\left(\Omega_p,z_p,\bar{z}_p\right) \, . \label{115}
\end{align}
The result of the $\Omega$-integral is given below ($u\neq u_p$):
\[\lim\limits_{\delta\rightarrow0^+}\int\limits_{-\infty}^\infty \frac{d\Omega}{2\pi}\frac{e^{-i\Omega\left(u-u_p\right)}}{(\Omega+i\delta)^{k-m+1}\left(\Omega_p-\Omega-i\delta\right)}=
\begin{cases}
\theta\left(u-u_p\right)\sum\limits_{r=m}^{k}\frac{\left(u-u_p\right)^{k-r}}{i^{k-r+1}(k-r)!\text{ }\Omega_p^{r-m+1}} & , k\geq m\\
0 & , k<m
\end{cases}
\]
where we have used the fact that the integrand in \eqref{115} as a function of $\Omega$ is not singular\footnote{The reason is that the apparent singularity at $\Omega=\Omega_p-i\delta$ is absent in the original integrand \eqref{116} and is artificially generated by shifting the other integration-variable $\Omega_p$ in \eqref{115}.} at $\Omega=\Omega_p-i\delta$. Using this identity, the RHS is obtained as:
\begin{align}
&\lim\limits_{\delta\rightarrow0^+} i^{k}\partial_{u_p}^{-l}\int\limits_{-\infty}^\infty \frac{d\Omega}{2\pi}\frac{e^{-i\Omega\left(u-u_p\right)}}{(\Omega+i\delta)^k}\int\limits_{-\infty}^{\infty}\frac{d\Omega_p}{2\pi}{e^{-i\Omega_pu_p}}\frac{\bar{z}-\bar{z}_p}{{z-z_p}}\frac{\Omega_p^2}{(\Omega+i\delta)\left(\Omega_p-\Omega-i\delta\right)}\nonumber\\
&\hspace{50mm}\times\sum_{m=0}^\infty\left(\frac{\Omega+i\delta}{\Omega_p}\right)^{m}\frac{\left(\bar{z}-\bar{z}_p\right)^m}{m!}\bar{\partial}_p^m\text{ } a_+\left(\Omega_p,z_p,\bar{z}_p\right)\nonumber\\
=&\text{ }\partial_{u_p}^{-l}\left[\theta\left(u-u_p\right)\int\limits_{-\infty}^{\infty}\frac{d\Omega_p}{2\pi}{e^{-i\Omega_pu_p}}\sum_{m=0}^k\sum_{r=m}^k\frac{\left(\bar{z}-\bar{z}_p\right)^{m+1}}{\left(z-z_p\right)\cdot m!}\text{ }\frac{\left(u-u_p\right)^{k-r}}{\left(-i\Omega_p\right)^{r-1}(k-r)!}\text{ }\bar{\partial}_p^m a_+\left(\Omega_p,z_p,\bar{z}_p\right)\right]\nonumber\\
=&-i\partial_{u_p}^{-l}\left[\theta\left(u-u_p\right)\sum_{r=0}^k\sum_{m=0}^r\frac{\left(\bar{z}-\bar{z}_p\right)^{m+1}}{\left(z-z_p\right)\cdot m!}\text{ }\frac{\left(u-u_p\right)^{k-r}}{(k-r)!}\text{ }\bar{\partial}_p^m S^+_{r-1}\left(u_p,z_p,\bar{z}_p\right)\right]\nonumber\\
=&-i\theta\left(u-u_p\right)\sum_{r=0}^k\sum_{m=0}^r\frac{\left(\bar{z}-\bar{z}_p\right)^{m+1}}{z-z_p}\text{ }\frac{\left(u-u_p\right)^{k-r}}{(k-r)!}\frac{(r+1-m)_l}{l!\cdot m!}\text{ }\bar{\partial}_p^m S^+_{r+l-1}\left(u_p,z_p,\bar{z}_p\right)\label{117}
\end{align}
where we have used the following action of the inverse derivative operator $\partial_{u_p}^{-1}$:
\begin{align*}
\partial_{u_p}^{-1}\left[\theta\left(u-u_p\right)\frac{\left(u-u_p\right)^{k-r}}{(k-r)!}\text{ }S^+_{r-1}\left(u_p\right)\right]=\theta\left(u-u_p\right)\sum_{n=0}^{k-r}\frac{\left(u-u_p\right)^{n}}{n!}\text{ }S^+_{k-n}\left(u_p\right)
\end{align*}
and repeated it $l$-times. To go from the second to the third line above, we have used the correspondence \eqref{128}. Thus, the said correspondence leads from the 4D collinear splitting \eqref{116} to the following 3D Carrollian CFT position-space OPE:
\begin{align*}
S^+_k(u,z,\bar{z})S^+_l\left(u_p,z_p,\bar{z}_p\right)\sim-i\theta\left(u-u_p\right)\sum_{r=0}^k\sum_{m=0}^r\frac{\left(\bar{z}-\bar{z}_p\right)^{m+1}}{z-z_p}&\text{ }\frac{\left(u-u_p\right)^{k-r}}{(k-r)!}\text{ }\frac{(r+1-m)_l}{l!\cdot m!}\\
&\times\bar{\partial}_p^m S^+_{r+l-1}\left(u_p,z_p,\bar{z}_p\right)
\end{align*}
which is exactly the symmetry-derived OPE \eqref{31} under the retarded boundary condition. We have verified the above $S^+_0S^+_0$ OPE explicitly for the case when the LHS is extracted from the Carrollian correlator corresponding to the six-point MHV graviton amplitude in the 4D Einstein gravity.  

\medskip

Alternatively, one can use the series expansion \eqref{114}$_{2\bar{h}_p=-1}$ on the second line in \eqref{115}   and then the following identity (for $u\neq u_p$):
\[
\lim\limits_{\delta\rightarrow0^+}\int\limits_{-\infty}^\infty \frac{d\Omega}{2\pi}\frac{e^{-i\Omega\left(u-u_p\right)}}{(\Omega+i\delta)^{k-r+1}}=
\begin{cases}
\theta\left(u-u_p\right)\frac{\left(u-u_p\right)^{k-r}}{i^{k-r+1}(k-r)!} & , k\geq r\\
0 & , k<r
\end{cases}
\] 
to recover \eqref{117} and hence the OPE \eqref{31}. This calculation thus demonstrates the validity of the regularized sum in \eqref{114} in the distributional sense.

\medskip

For the special case $l=0$, \eqref{117} reduces to:
\begin{align*}
-i\theta\left(u-u_p\right)\sum_{r=0}^k\sum_{m=0}^r\frac{\left(\bar{z}-\bar{z}_p\right)^{m+1}}{\left(z-z_p\right)\cdot m!}\text{ }\frac{\left(u-u_p\right)^{k-r}}{(k-r)!}\text{ }\bar{\partial}_p^m\partial_{u_p}^{1-r} S^+_{0}\left(u_p,z_p,\bar{z}_p\right)
\end{align*}
where we have used the following relation: $\partial_{u_p}^{-r} S^+_{0}\left(u_p\right)=S^+_{r}\left(u_p\right)$ for $r\geq-1$, as seen from the correspondence \eqref{112}. This in turn leads to the special case of the OPE \eqref{38} with the conformal Carrollian primary $\Phi$ taken to be the graviton $S^+_0$ (so that $2\bar{h}=-1$). Extending the derivation of \eqref{117}$_{l=0}$ to the case of the Einstein-Yang-Mills collinear splitting \eqref{113}, one obtains exactly the OPE \eqref{38} if the following identifications, valid as Carrollian correlator/bulk massless $\mathcal{S}$-matrix insertions, are made (suppressing $z,\bar{z}$):
\begin{align}
&{\Phi}(u)=i\int\limits_{-\infty}^\infty \frac{d\Omega}{2\pi}\text{ }e^{-i\Omega u}\text{ }a_{s}(\Omega)\hspace{2mm}\Longrightarrow\hspace{2mm}\Phi_k(u):=\partial_{u}^{-k} \Phi(u)=i^{k+1}\int\limits_{-\infty}^\infty \frac{d\Omega}{2\pi}\text{ }\frac{e^{-i\Omega u}}{\Omega^k}\text{ }a_{s}(\Omega)\nonumber\\
\Longrightarrow\hspace{2mm}&ia_{s}(\Omega)=\int\limits_{-\infty}^\infty {du}\text{ }e^{i\Omega u}\text{ }\Phi(u):=\tilde{\Phi}(\Omega)\hspace{10mm}(\text{for }\Omega\in\mathbb{R}-\{0\})
\end{align}
where $a_{s}(\Omega>0)=a^{\text{out}}_{s}(\omega>0)$ and $a_{s}(\Omega<0)={a^{\text{in}}_{-s}}^{\dagger}(\omega>0)$ with $s$ being the helicity of the massless bulk field $\tilde{\Phi}$ in the `outgoing' convention.\footnote{Had we used the unilateral Fourier correspondence \cite{Donnay:2022aba,Donnay:2022wvx}, valid as correlator/$\mathcal{S}$-matrix insertion:
\begin{align*}
{\Phi}^{\epsilon=1}(u,z,\bar{z})=i\int\limits_{0}^\infty \frac{d\omega}{2\pi}\text{ }e^{-i\omega u}\text{ }a^{\text{out}}_{s}(\omega,z,\bar{z})\hspace{5mm};\hspace{5mm}{\Phi}^{\epsilon=-1}(u,z,\bar{z})=i\int\limits_{0}^\infty \frac{d\omega}{2\pi}\text{ }e^{i\omega u}\text{ }{a^{\text{in}}_{-s}}^{\dagger}(\omega,z,\bar{z})
\end{align*}
we would have not recovered the OPE \eqref{38} at finite $u$ because the lower limit of the $\omega_p$-integral would have shifted from $0$ in the second line to $\Omega$ in the last line of \eqref{115}, spoiling the derivation of \eqref{117}. (Compare with Footnote \ref{extendedFourier}.) However, the $u\rightarrow\infty$ limit of the OPE \eqref{38} can still be obtained using the original correspondence of \cite{Donnay:2022aba,Donnay:2022wvx} for the field $\Phi$.} The above identifications thus generalize the correspondence \eqref{128} for the graviton to bulk massless Einstein-Yang-Mills fields of other helicities. 

\medskip

A comment regarding the similar derivation in \cite{Mason:2023mti} of the conformal Carrollian OPE starting from the collinear splitting \eqref{88} of one outgoing positive helicity graviton and another (physically) outgoing Einstein-Yang-Mills particle, and using the unilateral Fourier correspondence \eqref{eq:AtoC} between the Carrollian correlator/massless amplitude of \cite{Donnay:2022aba,Donnay:2022wvx} is in order. Though the coefficient of the most singular term in the Carrollian graviton OPE $C_{zz}(u,z,\bar{z})C_{zz}(u_p,z_p,\bar{z}_p)$ so obtained was found divergent, the descendant OPE $\partial_u C_{zz}(\mathbf{x})\partial_{u_p} C_{zz}(\mathbf{x}_p)$ turned out to be well-defined (finite) \cite{Mason:2023mti}. Defining the Carrollian fields $C^+_{k\geq0}(u,z,\bar{z})$ below, in analogy with the $S^+_k$ fields \eqref{62}:
\begin{align*}
C^+_k(u,z,\bar{z})=-\frac{2\pi}{k!}\int\limits_{-\infty}^udu^\prime\text{ }(u-u^\prime)^k\text{ }\partial_{u^\prime}C_{zz}(u^\prime,z,\bar{z})
\end{align*}
one however finds, using the said finite, descendant OPE, that all such $C^+_k(\mathbf{x})C^+_l(\mathbf{x}_p)$ OPEs at finite $u$ diverge. The finiteness of the $S^+_kS^+_l$ OPEs at finite $u$ \eqref{31}, in view of the correspondence \eqref{62}, may then be attributed to a remarkable cancellation between the divergent pieces of the four OPEs: $C^+_kC^+_l$, $C^+_kD^+_l$, $D^+_kC^+_l$ and $D^+_kD^+_l$. 

\medskip

On the contrary, all the $C^+_{k(e)}(u\rightarrow\infty,z,\bar{z})\Phi(\mathbf{x}_p)$ `OPEs' were shown to be finite in \cite{Mason:2023mti}, with the object $C^+_{k(e)}$ defined analogously to \eqref{63} as:
\begin{align*}
-\frac{1}{2\pi}C^+_{k(e)}(u,z,\bar{z})=\frac{(-)^k}{k!}\int\limits_{-\infty}^udu^\prime\text{ }{u^\prime}^k\text{ }\partial_{u^\prime}C_{zz}(u^\prime,z,\bar{z})
\end{align*}
just as the $S^+_{k(e)}(u\rightarrow\infty,z,\bar{z})\Phi(\mathbf{x}_p)$ `OPEs' extracted from \eqref{38}. Following the algorithm of \cite{Mason:2023mti}, it can also be shown that the $C^+_{1-k(e)}(u\rightarrow\infty,z,\bar{z})C^+_{1-l(e)}(u_p\rightarrow\infty,z_p,\bar{z}_p)$ `OPE' is same (upto a constant factor) as the celestial conformally soft graviton OPE $H^k(z,\bar{z})H^l(z_p,\bar{z}_p)$ \cite{Guevara:2021abz} encoding the $Lw_{1+\infty}$ symmetry. This is to be contrasted with the conformal Carrollian OPE $S^+_kS^+_l$ \eqref{31} at finite $u, u_p$ (and also the $S^+_{k(e)}(u\rightarrow\infty,z,\bar{z})S^+_{l(e)}(u_p\rightarrow\infty,z_p,\bar{z}_p)$ `OPE') containing exactly the same quantum symmetry algebra, that was derived intrinsically from Carrollian symmetry arguments in \cite{Saha:2023abr} or above using the bilateral Fourier correspondence \eqref{112} on the Einstein-Yang-Mills graviton collinear splitting \eqref{116}.

\addcontentsline{toc}{section}{References}
\bibliographystyle{style}
\bibliography{Biblio}

\end{document}